\documentclass[12pt]{article}
\usepackage{amsmath,amssymb,epsfig,amsfonts}
\usepackage{graphicx,subfigure}
\usepackage{cite}

\makeatletter

\newcommand{\Pfaff}{A}
\newcommand{\bZ}{\mathbb{Z}}

\newcommand{\bP}{\mathbb{P}}

\newcommand{\bF}{\mathbb{F}}

\newcommand{\cE}{\mathcal{E}}
\newcommand{\cN}{\mathcal{N}}

\newcommand{\cR}{\mathcal{R}}

\newcommand{\cO}{\mathcal{O}}

\newcommand{\cL}{\mathcal{L}}
\newcommand{\cC}{\mathcal{C}}

\newcommand{\unnumsubsection}[1]{\refstepcounter{subsection}\section*{#1}}
\newcommand{\unnumsubsubsection}[1]{\refstepcounter{subsubsection}\section*{#1}}

\def\ov#1{{\overline{#1}}}

\def\unit{{1\kern-.65ex {\rm l}}}
\def\1{{1\kern-.65ex {\rm l}}}

\def\CA{{\cal A}}

\def\CC{{\cal C}}
\def\CD{{\cal D}}
\def\CE{{\cal E}}

\def\CG{{\cal G}}

\def\CJ{{\cal J}}

\def\CL{{\cal L}}

\def\CN{{\cal N}}
\def\CO{{\cal O}}

\def\CW{{\cal W}}
\def\CX{{\cal X}}
\def\CY{{\cal Y}}
\def\CZ{{\cal Z}}

\def\cs{c_{1,B_2}}
\def\cb{c_{1,B_3}}
\def\pfaff{\text{Pfaff}(\Sigma,D_{-,V_{\Sigma}})}
\def\pfaffHet{\text{Pfaff}(c_{Het},D_{-,\CL_{\Gamma}})}

\newcount\hour \newcount\minute
\hour=\time \divide \hour by 60
\minute=\time
\def\now{%
\ifnum \hour<13
  \ifnum \hour=0 \advance \hour by 12 \number\hour:\else \number\hour:\fi%
     \ifnum \minute<10 0\fi%
     \number\minute%
\ A.M.%
\else \advance \hour by -12 \number\hour:%
  \ifnum \minute<10 0\fi%
  \number\minute%
  \ P.M.%
\fi%
}

\makeatother

\begin{document}

\baselineskip=18pt  
\numberwithin{equation}{section}  
\allowdisplaybreaks  

\thispagestyle{empty}

\vspace*{-3.5cm} 
\begin{flushright}
{\tt NSF-KITP-12-170}\\
{\tt UPR-1240-T} \\
{\tt EFI-12-23}
\end{flushright}

\vspace*{0.3cm} 
\begin{center}
 {\LARGE On Seven-Brane Dependent \\ Instanton Prefactors in F-theory}
 \vspace*{1.0cm}

Mirjam Cveti{\v c}$^{1,2}$, Ron Donagi$^3$, James Halverson$^{1,4}$, and  Joseph~Marsano$^{5,6}$ \\
\vspace*{1.0cm} 

{\small
$^1$ Department of Physics and Astronomy, University of Pennsylvania\\
209 S. 33rd St, Philadelphia, PA 19104 USA \\ \vspace{.25cm} 

$^2$ Center for Applied Mathematics and Theoretical Physics\\
University of Maribor, Maribor, Slovenia \vspace{.25cm} 

$^3$ Department of Mathematics, University of Pennsylvania\\
209 S. 33rd St, Philadelphia, PA 19104-6395 USA \\ \vspace{.25cm} 

$^4$ Kavli Institute for Theoretical Physics, University of California\\
Santa Barbara, CA 93106-4030 USA \\ \vspace{.25cm} 

$^5$ Enrico Fermi Institute, University of Chicago\\
5640 S Ellis Ave, Chicago, IL 60637, USA \\ \vspace{.25cm}

$^6$ Sunrise Futures LLC \\
30 S Wacker Suite 1706, Chicago, IL 60606
}
 \vspace*{.9cm} 

\noindent {\footnotesize \texttt{cvetic@cvetic.hep.upenn.edu, donagi@math.upenn.edu,\\jim@kitp.ucsb.edu, marsano@theory.caltech.edu}}
\vspace{.4cm}
\end{center}
\abstract{We study the moduli-dependent prefactor of M5-instanton
  corrections to the superpotential in four-dimensional F-theory
  compactifications. In light of the M-theory and type IIb limits and
  also heterotic duality, we propose that the explicit moduli
  dependence of the prefactor can be computed by a study of zero modes
  localized at intersections between the instanton and
  seven-branes. We present an instanton prefactor in an $E_6$ F-theory
  GUT which does not admit a heterotic dual and show that it vanishes
  if and only if a point of $E_8$ enhancement is present in the
  instanton worldvolume. More generically, we discuss the relationship
  between points of $E_8$ and superpotential zeroes and give
  sufficient conditions for such a point to cause a zero, even for an
  $SU(5)$ GUT. We scan a large class of compactifications for
  instanton physics and demonstrate that many instantons have the same
  prefactor structure. We discuss the associated implications and
  complications for moduli stabilization.  We present an explicit
  resolution and construction of G-flux in a generic $E_6$ GUT and
  identify a global compactification of the local model spectral cover
  which happens to facilitate prefactor computations. Via a Leray
  spectral sequence, we demonstrate the relationship between
  right-movers of heterotic worldsheet instantons, 3-3 strings of
  euclidean D3 instantons, and the Fermi zero modes of
  M5-instantons.}

\clearpage
\tableofcontents 

\section{Introduction} 

The importance of instanton corrections in string compactifications
was already recognized in the early days of the heterotic string
\cite{Wen:1985jz,Dine:1986zy,Dine:1987bq}, when it was found that
hierarchical Yukawa couplings in $E_6$ grand unified theories can be
generated by non-perturbative worldsheet instanton corrections to the
superpotential. In type II theories, euclidean D-brane instanton
corrections to the superpotential play a crucial role in the
stabilization of moduli \cite{Kachru:2003aw,Balasubramanian:2005zx}
and can generate charged matter couplings which are forbidden in
string perturbation theory by anomalous $U(1)$ symmetries
\cite{Blumenhagen:2006xt,Ibanez:2006da,Florea:2006si}.  Historically,
it has often been the case that an understanding of instanton effects
in one corner of the landscape sheds light on instanton effects in
another via duality or a special limit.  This includes, for example,
the rederivation of heterotic worldsheet instanton corrections via
duality with euclidean D1-instanton corrections\cite{Witten:1999eg}
and an understanding of euclidean D3-instantons in four-dimensional
compactifications of F-theory via three-dimensional compactifications
of M-theory \cite{Witten:1996bn}.

Over the last four years there has been renewed interest in $\cN=1$,
$d=4$ compactifications of F-theory, initiated by the possibility of
obtaining semi-realistic grand unified theories
\cite{Donagi:2008ca,Beasley:2008dc}. This approach to GUT model
building uniquely combines two advantages of the heterotic and type II
superstrings, giving rise to interesting GUT physics through the
presence of exceptional gauge structures while localizing the gauge
degrees of freedom in the internal space.  In addition to many works
on local model-building, there have been many attempts to construct
global models \cite{Andreas:2009uf, Marsano:2009ym, Collinucci:2009uh,
  Blumenhagen:2009up, Marsano:2009gv, Blumenhagen:2009yv,
  Marsano:2009wr, Grimm:2009yu, Cvetic:2010rq, Chen:2010ts,
  Chen:2010tp, Chung:2010bn,
  Chen:2010tg,Knapp:2011wk,Knapp:2011ip,Marsano:2012yc} and properly
understand chirality inducing $G_4$-flux \cite{Marsano:2010ix,
  Marsano:2011nn, Braun:2011zm, Marsano:2011hv, Krause:2011xj,
  Grimm:2011fx, Braun:2012nk, Kuntzler:2012bu, Krause:2012he,
  Collinucci:2010gz, Collinucci:2012as} and abelian symmetries
\cite{Grimm:2010ez, Dolan:2011iu, Marsano:2011nn, Grimm:2011tb,
  Morrison:2012ei}. A proper understanding of chirality inducing flux
and abelian symmetries is crucial for the understanding of instanton
physics in the weakly coupled type IIb limit, and given the recent
progress on these issues in F-theory one would hope to understand
F-theoretic instantons in this light.

In this work we study M5-instantons\footnote{We choose to call these
M5-instantons, since their behavior is most fundamentally captured by
M5-instantons in the defining $d=3$ M-theory compactification. 
The instantons in F-theory are actually euclidean D3's, and the relationship to
M5-instantons can be made concrete in the geometry. } in F-theory, focusing
on the dependence of the instanton prefactor on seven-brane moduli.
Lacking a fundamental definition of F-theory, any understanding of
F-theoretic physics must be obtained by comparison to the M-theory limit, and
also to a heterotic dual or the IIb limit when they exist. This applies to
the study of instanton corrections, and to the authors'
knowledge all works on the subject have utilized this perspective.
This is particularly evident in the early work on the subject.  For
example, in \cite{Witten:1999eg} the well known arithmetic genus
constraint $\chi(D,\cO_D)=1$ on divisors wrapped by instantons which
correct the superpotential was derived via an understanding of
M5-instantons in $d=3$ compactifications of M-theory. It was found in
\cite{Donagi:1996yf} that in a very special Calabi-Yau fourfold an
infinite number of divisors satisfying the arithmetic genus constraint
exist, giving rise to an infinite number of superpotential corrections
which exhibit an $E_8$ symmetry\footnote{This is not to be confused
with the points of $E_8$ enhancement discussed in this work.}. It also is known
\cite{Katz:1996th,Diaconescu:1998ua} that M5-instantons in $d=3$
M-theory play a crucial role in recovering the physics of $d=4$
supersymmetric QCD in the F-theory limit. Via the M-theory limit,
it was shown in \cite{Ganor:1996pe} that superpotential zeroes occur
when the moduli of spacetime filling D3-branes are tuned such that
the D3-brane comes into contact with the instanton, due to the
appearance of additional zero modes. The appearance of zero
modes at codimension one subloci in moduli space will be the salient feature determining
instanton physics in this paper, where the modes of relevance are
essentially ED3-7 zero modes from the type IIb point of view.

Since the appearance of \cite{Donagi:2008ca,Beasley:2008dc}, there
have been many works studying instantons in F-theory. In
\cite{Blumenhagen:2010ja} it was shown how fermionic instanton zero
modes in type IIb uplift to F-theory and the relationship of these
modes to the arithmetic genus. Spectral cover techniques were used to
describe instanton intersections with D7-branes.  In
\cite{Cvetic:2010rq} an instanton in a semi-realistic type IIb GUT was
uplifted to F-theory and the cohomology determining the arithmetic
genus was studied. Instantons in F-theory were discussed via dualities
and limits in \cite{Donagi:2010pd} and it was shown that the
worldvolume theory of an M5-brane instanton reduces to an MSW CFT
\cite{Maldacena:1997de} if it is a particular surface fibration over a
curve.  In \cite{Grimm:2011dj} it was shown that instanton flux in
F-theory can alleviate the tension \cite{Blumenhagen:2007sm} which can
occur between chiral matter instanton prefactors and moduli
stabilization in the type IIb limit. In \cite{Marsano:2011nn}
$G_4$-flux and abelian symmetries were discussed in the spectral
divisor formalism and some associated aspects of instanton physics
were addressed. In \cite{Cvetic:2011gp} instanton zero modes in
F-theory were studied via anomaly inflow and $(p,q)$-string junctions,
and (in examples) some physics of the moduli dependent Pfaffian
prefactor were understood via heterotic duality. It is known that
worldvolume fluxes on the instanton can lift fermionic zero modes
which would prevent the instanton from contributing to the
superpotential \cite{Bianchi:2011qh}. In \cite{Kerstan:2012cy} it was
shown that comparison to the type IIb limit can determine which theta
divisor on the intermediate Jacobian corresponds to the prefactor of
an M5-instanton correction.

In this paper we will study the computation and physical understanding
of the moduli-dependent instanton prefactor via dualities and limits
of F-theory. We will cast the discussion of instanton prefactors into
the language of the M-theory and heterotic duals \cite{Vafa:1996xn},
and also the weakly coupled type IIb limit.  We will not address the
dependence of the instanton prefactor on spacetime filling D3-branes
\cite{Ganor:1996pe,Baumann:2006th}, but instead will focus solely on
the seven-brane moduli dependence. While it is important to consider
any moduli dependence in a string compactification, we find the
seven-brane moduli dependence to be particularly interesting since it
gives a connection between the effective potential on moduli space and
four-dimensional gauge theories.

Let us briefly review the picture of ED3-instantons in type IIb, since
much of the intuition is expected to carry over to F-theory.  Consider
a euclidean D3-brane instanton wrapping a divisor $D$ with associated
K\" ahler modulus $T$ in a type IIb orientifold compactification on a
Calabi-Yau threefold. Under certain circumstances, the ED3 can correct
the superpotential, and the structure of the correction depends
heavily on the fermionic instanton zero modes.  At a given
point in the D7-brane moduli space of a type IIb compactification, the
coupling of fermionic ED3-D7 zero modes to matter fields charged under
the four-dimensional gauge group can generate superpotential
corrections with prefactors that contain the charged fields. The
correction takes the schematic form \cite{Blumenhagen:2006xt,
  Ibanez:2006da, Florea:2006si}
\begin{equation} \Delta W \sim \prod_i \Phi_i^{k_i} \, e^{-T}
\end{equation} where the $\Phi_i$ are chiral supermultiplets charged
under $G_{4d}$. Such effects are
crucial for phenomenological considerations, since
instantons can generate leading order contributions to
couplings which are forbidden in perturbation theory by
anomalous $U(1)$ symmetries. For example, at weak coupling the
$10_M\, 10_M\, 5_H$ of an $SU(5)$ GUT and Majorana right-handed
neutrino mass terms are always forbidden, but can be generated by instantons
\cite{Blumenhagen:2006xt, Ibanez:2006da, Blumenhagen:2007zk}. While 
corrections with matter prefactors are very interesting, we will not study them here.

In the absence of these ED3-D7 zero modes, the prefactor is a
constant determined by the vacuum expectation values of the
relevant moduli, including those determining the structure of
seven-branes. The superpotential correction 
takes the schematic form
\begin{equation} \Delta W \sim A\, e^{-T},
\end{equation} 
and in this paper we will study the prefactor\footnote{In type IIb and
  F-theory, the prefactor is in fact the Pfaffian of a chiral Dirac
  operator representing one loop fluctuations around the instanton
  background, but we will utilize an equivalent algebraic
  description\cite{Buchbinder:2002ic} which seems to be more
  tractable.} $A$ as a function of seven-brane moduli, denoted
collectively as $\phi$. The functional form of $A(\phi)$ is determined
by the spectrum of zero modes, which can change upon movement in
moduli space as additional modes become massless. This can be studied
concretely via techniques in algebraic geometry, where the modes are
counted by bundle cohomology and the corresponding Hodge numbers can
jump at subloci in moduli space. It goes without saying that
understanding this structure is important for the physics of
four-dimensional string compactifications. 

We will be interested in the study of $A(\phi)$ for M5-instantons in
F-theory. Qualitatively, there are two issues which we find
particularly interesting, and we will focus on those.  The first is to
identify the physics which determines $A$ in a class of GUT
compactifications, and in the process to give a method for the
explicit computation of $A$ as a function of seven-brane moduli, which
are geometrized in F-theory as complex structure moduli. We will see
that the relevant zero modes are counted by the cohomology of a line
bundle $\cL_A$ on a distinguished spectral curve
$c_{loc}$. From a type IIb perspective, these zero modes can be
thought of as ED3-D7 strings localized at curves where the instanton intersects
the GUT seven-brane.  Then the 7-brane dependent prefactor $A(\phi)$
has a zero whenever $\cL_A$ obtains a section. A non-trivial
polynomial $A(\phi)$ corresponds to the situation where $\cL_A$ does
not have a section for generic moduli, but obtains a section in
codimension one, namely on the locus $A(\phi)=0$.  We will see that
this prescription matches nicely with expectations from heterotic
duality and the type IIb limit, and we will also discuss the
relationship to the M5-brane worldvolume theory.

The second issue we would like to discuss regards the structure of
$A$. In all known examples, the polynomial $A$ takes the schematic
form
\begin{equation} A = \prod_i f_i^{k_i}
\end{equation} for some other polynomials $f_i$ and integers $k_i$, so
that $A=0$ is a reducible codimension one subvariety in moduli
space. We would like to study its components defined by $f_i=0$ and
address whether these loci correspond to interesting seven-brane
physics. Such a correspondence should not be surprising, since zeroes
of $A$ occur when extra modes become light, which often occurs at
points of singularity enhancement in seven-brane moduli space. One
aspect we will study is the connection between superpotential
zeroes and points of $E_8$ enhancement\footnote{It is known \cite{Esole:2011sm} that
  one must be careful in the application of Kodaira's codimension one
  classification of singularities to codimension three
  singularities. The ``$E_8$'' points here correspond to specific
  tunings of moduli VEVs, rather than the appearance of an extended $E_8$
  Dynkin diagram after resolution of the codimension three singularity.}, which are of 
phenomenological interest \cite{Heckman:2009mn}. It would
also be interesting to address the powers $k_i$, though we will only say a few words
about this here.  Some of these issues have been discussed in
the dual heterotic
picture \cite{Curio:2008cm,Curio:2009wn,Curio:2010hd} and our thoughts
on the associated physics in F-theory have been influenced by
those works. However, here we are primarily interested in F-theory
physics and will take the seven-brane perspective.

The moduli dependence of the instanton prefactor $A$ enters the effective
potential on moduli space, and it would be interesting to study issues
of moduli stabilization taking this dependence into account.
For any given modulus $\phi$, it has been argued that the contributions
to $V(\phi)$ from $A$ are subleading compared to contributions from
the flux superpotential. However, for a given set of fluxes it is not
guaranteed that the flux superpotential contributes to $V(\phi)$, and
therefore contributions to the potential from $A$ may be relevant.
For these contributions, it is worth noting that known examples have 
$k_i\ge 2$, so that $f_i=0$ loci are often critical points of the correction,
and perhaps of the full effective potential. 

\subsection{Summary of Results and Paper Layout}
Since this work builds on many others, let us review the contributions
which the authors believe are new and put them into their proper
context. We will first focus on new insights of this work, and then
turn to illustrative examples and new technical progress.

\vspace{.3cm}
\noindent \emph{Conceptual Insights}
\vspace{.3cm}

The primary result of this paper is a conjecture for the computation
of M5-instanton prefactors as a function of complex structure moduli
in four-dimensional F-theory compactifications, analogous to the
heterotic work of \cite{Buchbinder:2002ic}, but not requiring a
heterotic dual\footnote{Some aspects of instantons in 
compactifications without a heterotic dual
were also discussed in \cite{Donagi:2010pd}.}. The complex structure moduli dependence determines
seven brane structure and therefore has interesting implications for
four dimensional gauge theories.  Our result is applicable to four
dimensional grand unified theories where the Higgsing of an $E_8$
gauge theory on the GUT seven-brane can be described by an $SU(n)$
spectral cover. This includes, for example, the relevant cases $G_{4d}
= E_6, SO(10),$ and $SU(5)$, and is the context for the work on global
and local F-theory GUTs over the last four years.  For simplicity we
will sometimes specify to the $E_6$ case. Our result will utilize a
spectral curve $c_{loc}$ which is an $n$-sheeted cover of a curve
$\Sigma\subset B_3$ where the instanton divisor intersects the GUT
stack. Since $SU(n)$ spectral covers describe the Higgsing of a single
local non-abelian gauge theory, our result likely has to be modified
if another non-abelian singularity intersects the GUT stack. However,
in such a case we still expect bundle cohomology on $\Sigma$ to play
an important role, though it may not admit a spectral curve
description. In this work we will not study $U(1)$ symmetries, though
they likely play an interesting role in instanton physics when they
exist.  The moduli-dependence of the prefactor which we study is
governed by the presence or absence of zero modes localized on
$\Sigma$.  They can be counted by studying moduli-dependent vector
bundle cohomology on $\Sigma$.  Alternatively, which will be our
method of choice, the zero modes are counted by line bundle cohomology
on $c_{loc}$. In this case, the seven-brane moduli dependence of the
prefactor emerges through the moduli in the defining equation of
$c_{loc}$. The cohomology which determines the prefactor is intuitive
from the heterotic and type IIb perspectives.  From the type IIb
perspective, the localized modes are naturally interpreted as ED3-7
string zero modes.  When a heterotic dual exists, on the other hand,
the prefactor is known to be determined by line bundle cohomology on
another curve $c_{Het}$.  As we demonstrate, $c_{Het}$ and $c_{loc}$
are isomorphic whenever a heterotic dual exists.  This allows us to
understand more directly how line bundle cohomology on $c_{loc}$
determines the partition function of the chiral 2-form on the M5
worldvolume through simple application of the cylinder map, which
provides a map from the Jacobian $\CJ(c_{loc})$ to the intermediate
Jacobian $\CJ(M5)$.  Modulo some issues related to the familiar
half-integral shift, this maps the Riemann $\theta$ divisor of
$\CJ_{c_{loc}}$ to a $\theta$ divisor of $\CJ(M5)$ that determines the
2-form partition function.  In the absence of a heterotic dual we
still have the curve $c_{loc}$ but we no longer have the cylinder map.
Nevertheless, we believe that (part of) the M5 partition function
continues to be related to cohomology on $c_{loc}$ in a way that we
make precise with spectral divisors.  While we do not have a proof of
this conjecture, we find it mathematically plausible and physically
very well-motivated, particularly from the heterotic and type IIb
perspectives.  In the $SO(10)$ and $SU(5)$ case, it would also be
important to study modes counted by cohomology of the associated
bundles $\wedge^k V$ on $\Sigma$. We leave this for future work.
 
Since instanton corrections to the superpotential are important for moduli stabilization, 
we systematically study prefactors across a large class of F-theory compactifications.
We perform a scan of all toric threefolds related to fine triangulations of
three-dimensional reflexive polytopes (that is, the entire $d=3$ Kreuzer-Skarke
list \cite{Kreuzer:1998vb}) and study each as the base $B_3$ of an F-theory
compactification with $G_{4d}=E_6$. In each case, we study instantons subject to
certain criteria, keeping track of the topological quantities which are sufficient
to determine the instanton prefactor and tabulating the results.
Interestingly, we find that there are $\cO(10^5)$ instanton
corrections in these compactifications, yet only $\cO(10)$ distinct prefactor structures.
Of these possible prefactors, only three have been explicitly computed, but these three 
make up over half of the prefactors in our scan. Two of the prefactors are identically
zero, a very common possibility with clear (and difficult) implications for K\" ahler  moduli stabilization. 
The other vanishes if and only if there exists a point of $E_8$ enhancement
in the instanton worldvolume\footnote{By this we mean that there exists a singular fiber,
naively \cite{Esole:2011sm} of $E_8$ type, above a point in the instanton worldvolume.}. It would be interesting, though very computationally intensive,
to compute the remaining prefactors. It would also be interesting to broaden the scan
by looking at different flux parameters, four-dimensional gauge groups, or base manifolds.

It is interesting that points of $E_8$ enhancement seem to play a special role in instanton
corrections, since they are known to be of phenomenological relevance \cite{Heckman:2009mn},
and we explore this relationship in an example-independent way. Note that the arguments we
present can be made in both the heterotic computation on $c_{Het}$ or the F-theory computation
on $c_{loc}$. Building on related arguments in the heterotic
case \cite{Curio:2009wn}, we write down sufficient conditions under which a point of $E_8$ enhancement
in the instanton worldvolume will cause a prefactor zero. Such conditions are useful because
they allow one to study prefactor zeroes caused by points of $E_8$ without having to explicitly
perform the tedious prefactor computation.  The connection between $E_8$ points and
superpotential zeroes may not be surprising, since enhanced symmetry points in moduli
space often exhibit additional light modes, which could cause the
zero. While this is true and evident in examples, the arguments we present proving the
conditions related to $E_8$ do not apply to points of $E_7$ or $E_6$, for example. 
Thus $E_8$ points seem to be
special. It would be interesting to explore these ideas further. 

\vspace{.3cm}
\noindent \emph{Illustrative Examples}
\vspace{.3cm}

We discuss three illustrative examples, emphasizing the computation of
the prefactor via the study of line bundle cohomology on $c_{loc}$ and
also studying the factorization and substructure of the
prefactor. All examples have $G_{4d}=E_6$. In the first example, we
study the instanton prefactor in an F-theory compactification with a
heterotic dual. The prefactor takes the form $\Pfaff \sim
f_\Lambda^4$, where $f_\Lambda$ is a polynomial which obtains a zero
if and only if a particular degree $0$ bundle $\Lambda$ obtains a
section. The order of vanishing of $f_\Lambda$, $4$, is also the
number of zero modes which become massless on the $f_\Lambda=0$
locus. This example was studied from an F-theory perspective in
\cite{Cvetic:2011gp}, but here we have understood aspects of the
\emph{entire} vanishing locus, as opposed to subloci within the
$\Pfaff=0$ locus. That a related bundle controls a polynomial factor in
$\Pfaff$ is the notion of `reduction' \cite{Curio:2008cm} from the
heterotic instanton literature. In the second example, we study an instanton
prefactor in an F-theory geometry which does not admit a heterotic
dual. There we have $\Pfaff \sim f_{E8}^4$, where $f_{E8}=0$ if and only
if there exists a point of $E_8$ enhancement in the instanton worldvolume.
In the third example we utilize the same base as in the second example,
but change the location of the GUT stack $B_2$. We perform a new prefactor
computation and find that the prefactor is identically zero for all relevant
moduli.
These examples demonstrate how the instanton prefactor can be computed in
terms of intrinsically F-theoretic data, even when a heterotic dual does not
exist.

\vspace{.3cm}
\noindent \emph{Technical Progress}
\vspace{.3cm}

One point of technical progress significantly aids the
computation. Let us discuss it in generality before discussing the
relevance for instanton computations. Consider an $E_8$ gauge theory
on a seven-brane wrapped on a four-cycle $B_2$ in the F-theory base
$B_3$. The Higgs bundle spectral cover is a spectral pair
$(\cC_{loc},\cN_{loc})$ which determines the Higgsing of the $E_8$
gauge theory to the GUT group. $\cC_{loc}$ is an $n$-sheeted cover of
$B_2$ describing the $SU(n)$ bundle that Higgses $E_8$ and $\cN_{loc}$
is a line bundle on it. In previous work, a non-compact $\cC_{loc}$
was utilized to study F-theory GUTs, but here we demonstrate a natural
way to realize a compact $\cC_{loc}$. In short, $\CC_{loc}$ is a
surface in the Calabi-Yau fourfold which can be realized as a complete
intersection of two divisors in the fourth exceptional divisor $E_4$
of the resolved ambient space. $E_4$ is an $\bF_1$-fibration over
$B_2$ which contains an ambient elliptic $CY_3$ $Z_{3,F}$ which
contains $\CC_{loc}$ as a divisor. When a heterotic dual exists,
$Z_{3,F}$ is isomorphic to the heterotic $CY_3$ $Z_{Het}$ and
$\CC_{loc}$ is isomorphic to the heterotic spectral cover $\CC_{Het}$.
That the fourth exceptional divisor $E_4$ plays a special role in
obtaining a nice compactification of an $SU(n)$ spectral cover is not
surprising: if it sat in the third exceptional divisor $E_3$, then an
$SU(4)$ gauge symmetry would admit a spectral cover description, which
it does not. If, on the other hand, $\cC_{loc}$ sat in the fifth
exceptional divisor $E_5$, this compactification of the spectral
cover would describe cases of $SO(10)$ gauge symmetry, but not
$SU(5)$. Simply put, this compactification of the spectral cover
begins to exist when the spectral cover description of the gauge
theory begins to exist.

Let us discuss why this is useful for instanton computations.  With
this nice compactification of $\cC_{loc}$ as a divisor in an ambient
elliptic Calabi-Yau threefold, we can intersect with the instanton
divisor to obtain the spectral curve of interest, $c_{loc}$, as a
divisor in an elliptic surface $\cE$.  Then the line bundle cohomology
on $c_{loc}$ which determines the prefactor can be studied using a
Koszul sequence from line bundles on $\cE$. This method is mathematically 
identical
to that of \cite{Buchbinder:2002ic}, even though computing cohomology
on $c_{loc}$ in this way does \emph{not} require the existence of a
heterotic dual. In fact, given an F-theory base $B_3$ and two divisors
wrapped by the GUT stack and instanton, it is trivial to compute the
topological quantities which determine the structure of $c_{loc}$ and
the line bundle $\cL_A$, the cohomology of which must be computed to
determine the prefactor.  This allows for an efficient
systematic scan, as discussed.

In the end, we arrive at a prescription for computing (part of) the moduli dependence
of the instanton prefactor in a straightforward way using well-established techniques.
From a practical point of view in a given example, studying line bundle cohomology
on the spectral curve $c_{loc}$ is a much more straightforward procedure than the
analogous procedure for the M5-brane worldvolume theory. The
latter would involve explicitly understanding
the intermediate Jacobian of the $M5$, its moduli dependence, its theta divisors, and
then picking the `right' theta divisor \cite{Witten:1996hc}.

\vspace{1cm}
\noindent \emph{Paper Layout}
\vspace{.3cm}

Let us discuss the layout of this paper.  The reader interested only
in new results about instanton prefactors can skip to sections
\ref{sec:instantons in F} and later.  

In section \ref{sec:instantons in F} we propose that the seven-brane
moduli dependence of a component of $A(\phi)$ can be computed via a
study of the cohomology of a line bundle $\cL_A$ on a spectral curve
$c_{loc}$, which exists even when there is not a heterotic dual. We
discuss the relation of this method to heterotic duals and also the
M-theory and type IIb limits. We briefly describe the topological
computations needed to be done in order to set up the prefactor
computation. We describe in detail how the compact realization of
$\cC_{loc}$ in a natural ambient space aids the computation. In
section \ref{sec:scan} we perform a scan for instanton physics in a
broad class of F-theory $E_6$ GUTs, utilizing the entire
Kreuzer-Skarke database of three dimensional toric varieties as
F-theory bases $B_3$.  In section \ref{sec:E8} we examine the
relationship between points of $E_8$ enhancement and superpotential
zeroes in generality.  In section \ref{sec:examples} we discuss in
detail three examples of instanton prefactors in $E_6$ GUTs, setting
up the prefactor computation as described in section
\ref{sec:computing}. The first example has a heterotic dual, while the
last two do not.  In appendix \ref{app:resolutions} we resolve an
$E_6$ GUT in F-theory and present the compact Higgs bundle spectral
cover as a surface in the resolved fourfold.  In appendix \ref{app:B3}
we give a non-toric description of the base geometry in the second and
third examples of section \ref{sec:examples}.

Sections
\ref{sec:het worldsheets}, \ref{sec:FM5}, and \ref{sec:hetf section} primarily
serve to remind the reader of important details of worldsheet
instantons in the heterotic string and M5-instantons in F-theory which
will be necessary for framing the new results in this work. In section
\ref{sec:het worldsheets} we review facts about heterotic worldsheet
instanton zero modes. The instanton prefactor includes the
Pfaffian of a chiral Dirac
operator which depends on the gauge bundle $V$ and therefore it's
moduli.  It can be computed in an algebraic description in terms of
the moduli which determine a spectral curve $c_{Het}$ and
appear as complex structure (7-brane) moduli in F-theory
duals. We discuss
the relationship between the prefactor and a theta divisor on the
Jacobian $\CJ(c_{Het})$. 
In section \ref{sec:FM5} we review the spectral divisor formalism for describing
gauge degrees of freedom in F-theory and present our
compact realization of the Higgs bundle spectral cover $\CC_{loc}$.
We review details of the M5-brane worldvolume theory and discuss the
importance of Fermi zero modes and the chiral 2-form for instanton
corrections.
In section \ref{sec:hetf section} we describe the relationship between
heterotic worldsheet instantons and their dual M5-instantons in F-theory.

\subsection{Notational Conventions}
\label{app:notation}

For convenience, let us collect the notation used throughout the paper.
We will address the notation of each section in order.

In section \ref{sec:het worldsheets} we use the following notation to
describe $d=4$ $\cN=1$ heterotic compactifications with worldsheet
instanton corrections. The heterotic Calabi-Yau threefold $Z_{Het}$ is
an elliptic fibration $\pi_{Het}:Z_{Het} \rightarrow B_2$. The
holomorphic vector bundle $V$ breaks one $E_8$ factor down to some
gauge group $G_{4d}$, often $E_6$ in examples. We realize $Z_{Het}$ as
a Weierstrass model, specified as a hypersurface in a bundle over
$B_2$ whose fiber is $\bP_{1,2,3}$ with
homogeneous coordinates $[v,x,y]$.  The bundle $V$ is specified by a
spectral cover $\CC_{Het}$ and a line bundle $\cN_{Het}$ on it. $\CC_{Het}$
is a multi-sheeted cover with covering map $p_{Het}:\CC_{Het}\rightarrow B_2$. 

We wrap a worldsheet instanton on a curve $\Sigma$
in $B_2$.  The
component of the instanton prefactor we study vanishes\footnote{Since $\Sigma$ is a $\bP^1$, we may write $\cO_\Sigma(-1)$
for $K_\Sigma^{1/2}$ throughout.}
if and only if $h^i(\Sigma,V|_\Sigma \otimes \cO_\Sigma(-1)) \ne
0$. $\cE\equiv \pi_{Het}^{-1} \Sigma $ is an elliptic surface, and
$V|_{\cE}$ is specified as a spectral curve $c_{Het} \equiv
\CC_{Het}|_\cE$ and a line bundle $\cN_{c,Het}\equiv
\cN_{Het}|_\cE$. $c_{Het}$ is a multi-sheeted cover with covering map
$p_{c,Het}:c_{Het} \rightarrow \Sigma$. The condition regarding
prefactor zeroes can be equivalently stated in terms of
$h^i(c_{Het},\cN_{c,Het}\otimes p^*_{c,Het}K_\Sigma^{1/2})\equiv h^i(c_{Het},\cL_A)$\footnote{Throughout,
we use $\cL_A$ to denote the line bundle on a spectral curve whose cohomology controls the prefactor $A$,
regardless of whether that curve is $c_{Het}$ or $c_{loc}$.}. This
cohomology can be computed via a Koszul sequence from $\cE$.
Moduli dependence enters through the defining equation $f_{c_{Het}}$
appearing as the injective map in the Koszul sequence.

In section \ref{sec:FM5} we use the following notation to describe
$d=4$ $\cN=1$ compactifications of F-theory with M5-instanton
corrections. The F-theory Calabi-Yau fourfold $Y_4$ is an elliptic
fibration $\pi:Y_4\rightarrow B_3$. We specify $Y_4$ as a hypersurface
in a $\bP_{1,2,3}$ bundle over $B_3$. We wrap a GUT 7-brane on a
divisor $B_2$ which is a component of the discriminant in $B_3$. We
study a spectral divisor $\CC_F$ inside $Y_4$ which is the cylinder in
the case of heterotic duality and is useful for specifying a
dictionary between local and global F-theory models.  A chirality
inducing M-theory flux $G_4$ can be determined from a choice of line
bundle $\cN_F$ on $\CC_F$.  More specifically, $c_1(\cN_F)$ yields a
$(1,1)$-form whose push forward into $Y_4$ determines $G_4$ up to a
specific half-integral shift that we describe in great detail.  The
line bundle $\CN_F$ itself can carry more information than its first
Chern class if $\CC_F$ is not simply connected and, in such cases,
this extra information pushes forward to extra data of $C_3$ that is
not determined by $G_4$.  $\CC_F$ is singular in $Y_4$ but becomes a
smooth divisor $\tilde\CC_F$ in the resolved fourfold $\tilde
Y_4$. The gauge theory on the GUT stack can be described in terms of a
Higgs bundle spectral cover $\CC_{loc}$ and a line bundle $\CN_{loc}$
on it. $\CC_{loc}$ is a multi-sheeted cover with covering map
$p_{loc}:\CC_{loc}\rightarrow B_2$ and can be obtained from the
spectral divisor as $\CC_{loc}=\tilde \CC_F|_{\pi^*B_2}$. The line
bundle $\CN_{loc}$ can be obtained via restriction
$\CN_{loc}\equiv\CN_F|_{\CC_{loc}}$.  We show that $\CC_{loc}$ can
also be realized as a divisor in an auxiliary elliptic Calabi-Yau
threefold $Z_{3,F}$ with projection map $\pi_{3,F}:Z_{3,F}\rightarrow
B_2$.  The threefold $Z_{3,F}$ and spectral cover $\CC_{loc}$ sit
inside the resolution of the GUT singular locus and are isomorphic to
the threefold $Z_{Het}$ and spectral cover $\CC_{Het}$ when a
heterotic dual exists. $Z_{3,F}$ and $\CC_{loc}$ also exist in the
absence of a heterotic dual, however.

We consider instantons in F-theory as a limit of instantons in $d=3$
$\cN=2$ compactifications of M-theory. In M-theory on $Y_4$, we
consider M5-instantons on divisor $\pi^*D3$ for surfaces $D3$ in
$B_3$, where the instanton in F-theory is a D3-instanton on
$D3$.   From the perspective of the M5-brane worldvolume theory,
the prefactor of an M5-instanton correction is the partition function $\CZ_\phi$
of the chiral 2-form field $\phi_2$. It is a $\theta$ function on the intermediate
Jacobian $\CJ(M5)$. 
Thought of as ED3-instantons in type IIb, we study the component of the prefactor controlled
by bundle cohomology on 3-7 intersection curves.

In section \ref{sec:hetf section} we discuss instantons under heterotic / F-theory duality.
In such a case, $B_3$ is a $\bP^1$ fibration over $B_2$, so that $Y_4$ is globally $K3$-fibered
over $B_2$. One must also take the stable degeneration limit, in which the $K3$-fibration
splits into two $dP_9$-fibrations glued along a common elliptic fibration, the heterotic
threefold $Z_{Het}$. We focus on one $dP_9$-fibration $Y_4^{'}$. Its base is $\bP^1$-fibered
 and the two sections (copies of $B_2$) are the GUT stack at $Z=0$ and the heterotic threefold base
$B_2$ at $W=0$, where $(Z,W)$ are the homogeneous coordinates on the $\bP^1$ fiber. In the case of heterotic / F-theory duality, the spectral divisor
$\CC_F$ is the cylinder $\CC_{cyl}$, which is a $\bP^1$ fibration over $\CC_{Het}=\CC_{loc}$. 
The associated projection map(s) is $p_{cyl}$. $\CC_{cyl}$ is equipped with a line bundle
$\cN_{cyl}$ and the pair $(\CC_{cyl},\cN_{cyl})$ maps to $(\CC_{Het},\cN_{Het})$ or
$(\CC_{loc},\cN_{loc})$ upon restriction to $W=0$ and $Z=0$, respectively. Thus the cylinder
and its line bundle give a map for translating data from the heterotic geometry to the
local $E_6$ geometry in F-theory.  

A worldsheet instanton on $\Sigma$ is dual to an M5-instanton on the pullback
of $\Sigma$ under the $K3$-fibration. Focusing on the zero modes of only the bundle $V$
corresponds to studying one of the $dP_9$ fibers in the stable degeneration limit, so we study $M5\equiv \rho^*\Sigma$
where $\rho:Y_4^{'}\rightarrow B_2$. $M5$ is itself an elliptic fibration 
$\pi_{M5}: M5 \rightarrow D3$
and in the case of duality $D3$ is $\bP^1$-fibered $\nu:D3\rightarrow \Sigma$.  Restricting
the cylinder, spectral covers, and associated bundles to the $M5$ gives a miniature version
of the cylinder, a surface $c_{cyl}$ with bundle $\cN_{M5,cyl}$ which restricts to pairs
of spectral curves
$(c_{Het},\cN_{c,Het})$ and $(c_{loc},\cN_{c,loc})$ at $W=0$ and $Z=0$ respectively.
$c_{cyl}$ is a multi-sheeted cover of $D3$ with $p_{M5}: c_{cyl}\rightarrow D3$. $c_{Het}$ and 
$c_{loc}$ are both multi-sheeted covers of $\Sigma$, the first in the heterotic geometry and
the inside the GUT singular locus. The $\cL_A$ cohomologies on $c_{loc}$ and $c_{Het}$ are isomorphic,
and thus the prefactor can be determined by studying cohomology on either curve.

In section \ref{sec:instantons in F} we discuss the physics and interpretation of
the moduli-dependent cohomology $h^i(c_{loc},\cN_{c,loc}\otimes
p^*_{c,loc}K_\Sigma^{1/2})\equiv h^i(c_{loc},\cL_A)$ and conjecture
that it (and similar cohomologies of associated bundles $\wedge^k V$)
can be used to determine the instanton prefactor in terms of algebraic
seven-brane moduli.  $[c_{loc}]$ is a
divisor in an elliptic surface $\cE\equiv \pi_{3,F}^{-1}\Sigma$ of
class $ns+rF$ where $n=rk(V)$, $s$ is the section of $\cE$, $r\equiv
\eta \cdot_{B_2} \Sigma$, $\eta = 6\cs + N_{B_2|B_3}$, and $F$ is the
fiber class of $\cE$. $\cL_A$ is a bundle on $c_{loc}$ and is obtained via
restriction from a line bundle (also called $\cL_A$, abusing notation)
on $\cE$ of first Chern class $c_1(\cL_A) = (\lambda + \frac{1}{2})n\,
s + [r(\frac{1}{2} - \lambda) + \chi(\frac{1}{2} + n\lambda)-1)]F$,
where $\chi \equiv \cs \cdot_{B_2} \Sigma$ and $\lambda$ is a flux parameter,
integral or half-integral depending on $n$. Given the data $\{n,\lambda,r,\chi\}$
the mathematical computation can go forward in a way identical to the
heterotic computations of \cite{Buchbinder:2002ic}.  $h^i(c_{loc},\cN_{c,loc}\otimes p^*_{c,loc}K_\Sigma^{1/2})\equiv h^i(c_{loc},\cL_A)$
exists whether or not there is a heterotic dual.

\section{Heterotic Worldsheet Instantons}
\label{sec:het worldsheets} 

We begin in this section by reviewing background material on worldsheet instantons in $E_8\times E_8$ heterotic strings.  This will allow us to set some notation and frame the discussion in a way that closely parallels our subsequent study of M5 instantons.

We consider the compactification of $E_8\times E_8$ heterotic strings on an elliptically fibered Calabi-Yau 3-fold $Z_{Het}$ with section over a base $B_2$
\begin{equation}\pi_{Het}:Z_{Het}\rightarrow B_2\end{equation}
For simplicity, we restrict our attention to one $E_8$ factor and introduce an $SU(3)$ bundle $V$ to break this $E_8$ down to $E_6$.  Insights into instantons gained in this case can be extended
to other GUT groups via standard techniques. Our interest is in the superpotential couplings generated by worldsheet instantons that wrap curves in $B_2$ with particular attention to their dependence on $V$.

The 3-fold $Z_{Het}$ can be realized as a Weierstrass model, which we specify in this paper as a hypersurface inside a $\mathbb{P}^2_{1,2,3}$ bundle over $B_2$.  Letting $[v,x,y]$ denote the weighted homogeneous coordinates of the $\mathbb{P}^2_{1,2,3}$ fiber, we write the defining equation as
\begin{equation}y^2 = x^3 + fxv^4 + gv^6.\label{zhet}\end{equation}
The bundle $V$ is constructed via a Fourier-Mukai transform from a spectral cover
\begin{equation}\CC_{Het}:\,\,\,a_0 v^3 + a_2 vx + a_3 y=0\label{chet}\end{equation}
and a choice of line bundle $\CN_{Het}$ on $\CC_{Het}$.  In writing \eqref{zhet} and \eqref{chet} we have introduced a number of holomorphic sections on $Z_{Het}$ which are associated to the bundles
\begin{equation}\begin{array}{c|c}
\text{Section} & \text{Bundle} \\ \hline
v & \CO(\sigma_{Het}) \\
x & \CO(2[\sigma_{Het}+\cs]) \\
y & \CO(3[\sigma_{Het}+\cs]) \\
a_0 & \CO(\eta) \\
a_2 & \CO(\eta -2\cs) \\
a_3 & \CO(\eta-3\cs).
\end{array}\label{hetbundles}\end{equation}
Here, $\sigma_{Het}$ is the divisor class of the section of $Z_{Het}$, $\cs$ is shorthand for the anti-canonical class of $B_2$, and $\eta$ is a divisor class on $B_2$ that we are free to choose subject to the condition that all bundles in \eqref{hetbundles} actually admit holomorphic sections.  Note that we do not notationally distinguish between divisor classes on $B_2$ or their pullbacks to $Z_{Het}$.

To ensure that the bundle $V$ obtained from \eqref{chet} and $\CN_{Het}$ has structure group $SU(3)$ rather than $U(3)$ one must impose the `traceless' condition
\begin{equation}c_1(p_{Het*}\CN_{Het}) = 0\label{traceless}\end{equation}
where $p_{Het}$ is the covering map
\begin{equation}p_{Het}: \CC_{Het}\rightarrow B_2.\end{equation}
By Grothendieck-Riemann-Roch,
\begin{equation}c_1(p_{Het*}\CN_{Het}) = p_{Het*}c_1(\CN_{Het}) - \frac{1}{2}p_{Het*}r_{Het}\end{equation}
where $r_{Het}$ is the ramification divisor of the covering $p_{Het}$
\begin{equation}r_{Het} = p_{Het}^*\cs - c_1(\CC_{Het})\end{equation}
so bundles $\CN_{Het}$ satisfying \eqref{traceless} can be associated to divisors $\gamma_{Het}$ satisfying
\begin{equation}p_{Het*}\gamma_{Het}=0\label{tracelessgamma}\end{equation}
according to
\begin{equation}c_1(\CN_{Het}) = \lambda\gamma_{Het} + \frac{r_{Het}}{2}\label{Nhetdef}\end{equation}
where $\lambda$ will be a half-integer whenever $r_{Het}/2$ is a half-integral class.  The spectral cover $\CC_{Het}$ is simply connected in general.  In this case, there exists a unique divisor class $\gamma_{Het}$ satisfying \eqref{tracelessgamma}
\begin{equation}\gamma_{Het} = 3\sigma_{Het} - (\eta-3\cs)\label{gammahet}\end{equation}
Given our Calabi-Yau $Z_{Het}$ and bundle $V$, we pick a distinguished curve $\Sigma$ in $B_2$ and investigate the superpotential correction generated by worldsheet instantons wrapping $\Sigma$.  This correction will vanish if the instanton does not have the right Fermi zero mode structure.  In general, one has two types of zero modes; vector-like pairs of left- and right-movers that do not couple to $V$ and left-movers that do couple to $V$.  We address each of these in turn.

\subsection{Fermi Zero Modes that Don't Couple to $V$}

For completeness, first recall the structure of the right-moving Fermi zero modes, which don't couple to $V$.  Following \cite{Witten:1999eg,Beasley:2005iu,Donagi:2010pd}, one can establish that these modes are counted by
\begin{equation}2\times\left[h^0(\Sigma,N_{\Sigma/Z_{Het}}) + h^0(\Sigma,\CO) + h^0(\Sigma,K_{\Sigma})\right]\end{equation}
with $N_{\Sigma/Z_{Het}}$ denoting the normal bundle of $\Sigma$ inside the heterotic Calabi-Yau $Z_{Het}$.  In this paper we always take $\Sigma$ to be a curve that sits in the section $B_2$ of the elliptically fibered 3-fold $Z_{Het}$.  In that case
\begin{equation}N_{\Sigma/Z_{Het}} = N_{\Sigma/B_2}+K_{B_2}|_{\Sigma} = N_{\Sigma/B_2}+K_{\Sigma}\otimes N_{\Sigma/B_2}^{-1}\end{equation}
and the Fermi zero mode spectrum is counted by
\begin{equation}2\times\left[h^0(\Sigma,N_{\Sigma/B_2}) + h^1(\Sigma,N_{\Sigma/B_2}) + h^0(\Sigma,\CO) + h^1(\Sigma,\CO)\right]\end{equation}
where we used Serre duality to relate $h^0(\Sigma,K_{\Sigma}\otimes N_{\Sigma/B_2}^{-1})=h^1(\Sigma,N_{\Sigma/B_2})$ and $h^0(\Sigma,K_{\Sigma}) = h^1(\Sigma,\CO)$.  The contribution from $h^0(\Sigma,\CO)=1$ is just the ordinary pair of `universal' Fermi zero modes that one expects from an instanton that generates a superpotential coupling.  To ensure that a superpotential coupling is truly generated, however, the remaining cohomologies must vanish
\begin{equation}0 = h^0(\Sigma,N_{\Sigma/B_2}) = h^1(\Sigma,N_{\Sigma/B_2}) = h^1(\Sigma,\CO).\end{equation}  The condition $h^1(\Sigma,\CO)=h^0(\Sigma,K_{\Sigma})$ implies that $\Sigma$ must be a $\mathbb{P}^1$ while the remaining two terms indicate that the normal bundle of $\Sigma$ in $B_2$ must have vanishing cohomologies.  On $\mathbb{P}^1$, this is only true for $N_{\Sigma/B_2}=\CO(-1)$.  Since the degree of the normal bundle of any $\mathbb{P}^1$ inside a Calabi-Yau 3-fold is $-2$, we avoid extra right-handed Fermi zero modes only when
\begin{equation}\Sigma = \mathbb{P}^1\qquad N_{\Sigma/Z_{Het}} = \CO(-1) \oplus \CO(-1).\end{equation}
The second condition will be true whenever
\begin{equation}N_{\Sigma/B_2}= \CO(-1).\end{equation} Note that such
an instanton will always be isolated in the sense that
$N_{\Sigma/Z_{Het}}$ has no global sections. Such instantons do not
have non-universal bosonic zero modes, which would include bosonic
deformation modes transverse to the instanton.

\subsection{Fermi Zero Modes that Couple to $V$: the Pfaffian}
\label{subsec:hetpfaff}

We turn now to the worldsheet fermions that couple to $V$.  Our interest in this paper is to study the dependence of instanton-generated couplings on the moduli of $V$.  Integrating out worldsheet fermions that couple to $V$ yields a prefactor proportional to the Pfaffian of the Dirac operator $D_-$ on $\Sigma$
\begin{equation}W\sim \pfaff e^{i\int_{\Sigma}B}\label{HetPrefactor}\end{equation}
associated to the bundle 
\begin{equation}V_{\Sigma} = V|_{\Sigma}.\end{equation}
It is well-known that only the product $\pfaff e^{i\int_{\Sigma}B}$ is gauge invariant; the individual terms of the product are not.  Rather, $B$ shifts under a gauge transformation parametrized by $\epsilon$ according to
\begin{equation}B\rightarrow B + \text{tr}\epsilon F\label{Btransf}\end{equation}
while the Pfaffian transforms in the right way to cancel this.  This is the famous statement that the Pfaffian is not a function but rather a section of a suitable line bundle on the moduli space of gauge fields $V_{\Sigma}$.

We can describe $V_{\Sigma}$ and its moduli nicely by restricting
the spectral data of $V$ to the elliptic fibration $\CE$ over $\Sigma$
\begin{equation}\CE = \pi_{Het}^*\Sigma\qquad \pi_{\CE}:\CE\rightarrow \Sigma.\end{equation}
Defining 
\begin{equation}c_{Het} = p_{Het}^*\Sigma\qquad \CN_{c,Het} = \CN_{Het}|_{c_{Het}}\qquad p_{c,Het}:c_{Het}\rightarrow B_2\end{equation}
we have that $V_{\Sigma}$ is simply the pushforward of $\CN_{c,Het}$ 
\begin{equation}V_{\Sigma} = p_{c,Het*}\CN_{c,Het}.\end{equation}
In general, $\pfaff$ depends on the choice of curve $c_{Het}$ in $\pi_{Het}^*\Sigma$ as well as the choice of line bundle $\CN_{c,Het}$ on $c_{Het}$.  It vanishes whenever $D_{-,V_{\Sigma}}$ admits zero modes or, equivalently, whenever $H^0(\Sigma,K_{\Sigma}^{1/2}\otimes V_{\Sigma})$ is nontrivial.  Cohomologies of $K_{\Sigma}^{1/2}\otimes V_{\Sigma}$ can be related to cohomologies on $c_{Het}$
\begin{equation}H^p(\Sigma,K_{\Sigma}^{1/2}\otimes V_{\Sigma}) = H^p(c_{Het},p_{c,Het}^*K_{\Sigma}^{1/2}\otimes \CN_{c,Het}).\end{equation}
by a Leray spectral sequence.
Note that there is no ambiguity in the choice of spin structure $K_{\Sigma}^{1/2}$ since we always take $\Sigma=\mathbb{P}^1$.  Further, since $c_1(V_{\Sigma})=0$ we have that
\begin{equation}0 = \chi(\Sigma,K_{\Sigma}\otimes V_{\Sigma}) = \chi(c_{Het},p_{c,Het}^*K_{\Sigma}^{1/2}\otimes \CN_{c,Het})\end{equation}
so that the bundle $p_{c,Het}^*K_{\Sigma}^{1/2}\otimes \CN_{c,Het}$ has degree $g_{c_{Het}}-1$.  The full space of possible $[p_{c,Het}^*K_{\Sigma}^{1/2}\otimes\CN_{c,Het}]$'s, then comprises the degree $g-1$ Jacobian $\CJ_{g-1}(c_{Het})$ of the covering curve $c_{Het}$.  In general, very few bundles on $\CJ_{g-1}(c_{Het})$ will arise from the restriction of a bundle $\CN_{Het}$ on $\CC_{Het}$.  Let us forget about this for a moment, though, and consider the Pfaffian of a generic $SU(3)$ bundle $V_{\Sigma}$ on $\Sigma$ constructed from a fixed cover $c_{Het}$ and an arbitrary choice of degree $g_{c_{Het}}-1$ bundle $p_{c,Het}^*K_{\Sigma}^{1/2}\otimes\CN_{c,Het}$.  We consider first the dependence of this Pfaffian on $p_{c,Het}^*K_{\Sigma}^{1/2}\otimes\CN_{c,Het}$ before using the fact that Jacobians vary holomorphically in families to see how it depends on the moduli of $c_{Het}$.

It is well known that the dependence of $\pfaff$ on $\CN_{c,Het}$ is uniquely fixed by its vanishing locus, namely those choices of $\CN_{c,Het}$ such that $\CN_{c,Het}\otimes p_{c,Het}^*K_{\Sigma}^{1/2}$ admits holomorphic sections.  We will review this uniqueness result explicitly in a moment but first let us use it to relate the $\CN_{c,Het}$-dependence of $\pfaff$, which is a partition function of free fermions coupled to a non-Abelian vector bundle $V_{\Sigma}$, to the more familiar problem of a partition function of free fermions coupling to an Abelian bundle.  If we choose a spin structure $K_{c_{Het}}^{1/2}$ on $c_{Het}$, we can rewrite the degree $g_{c_{Het}}-1$ bundle of interest as
\begin{equation}\CN_{c,Het}\otimes p_{c,Het}^*K_{\Sigma}^{1/2} = K_{c_{Het}}^{1/2}\otimes \CL_{\Gamma}\label{NchetLgamma}\end{equation}
where
\begin{equation}\begin{split}\CL_{\Gamma} &= \CN_{c,Het}\otimes p_{c,Het}^*K_{\Sigma}^{1/2}\otimes K_{c_{Het}}^{1/2} \\
&= \CN_{c,Het}\otimes \CO(-R_{Het}/2).\end{split}\label{LGammaDef}\end{equation}
Here, $R_{Het}$ is the ramification divisor of the covering $p_{c,Het}$ and $\CO(-R_{Het}/2)$ is a square root of $\CO(-R_{Het})$ that is determined by the choice of spin structure $K_{c_{Het}}^{1/2}$.  All we have done here is use $K_{c_{Het}}^{1/2}$ to define a map to the degree $0$ part of the Jacobian
\begin{equation}\CJ_{g-1}(c_{Het})\rightarrow \CJ_0(c_{Het})\end{equation}
that takes 
\begin{equation}\CN_{c,Het}\otimes p_{c,Het}^*K_{\Sigma}^{1/2}\mapsto\CL_{\Gamma}\end{equation}
where we emphasize that $\CL_{\Gamma}$ has degree 0.  Roughly speaking, one can think of $\CL_{\Gamma}$ as the restriction of the $\lambda\gamma_{Het}$ piece of $\CN_{Het}$ \eqref{Nhetdef} to $c_{Het}$.  Because $\lambda\gamma_{Het}$ and $r_{Het}/2$ are both non-integral in general, the separation of $c_1(\CN_{Het})$ in \eqref{Nhetdef} cannot be lifted to a split of $\CN_{Het}$ into the tensor product of line bundles $\CO(\lambda\gamma_{Het})\otimes\CO(r_{Het}/2)$.  Upon restricting to $c_{Het}$, though, $\gamma|_{c_{Het}}\equiv \Gamma_{Het}$ and $r_{Het}|_{c_{Het}}=R_{Het}$ are even divisors whose corresponding line bundles, $\CO(\Gamma_{Het})$ and $\CO(R_{Het})$, admit square roots, $\CO(\Gamma_{Het}/2)$ and $\CO(R_{Het}/2)$.  There are $2^{2g_{c_{Het}}}$ choices for how we take these square roots corresponding to the $2^{2g_{c_{Het}}}$ choices of spin structure $K_{c_{Het}}^{1/2}$.

Now, the $\CL_{\Gamma}$-dependence of our desired Pfaffian is uniquely fixed by the fact that it vanishes when $H^0(c_{Het},K_{c_{Het}}^{1/2}\otimes \CL_{\Gamma})$ is nontrivial.  This is precisely the vanishing locus of the partition function of a free fermion on $c_{Het}$ coupled to $\CL_{\Gamma}$ and hence
\begin{equation}\pfaff \sim \pfaffHet\end{equation}
where it is understood that the map from $V_{\Sigma}$ to $(c_{Het},\CL_{\Gamma})$ depends on the choice of spin structure on $c_{Het}$.  

Given this equivalence, we could have addressed the problem of
studying $\pfaff$ by computing $\pfaffHet$ directly.  The latter
object is well studied and depends on $\CL_{\Gamma}\in\CJ_0(c_{Het})$
not as a holomorphic function but as a holomorphic section of a line
bundle $\hat{L}$ on $\CJ_0(c_{Het})$.  This reflects the fact that
$\pfaffHet$ is not gauge invariant, a welcome property because it must
violate gauge invariance in the right way to cancel the anomalous
transformation law of the $B$-field \eqref{Btransf}.  On these grounds
alone one can argue that $\frac{1}{2\pi}c_1(\hat{L})=\omega$ must
yield a principal polarization \cite{Witten:1996hc} of the Abelian variety $\CJ_0(c_{Het})$.
This is almost enough to determine $\hat{L}$ but not quite because
$\CJ_0(c_{Het})$ is not simply connected.  One must further specify
the holonomies of $\hat{L}$ around the 1-cycles of $\CJ_0(c_{Het})$
and, as discussed in \cite{Witten:1996hc}, there is no unique way to
do this.  Instead, there are $2^{2g_{c_{Het}}}$ choices of bundle
$\hat{L}_m$ on $\CJ_0(c_{Het})$ with
$\frac{1}{2\pi}c_1(\hat{L}_m)=\omega$, one for each choice of spin
structure on $c_{Het}$.  Each of these $2^{2g_{c_{Het}}}$ bundles,
$\hat{L}_m$, admits a unique holomorphic section $\theta_m$ whose
vanishing specifies a theta divisor $\Theta_m$.  It remains to
determine which line bundle is the `right' one.  Given that,
$\pfaffHet$ will be proportional to the corresponding $\theta_m$.

As we review in section \ref{subsubsec:chiral2form}, one encounters a completely analogous problem when studying the partition function of the chiral 2-form on an M5 instanton.  A general approach for determining the `right' $\theta_m$, applicable both for this situation and our heterotic Pfaffians, is outlined in \cite{Witten:1996hc}.  While straightforward, this procedure is not easy to apply in practice.  For our heterotic worldsheet instantons it is actually unnecessary because we have more information available than the transformation properties of the desired partition function: we know its vanishing locus precisely and in a way that doesn't have an explicit dependence on spin structure.  As we discuss in section \ref{subsec:conjecture}, it is our hope that similar considerations can apply to M5 instantons as well. 

Let us review why the Pfaffian doesn't have an explicit dependence on a choice of spin structure.
Recall that $\pfaff$ vanishes whenever $H^0(c_{Het},\CN_{c,Het}\otimes p_{c,Het}^*K_{\Sigma}^{1/2})$ is nontrivial.  The collection of degree $g_{c_{Het}}-1$ line bundles $\cL_A=\CN_{c,Het}\otimes p_{c,Het}^*K_{\Sigma}^{1/2}$ with section is remarkable in that it is a theta divisor of the Abelian variety $\CJ_{g-1}(c_{Het})$.  Because of its special origin, this theta divisor is distinguished among the others on $\CJ_{g-1}(c_{Het})$ and it is typically referred to as the Riemann Theta divisor, $\Theta_R$.  Like all theta divisors, it specifies a line bundle on $\CJ_{g-1}(c_{Het})$ whose first Chern class is a principal polarization and admits a unique holomorphic section $\theta_R$ that vanishes along $\Theta_R$.  The $\CN_{c_{Het}}$-dependence of $\pfaff$ is therefore determined entirely by its vanishing locus up to an overall constant
\begin{equation}\pfaff \sim \theta_R(\CN_{c,Het}\otimes
  p_{c_{Het}}^*K_{\Sigma}^{1/2}).\end{equation}
When we rewrite our bundle in terms of $\CL_{\Gamma}$ as in \eqref{NchetLgamma}, all we do is obtain translates of $\theta_R$ on $\CJ_0(c_{Het})$
\begin{equation}\pfaff \sim \theta_R(K_{c_{Het}}^{1/2}\otimes \CL_{\Gamma}) \equiv \theta_{-K_{c_{Het}}^{1/2}}(\CL_{\Gamma})\end{equation}
The translate $\theta_{-K_{c_{Het}}^{1/2}}$ defines a theta divisor on $\CJ_0(c_{Het})$ and is the `right' $\theta_m$ for the choice of spin structure $K_{c_{Het}}^{1/2}$.  Said differently, given a choice of spin structure the Pfaffian is specified by the theta divisor $\Theta_m$ of $\CJ_0(c_{Het})$ which is a translate of $\Theta_R$ by $K_{c_{Het}}^{1/2}$.

We see that the introduction of a spin structure on $c_{Het}$ is completely unnecessary to determine the dependence of $\pfaff$ on the choice of line bundle on $c_{Het}$.  Rather, it is an auxiliary operation that arises when we try to mimic the separation \eqref{Nhetdef} at the level of line bundles on $c_{Het}$.  Even then, however, the Pfaffian is ultimately determined entirely by its vanishing structure which makes no direct reference to any of this.  One hopes that M5 instantons can be understood in a similar way, allowing us to get a handle on vanishing properties without following through the general procedure of \cite{Witten:1996hc}.  An approach like this is guaranteed to work when a heterotic dual exists and we would like to suggest that it may apply more generally as well.

\subsection{Moduli Dependence of the Pfaffian}
\label{subsec:pfaffbdo}

Our previous considerations focused on the vanishing structure of the
Pfaffian and determining the right $\theta$ function for controlling
its variation as one moves in the moduli space of bundles on
$c_{Het}$.  In general, only a discrete set of bundles $\CN_{c,Het}\in
\CJ_{g-1}(c_{Het})$ will arise as the restriction of bundles
$\CN_{Het}$ on $\CC_{Het}$.  This makes the previous discussion useful
for comparisons with M5 instantons but not very relevant for
determining the moduli dependence of the instanton prefactor.  The
latter requires an understanding of how $\pfaff$ depends on the
algebraic moduli appearing in the defining equation of $c_{Het}$ and
it is this dependence that was computed in a number of examples in
\cite{Buchbinder:2002ji,Buchbinder:2002ic,Buchbinder:2002pr}.  As
explained in \cite{Donagi:2010pd}, the Pfaffian of a bundle $\CL_A$ on
a curve $c_{Het}$ can be obtained as the determinant of a map $f$
appearing in an exact sequence of the form
\begin{equation}0\rightarrow H^0(c_{Het},\CL_A)\rightarrow W_1\xrightarrow{f}W_2\rightarrow H^1(c_{Het},\CL_A)\rightarrow 0\label{fsequence}\end{equation}
In \cite{Buchbinder:2002ji,Buchbinder:2002ic,Buchbinder:2002pr}, sequences of this type were obtained by studying the Koszul complex relating line bundles on $c_{Het}$ to those on $\CE=\pi_{Het}^*\Sigma$, the elliptic fibration over $\Sigma$ that contains $c_{Het}$.
A Koszul approach is made possible because our bundle of interest
\begin{equation}\CL_A = \CN_{c,Het}\otimes p_{c,Het}^*K_{\Sigma}^{1/2}\end{equation}
is the restriction of a line bundle $\CL_{\CE}$ on $\CE$
\begin{equation}\CL_{\CE} = \CO_{Z_{Het}}(\lambda\gamma+r/2)|_{\CE}\otimes \pi_{\CE}^*K_{\Sigma}^{1/2}. 
\end{equation}
The Koszul sequence
\begin{equation}
  0 \rightarrow \CL_A \otimes \cO(-c_{Het}) \xrightarrow{f_{c_{Het}}} \CL_A \rightarrow \CL_A|_{c_{Het}} \rightarrow 0
\end{equation}
and leads to the long exact cohomology sequence
\begin{equation}\begin{split}0 &\rightarrow H^0(\CE,\CL_{\CE}\otimes \CO(-c_{Het}))\rightarrow H^0(\CE,\CL_{\CE})\rightarrow H^0(c_{Het},\CL_c) \\
&\rightarrow H^1(\CE,\CL_{\CE}\otimes \CO(-c_{Het}))\xrightarrow{f} H^1(\CE,\CL_{\CE})\rightarrow H^1(c_{Het},\CL_c) 
\\
&\rightarrow H^1(\CE,\CL_{\CE}\otimes \CO(-c_{Het}))\rightarrow H^1(\CE,\CL_{\CE})\rightarrow 0
\end{split}\end{equation}
When $H^0(\CE,\CL_{\CE})=0$ and
$H^1(\CE,\CL_{\CE}\otimes\CO(-c_{Het}))=0$ we have a sequence of
precisely the form \eqref{fsequence} and the moduli dependence of
$\pfaffHet$ is captured by the determinant of the map $f$.  A matrix
representative for $f$ is often straightforward to compute so one can
obtain very explicit expressions.  Intriguingly, these often exhibit a
very high degree of structure that we would like to better
understand. To us, the structures are indicative of beautiful physics,
and we will study them in later sections.

\section{F-theory and M5 Instantons}
\label{sec:FM5}

We now turn to F-theory compactified on an elliptically fibered
Calabi-Yau 4-fold $Y_4$ with section over a base $B_3$
\begin{equation}\pi:Y_4\rightarrow B_3.\end{equation}
We review and clarify the spectral divisor formalism for constructing
$G$-fluxes in these compactifications with an eye toward properly
specifying the 3-form, $C_3$.  This requires keeping track of more
than the homology class of a holomorphic surface that is Poincare dual
to $G_4$.  We emphasize the specification of a line bundle on the
spectral divisor as a starting point and carefully construct the
bundle by which it should be twisted to yield a properly quantized
$G$-flux.  To make precise how these $G$-fluxes communicate to the
local GUT model on $B_2$, we also describe a careful study of the
compact surface that emerges from $Y_4$, or more properly its
resolution, and plays the role of the Higgs bundle spectral cover
$\cC_{loc}$.  Previous work \cite{Marsano:2011hv} has shown how the
non-compact spectral cover of local models arises from the restriction
of the spectral divisor to $\pi^*B_2$.  We study the full compact
surface $\cC_{loc}$ and demonstrate that it is naturally described as
a hypersurface in an auxiliary elliptically fibered Calabi-Yau 3-fold
that is isomorphic to the heterotic Calabi-Yau when a heterotic dual
exists.  We will see in subsequent sections that this auxiliary space
nicely facilitates instanton computations. Following this discussion,
we review some background on M5 instantons in preparation for
subsequent sections.

\subsection{F-theory Preliminaries}
\label{subsec:FM5prelim}
As usual, we think of F-theory as a limit of M-theory in which the
elliptic fiber of $Y_4$ is shrunk to zero size while holding its
complex structure modulus fixed.  In M-theory, the instantons of
interest come from Euclidean M5's that wrap vertical divisors of
$Y_4$, that is divisors of the form $\pi^*D3$ for surfaces $D3$ in
$B_3$. In taking the F-theory limit by taking two torus one-cycles to
zero size in succession, the M5-instanton becomes a D4-instanton in
type IIA and then a D3-instanton in F-theory\footnote{Recall that
  throughout we call the instantons in F-theory $M5$-instantons to
  emphasize their M-theoretic origin, keeping in mind that they are
  really D3-branes.}. As the notation suggests, these descend to
D3-instantons in the IIB limit while they often correspond to heterotic
worldsheet instantons when a heterotic dual exists.  Before getting to
M5's, however, let us describe our setup more carefully and set some
notation.

As with the heterotic 3-fold $Z_{Het}$, we will realize our F-theory 4-fold $Y_4$ as a hypersurface inside a $\mathbb{P}^2_{1,2,3}$ bundle over the base.  We denote the ambient $\mathbb{P}^2_{1,2,3}$ bundle by $W_5$ and restrict our attention to 4-folds in $W_5$ that exhibit an $E_6$ singularity along a surface $B_2$ in $B_3$.  When a heterotic dual exists, $B_2$ will be identified with the base of $Z_{Het}$.
Letting $[v,x,y]$ denote the weighted homogeneous coordinates of the $\mathbb{P}^2_{1,2,3}$ fiber, we write the defining equation of $Y_4$ in $W_5$ as
\begin{equation}y^2 = x^3 + fx z^4v^4 + gz^6v^6 +  z^2v^3\left[b_0z^3v^3 + b_2 zv x + b_3 y\right]\label{Y4def}\end{equation}
where the various objects that appear are holomorphic sections of the indicated bundles
\begin{equation}\begin{array}{c|c}
\text{Section} & \text{Bundle} \\ \hline
v & \CO(\sigma) \\
x & \CO(2\sigma + 2\cb) \\
y & \CO(3\sigma + 3\cb) \\
z & \CO(B_2) \\
f & 4\CO(\cb-B_2) \\
g & 6\CO(\cb-B_2) \\
b_m & \CO([6-m]\cb-[5-m]B_2)
\end{array}\end{equation}
Here $\sigma$ is the section of $Y_4$ and $\cb$ is shorthand for an anti-canonical divisor on $B_3$.  Note that we do not distinguish between divisors on $B_3$ and their pullbacks to $Y_4$.

A complete specification of an F-theory compactification requires not just $Y_4$ but also a configuration of the M-theory 3-form, $C_3$.  One typically provides the field strength $G_4$ associated to this 3-form and focuses on supersymmetric vacua where $G_4$ is harmonic of type $(2,2)$ and primitive.  In practice, these $G$-fluxes are specified by giving their Poincare dual holomorphic surfaces in $Y_4$ or, more carefully, its Calabi-Yau resolution $\tilde{Y}_4$.  Lorentz invariance requires that $G_4$ have `one leg on the fiber' which is to say that it integrates to zero over any surface that contains the elliptic fiber or sits inside the section.

Our fourfold $Y_4$ is singular along $B_2$ and the physics of that
singularity is captured by an 8-dimensional $E_6$ gauge theory on
$\mathbb{R}^{3,1}\times B_2$.  The $G$-fluxes of interest are the ones
that control the spectrum of this theory.  Describing them can be
complicated because the Poincare dual surfaces are smooth in
$\tilde{Y}_4$ but degenerate in the singular limit
$\tilde{Y}_4\rightarrow Y_4$.  A formalism has been developed for
simplifying the analysis based on the notion of a \emph{spectral
  divisor} \cite{Marsano:2010ix,Marsano:2011nn,Marsano:2011hv}.  For
our $Y_4$ \eqref{Y4def}, we define the spectral divisor $\CC_{F}$ as
the hypersurface of $Y_4$ given by{\footnote{  
 If we define $\CC_F$ by \eqref{specdivdef} one must take some care.  When we resolve the singularities of $Y_4$, the proper transform of \eqref{specdivdef} inside the resolution is actually reducible and the spectral divisor corresponds to a particular component.  To avoid this difficulty, we can alternatively define $\CC_F$ via the equation $y^2=x^3+fxz^4v^4+gz^6v^6$.  The proper transform of this is irreducible and corresponds precisely to the spectral divisor.  This latter equation is therefore easier to work with though it obscures the dependence of $\CC_F$ on the moduli $b_m$.}}

\begin{equation}\CC_{F}:\qquad b_0z^3v^3+b_2 zvx + b_3y=0. \label{specdivdef}\end{equation}
This is a
3-sheeted cover of $B_3$ inside $Y_4$ that is singular at $z=0$ where
the sheets come together.  The sheets are separated when the
singularity is resolved and the resulting divisor, $\tilde{\CC}_{F}$,
is smooth inside $\tilde{Y}_4$.  The usefulness of $\CC_{F}$, at least
initially, is that it provides a very efficient probe of the resolved
cycles in $\tilde{Y}_4$.  Given a (possibly singular) holomorphic
surface $\CG$ inside $\CC_{F}$ before the resolution, we know exactly
what the intersections of its proper transform $\tilde{\CG}$ will be
with the surfaces of $\tilde{Y}_4$ that degenerate in the blow-down
limit $\tilde{Y}_4\rightarrow Y_4$.  We can indirectly specify a
$G$-flux in $\tilde{Y}_4$, then, by specifying $\CG$ in $Y_4$; this
allows many chirality computations to be performed without having to
work with the resolved geometry explicitly.  One must take care,
however, that $\CG$ is quantized properly in light of the flux
quantization rule \cite{Witten:1996md}
\begin{equation}G_4 + \frac{1}{2}c_2(\tilde{Y}_4)\in H^{2,2}(\tilde{Y}_4,\mathbb{Z})\label{quantrule}\end{equation}
In particular, $c_2(\tilde{Y}_4)$ is not an even class in our $\tilde{Y}_4$ so $\CG$ must be a non-integral surface class of the right type.

Often we are interested in going beyond chirality computations and for
this we need a more careful specification of $G$-flux and, indeed,
$C_3$.  For that, we would like to start not with the homology class
of a holomorphic surface\footnote{Which happens to be a divisor in
  $\CC_{F}$, since $\CC_{F}$ is a threefold.} $\CG$ in $\CC_{F}$ but
instead a line bundle $\CN_F$ on $\tilde{\CC}_{F}$ (or the
corresponding sheaf on $\CC_{F}$).  If we let $\iota_F$ denote the
embedding map
\begin{equation}\iota_F:\tilde{\CC}_{F}\rightarrow \tilde{Y}_4\end{equation}
and $A_F$ the connection of $\CN_F$, we can obtain a 3-form from the
push-forward $\iota_{F*}A_F$.  This can't be our final $C_3$ because
the resulting $G$-flux would be an integral $(2,2)$-form that will not
satisfy \eqref{quantrule} for the resolution $\tilde{Y}_4$ of our
$Y_4$ \eqref{Y4def}.  We can consider, however, the ramification
divisor $r_F$ of the covering
\begin{equation}p_{F}:\tilde{\CC}_{F}\rightarrow B_2\end{equation}
and the bundle $\CO_{\tilde{\CC}_{F}}(r_F)$.  As we show in Appendix \ref{app:resolutions}, we should consider not $\CO_{\tilde{\CC}_F}(r_F)$ but rather a simple twist of this bundle that we denote by $\CL_{\hat{r}}$ \eqref{Lrhatdef}.
This bundle will have a connection $A_{\hat{r}}$ from which we can obtain a properly quantized $G$-flux as
\begin{equation}C_3 = \iota_{F*}\left(A_F - \frac{1}{2}A_{\hat{r}}\right)+\frac{1}{2}C_3^{(0)}\quad\leftrightarrow\quad G_4\sim \iota_{F*}\left(c_1(\CN_F)-\frac{1}{2}c_1(\CL_{\hat{r}})\right) + \frac{1}{2}G_4^{(0)}.\label{C3G4fromLocal}\end{equation}
Here, $G_4^{(0)}$ is a Cartan flux that corresponds to a particular worldvolume gauge field background \eqref{cartancorrectionterm} and $C_3^{(0)}$ is a corresponding potential.  As discussed in Appendix \ref{app:resolutions}, it is often necessary to add further integral worldvolume $G$-fluxes to \eqref{C3G4fromLocal} to preserve the $E_6$ gauge group.

For $C_3$ specified as in \eqref{C3G4fromLocal}, the spectral divisor describes precisely how the flux data impacts the $E_6$ gauge theory on $\mathbb{R}^{3,1}\times B_2$.  Recall that the holomorphic data of the gauge theory consists of a Higgs bundle spectral cover, $\CC_{loc}$ and a line bundle $\CN_{loc}$ on $\CC_{loc}$.  More specifically, $\CC_{loc}$ is a 3-sheeted cover of $B_2$ that lives in the total space of $K_{B_2}$ or some suitable compactification thereof.  Introducing the projection map
\begin{equation}p_{loc}:\CC_{loc}\rightarrow B_2
\end{equation}
the data $(\CC_{loc},\CN_{loc})$ determine the configuration of an adjoint scalar $\Phi_{loc}$ and adjoint gauge bundle $V_{loc}$ of the gauge theory on $B_2$ as \cite{Donagi:2009ra}
\begin{equation}\Phi_{loc}\sim p_{loc*}s\qquad V_{loc}\sim p_{loc*}\CN_{loc}\end{equation}
where $s$ is a normal coordinate to $B_2$ in the ambient 3-fold.

The Higgs bundle spectral cover emerges from $\tilde{Y}_4$ as the restriction of $\tilde{\CC}_F$ to (the proper transform of) $\pi^*B_2$
\begin{equation}\CC_{loc} = \tilde{\CC}_F|_{\pi^*B_2}\end{equation}
while the bundle $\CN_{loc}$ is just the restriction of $\CN_F$
\begin{equation}\CN_{loc} = \CN_F|_{\CC_{loc}}\end{equation}
To preserve the $E_6$ gauge group, $\CN_{loc}$ must satisfy $c_1(p_{loc*}\CN_{loc})=0$.  One can make a decomposition of $c_1(\CN_{loc})$ that is similar to \eqref{Nhetdef}
\begin{equation}c_1(\CN_{loc}) = \lambda\gamma_{loc} + \frac{r_{loc}}{2}.\label{Nlocgammaloc}\end{equation}
The bundle $\CL_{\hat{r}}$ that we obtain by twisting $\CO_{\CC_F}(r_F)$ in \eqref{Lrhatdef} is constructed so that
\begin{equation}\CL_{\hat{r}}|_{\CC_{loc}} = \CO_{\CC_{loc}}(r_{loc})\end{equation}
which means 
the divisor $\gamma_{loc}$ from \eqref{Nlocgammaloc} is identified with the holomorphic surface $\CG$ that is Poincare dual to the $G$-flux in \eqref{C3G4fromLocal} (up to Cartan corrections)
\begin{equation}\gamma_{loc} = \CG|_{\CC_{loc}}.\end{equation}

\subsubsection{A Comment on the Compact Realization of $\CC_{loc}$ as $\CC_{F}|_{\pi^*B_2}$}
\label{subsubsec:higgscomment}

Before proceeding, let us note that we provide a complete description
of the surface $\CC_{loc}$ in Appendix \ref{app:resolutions}.  There,
we demonstrate that $\CC_{loc}$ can be viewed as a complete
intersection inside an $\mathbb{F}_1$-fibration over $B_2$ that we
denote by $E_4$ \footnote{We resolve $Y_4$ by performing a series of blow-ups in the ambient space $W_5$ and passing to the 
proper transform.  Above generic points on $B_2$, all three sheets of $\tilde{\CC}_F$ in $\tilde{Y}_4$ meet the same 
node, say $C$, of the singular fiber.  The divisor obtained by fibering $C$ over $B_2$ is the restriction to $\tilde{Y}_4$ of an exceptional divisor (denoted $E_4$ in Appendix \ref{app:resolutions}) of the blown-up ambient space.  The $\mathbb{F}_1$-fibration over $B_2$ is this exceptional divisor and the two defining equations are the defining equations of $\CC_F$ as a complete intersection in the ambient 5-fold.
}.  
Letting $[u,q]$ denote coordinates on the fiber and $[W,X]$ coordinates on the base, the $\mathbb{F}_1$-fibration is specified by promoting $u,q,W,X$ to sections of the indicated bundles
\begin{equation}\begin{array}{c|c}
\text{Section} & \text{Bundle on }E_4 \\ \hline
W & f \\
X & f + 2\cs \\
u & b \\
q & b + f + 3\cs.
\end{array}\label{F1fibration}\end{equation}
The defining equations of $\CC_{loc}$ inside $E_4$ are
\begin{equation}\begin{split}Wq^2 &= u^2\left(X^3 + W^2X f + W^3 g\right) \\
0 &= b_3q + b_2uX + b_0uW.
\end{split}\label{clocF1}\end{equation}
Above generic points in $B_2$, the first equation defines an anti-canonical curve in the $\mathbb{F}_1$ fiber, which is just an elliptic curve.  Closer inspection tells us, in fact, that this equation specifies an elliptically fibered Calabi-Yau 3-fold $Z_{3,F}$ with base $B_2$ and section.  As we review in section \ref{subsec:hetf}, this is isomorphic to the heterotic 3-fold when a heterotic dual exists.  We stress, however, that the appearance of this (auxiliary) Calabi-Yau 3-fold does not depend on the existence of a heterotic dual.  We can map its realization to the more conventional one as a hypersurface in a $\mathbb{P}^2_{1,2,3}$ bundle over $B_2$ by implicitly defining new coordinates $v,x,y$ through
\begin{equation}\begin{split}W &\equiv v^2 \\
X &\equiv x \\
q &\equiv y \\
u &\equiv v.
\end{split}\label{3foldmap}\end{equation}
This is the Higgs bundle spectral cover of the $E_6$ gauge theory on $\mathbb{R}^{3,1}\times B_2$.  It has been known for a while that the non-compact Higgs bundle spectral cover, which can be viewed as a non-compact hypersurface in the total space of $K_{B_2}$, emerges from $\tilde{\CC}_F|_{\pi^*B_2}$ \cite{Marsano:2011nn,Marsano:2011hv} but the full compact  surface that appears hasn't been investigated before{\footnote{For most computations in the $E_6$ gauge theory, the compact surface will not matter as the non-compact spectral cover contains all the relevant information.  To describe instanton vanishing loci, however, details of how the non-compact spectral cover is compactified become very important.  Most local model studies use a compactification that sits in $\mathbb{P}(\CO\oplus K_{B_2})$ but the actual surface that appears in $\tilde{Y}_4$ is not isomorphic to this.  It is, however, isomorphic to the heterotic spectral cover when a heterotic dual exists.}}.

Though the equations \eqref{clocF1} appear in the explicit $E_6$
resolution of appendix \ref{app:resolutions}, this compact realization
of $\CC_{loc}$ begins to exist once $E_4$ begins to exist. In particular,
it applies to the cases of $SU(5)$ and $SO(10)$ gauge symmetry.
Utilizing \eqref{3foldmap}, the appropriate generalization of \eqref{clocF1}
to write the defining equation for $\CC_{loc}$ inside $Z_{3,F}$ is
\begin{equation}
f_{\CC_{loc}} = \sum_{q=1}^{n} b_q \, v^{n-q}x^{n_x}y^{n_y}
\label{eqn:clocgeneral}\end{equation}
where $q=2n_x + 3n_y$ and $G=E_6,SO(10), SU(5)$ correspond to $n=3,4,5$. Recall that
the flux parameter $\lambda$ is half-integral for $n$ odd and integral for $n$ even.

That the fourth exceptional divisor $E_4$ plays a special role in obtaining a
nice compactification of an $SU(n)$ spectral cover is not surprising: if it sat in
the third exceptional divisor $E_3$, then an $SU(4)$ gauge symmetry would admit
a spectral cover description, which it does not. If, on the other hand, $\cC_{loc}$
sat in the fifth exceptional divisor $E_5$ then this compactification of the spectral
cover would describe cases of $SO(10)$ gauge symmetry, but not $SU(5)$. Simply
put, this compactification of the spectral cover begins to exist when the spectral
cover description of the gauge theory begins to exist.

\subsection{M5 Instantons}

We now review some properties of M5 instantons and the conditions under which they to generate superpotential couplings.  For this, we consider Euclidean M5's wrapping vertical divisors of $Y_4$,  that is divisors of the form $\pi^*D3$ for surfaces $D3$ in $B_3$. In the IIB limit, these descend to D3-instantons while they correspond to heterotic worldsheet instantons when a heterotic dual exists.  For the purpose of this paper, we are interested in two types of fields on the M5: worldvolume fermions and the chiral 2-form $\phi_2$.  
The former play a similar role to the right-moving fermions on the heterotic worldsheet in that our only concern is whether the zero mode structure is appropriate for a superpotential coupling.  The 2-form, on the other hand, is trickier to study but yields a partition function that captures the instanton dependence on $C_3$ flux and the complex structure moduli.  We address each of these fields in turn.

\subsubsection{Fermi Zero Modes}

The spectrum of Fermi zero modes on an M5 instanton is well-known (see for instance \cite{Donagi:2010pd})
\begin{equation}2\times\left[h^{0,0}(M5) + h^{0,1}(M5) + h^{0,2}(M5) + h^{0,3}(M5)\right].\end{equation}
In the absence of fluxes which lift some of the Fermi zero modes, the instanton must give rise to
the 2 `universal' modes that arise from $h^{0,0}(M5)=1$ in order to generate a superpotential
correction. In particular, we must have 
\begin{equation}h^{0,1}(M5)=h^{0,2}(M5)=h^{0,3}(M5)=0.\label{M5cohomconds}\end{equation}
Note that Witten's arithmetic genus condition $\chi(M5,\CO_{M5})=1$ \cite{Witten:1996bn} is a necessary but not sufficient condition for this.

The M5-instantons we study have a special structure. They are `vertical' M5-instantons which
are elliptic fibrations over a base surface that we are denoting by $D3$
\begin{equation}P:M5\rightarrow D3.\end{equation}
It is often helpful to rephrase \eqref{M5cohomconds} in terms of cohomologies on $D3$ which, as the notation suggests, we think of as the surface wrapped by a D3-instanton in the IIB picture.  We do this with the Leray spectral sequence that starts with
\begin{equation}E_2^{ij}=H^i(S,R^jP_*\CO_{M5})\end{equation}
and terminates on the second page \cite{Bianchi:2011qh}.   Using explicit results for the direct images $R^jP_*\CO_{M5}$
\begin{equation}R^0P_*\CO_{M5}=\CO_{D3}\qquad R^1P_*\CO_{M5} = N_{D3/M5}=K_{D3}\otimes N_{D3/B_3}^{-1}\end{equation}
we can rewrite $h^p(M5,\CO_{M5})$ for $p=1,2,3$ as
\begin{equation}\begin{split}h^1(M5,\CO_{M5}) &= h^0(D3,K_{D3}\otimes N_{D3/B_3}^{-1})+h^1(D3,\CO_{D3}) \\
&= h^2(D3,N_{D3/B_3})+h^1(D3,\CO_{D3}) \\
h^2(M5,\CO_{M5}) &= h^2(D3,K_{D3}\otimes N_{D3/B_3}^{-1})+h^2(D3,\CO_{D3}) \\
&= h^1(D3,N_{D3/B_3})+h^2(D3,\CO_{D3}) \\
h^3(M5,\CO_{M5}) &= h^2(D3,K_{D3}\otimes N_{D3/B_3}^{-1}) \\
&= h^0(D3,N_{D3/B_3}).
\end{split}\end{equation}
To avoid fermi zero modes, then, the divisor $D3$ in $B_3$ must be rigid in the sense that
\begin{equation}\label{eqn:D3 W cont 1} h^1(D3,\CO_{D3})=h^2(D3,\CO_{D3})=0\end{equation}
and further the normal bundle of $D3$ inside the base $B_3$ of $Y_4$ must have no cohomology
\begin{equation}\label{eqn:D3 W cont 2} h^m(D3,N_{D3/B_3})=0\qquad m=0,1,2\end{equation}
For the first condition to hold, $D3$ should be a $dP_n$ or $\mathbb{F}_m$ surface.  It is sometimes possible that Fermi zero modes can be lifted in the presence of a nontrivial $G$-flux through a coupling of the form $\theta(\Gamma G)\theta$ \cite{Saulina:2005ve,Kallosh:2005gs,Lust:2005cu,Tsimpis:2007sx}.  One can think of such a $G$ as inducing a map $G:\Omega^{0,2}(M5)\rightarrow \Omega^{2,0}(M5)$ whose kernel corresponds to the unlifted zero modes.  We will restrict ourselves to instantons with $h^{0,2}(M5)=0$ in this paper, and so will not appeal to such a mechanism.

\subsubsection{Chiral 2-form}\label{subsubsec:chiral2form}

We now turn to the partition function of the chiral 2-form field on the M5-brane worldvolume.
We call this partition function $\CZ_{\phi}$.  Since no covariant Lagrangian formulation is known, it can be tricky to compute this quantity directly.  Following Witten \cite{Witten:1996hc}, however, we can make significant progress by looking at the transformation properties of $\CZ_{\phi}$ under bulk gauge transformations.  In general, the partition function of our M5 instanton will be proportional to
\begin{equation}e^{i\tau_{M5}\int_{M5}C_6}\CZ_{\phi}\label{M5partfcn}\end{equation}
where $C_6$ is the 6-form potential dual to the $M$-theory 3-form, $C_3$.  Just as with \eqref{HetPrefactor}, the product $e^{i\tau_{M5}\int_{M5}C_6}\CZ_{\phi}$ is invariant under bulk gauge transformations $C_3\rightarrow C_3+d\Lambda_2$ but the individual terms in the product are not.  Rather, $C_6$ transforms as
\begin{equation}C_6\rightarrow C_6+\frac{1}{2}\Lambda_2\wedge G_4\end{equation}
and the 2-form partition function $\CZ_{\phi}$ transforms in the opposite way due to the coupling
\begin{equation}S_{M5}\sim \int_{M5}\left(-\frac{1}{2}\phi_2\wedge G_4+\ldots\right)\label{M5G4coupling}\end{equation}
and the transformation law of $\phi_2$
\begin{equation}\phi_2\rightarrow\phi_2+\Lambda_2.\end{equation}
Because of this, the partition function $\CZ_{\phi}$ is not a function at all; it is actually a holomorphic section of a line bundle on the moduli space of $C_3$'s.  The coupling \eqref{M5G4coupling} induces a source for $\phi_2$ so the restriction of $G_4$ to the instanton must be\footnote{In the absence of M2-brane insertions. See section \ref{subsec:hetFinsts} for a brief discussion.} cohomologically zero for $\CZ_{\phi}$ to have any hope of being nonzero
\begin{equation}G_4|_{M5}=0\text{ in }H^{2,2}(M5).\end{equation}
Correspondingly, the moduli space of allowed $C_3$'s consists of harmonic 3-forms modulo large gauge transformations.  This is the intermediate Jacobian of the M5 
\begin{equation}\CJ(M5) = H^3(M5,\mathbb{R})/H^3(M5,\mathbb{Z})\end{equation}
which is an Abelian variety.  The transformation law
$\CZ_{\phi}\rightarrow \CZ_{\phi}e^{-i\int_{M5}\frac{1}{2}\phi_2\wedge
  G_4}$ implies that $\CZ_{\phi}$ is a holomorphic section of a line
bundle on $\CJ(M5)$ whose first Chern class is a principal
polarization.  This is almost enough to fix the $C_3$-dependence of
$\CZ_{\phi}$ because it means that $\CZ_{\phi}$ is proportional to a
$\theta$ function on $\CJ(M5)$.  However, there is a $\theta$ function
for each principal polarization, and it remains to determine which
theta function is the `right' one. Witten introduced a general procedure
for determining this for given
spin structures on $M5$ and $\tilde{Y}_4$ \cite{Witten:1996hc} but
applying it in practice is quite subtle{\footnote{This problem has
    received further attention in recent work related to D3-brane
    instantons in orientifold models and their F-theory lifts
    \cite{Kerstan:2012cy}.}}.  If the right $\theta$ function can be
found, however, we can in principle study the dependence of
$\CZ_{\phi}$ on the complex structure moduli of $Y_4$ (or at least
their restrictions to $M5$) by using the fact that $\CJ(M5)$ varies
holomorphically in holomorphic families{\footnote{This follows because
    the condition $h^{3,0}(M5)=0$ allows us to define a complex
    structure on $\CJ(M5)$ without ambiguity (i.e. there is no
    distinction between the Griffiths and Weil Jacobians).}}.

Though straightforward in principle, a careful study of $\CZ_{\phi}$ along these lines is difficult in practice.  We believe it is helpful to keep in mind the analogy to heterotic worldsheet instantons, where the problem of `choosing the right $\theta$ function' also arises.  There, we encountered the Pfaffian of a Dirac operator over fermions on $c_{Het}$ coupled to a degree zero line bundle $\CL_{\Gamma}$.  As with $\CZ_{\phi}$, the anomalous gauge transformation of $\pfaffHet$ implied that it was proportional to a $\theta$ function on an Abelian variety, the Jacobian $\CJ_0(c_{Het})$.  Determining the `right' $\theta$ function could have been tricky but we were aided by the fact that we knew the \emph{exact} vanishing locus of $\pfaffHet$.  This is because we could identify the relevant Fermi zero modes, which cause the prefactor to vanish, in a way that didn't depend on a choice of spin structure on $c_{Het}$.  This extra information of the vanishing locus fixed $\pfaffHet$ completely and provided a prescription for determining the proper $\theta$ function to associate with each choice of spin structure. 

Pursuing a similar strategy for M5's is difficult because we do not \emph{a priori} know the precise vanishing locus in general.  Unlike the case of heterotic worldsheet instantons, the vanishing is not due to Fermi zero modes but rather to the physics of the chiral 2-form.  Nevertheless, we might hope to gain some intuition from physics and, in particular, our expectations from the IIB limit.  Even though it is not directly attributable to Fermi zero modes, we expect 
the vanishing structure of $\CZ_{\phi}$ to be captured by the physics of 3-7 zero modes in IIB that arise from open strings with one end on the D3 and the other end on a stack of 7-branes.  This physics, in turn, should be controlled by cohomologies that can be interpreted in terms of fermions localized near $B_2$ in $Y_4$.  

As we will discuss in section \ref{subsec:conjecture}, this is exactly what
happens when a heterotic dual exists!  Even though heterotic
computations involve line bundle cohomologies on curves `away from the
$E_6$ singularity' in the usual sense, they can be reformulated isomorphically in the
local geometry near the singularity.  This reformulation does not
explicitly involve the structures necessitated by the heterotic dual
in any way and suggests that the physics behind the vanishing is
effectively captured by the local geometry near $B_2$, as we would
expect from a 3-7 interpretation.  This leads to a natural conjecture
for the vanishing locus of certain M5 partition functions in F-theory
compactifications without heterotic duals that is simple and allows
for straightforward computation.

\section{Heterotic/F-theory Duality}
\label{sec:hetf section}

Having reviewed both worldsheet instantons in the heterotic string and
M5-instantons in F-theory, let us now discuss how these effects are
related via duality.
We will begin by discussing aspects of heterotic/F-theory duality,
including the relation of the `cylinder map' to the spectral divisor
formalism.  By intersecting with the M5-brane divisor, we then
restrict this story to the `miniature' version of the cylinder map
that relates heterotic worldsheet instantons to M5 instantons in
F-theory duals.  We use the compact realization of the local model
spectral cover inside the F-theory compactification to rephrase the
vanishing locus in terms of cohomology computations in the local
geometry near the $E_6$ singularity.  In this way, we motivate a
conjecture for the vanishing locus of the M5 instanton partition
function in a class of F-theory compactifications that do not admit
heterotic duals.

\subsection{Heterotic/F-theory Duality}\label{subsec:hetf}
To construct models with a heterotic dual, we specialize the 4-fold \eqref{Y4def} by taking the base 3-fold $B_3$ to be a $\mathbb{P}^1$-fibration over a surface $B_2$ specified by a line bundle $N$
\begin{equation}B_3 = \mathbb{P}(\CO\oplus N).\label{B3type}\end{equation} 
We introduce the notation $s$ for the divisor class of the section of $B_3$ corresponding to the `$\CO$' factor and let $[Z,W]$ denote homogeneous coordinates on the fiber
\begin{equation}\begin{array}{c|c}\text{Section} & \text{Bundle} \\ \hline
Z & \CO(s) \\
W & \CO(s+t)
\end{array}\end{equation}
where $t$ is a divisor on $B_2$ and, as usual, we do not notationally distinguish between divisors on $B_2$ and their pullbacks to $B_3$.  With this notation, the divisor we called $B_2$ in section \ref{sec:FM5} is referred to as $s$ in order to distinguish it from the other section $s+t$ of $B_3$, which is also isomorphic to $B_2$.

The 4-fold $Y_4$ that results from \eqref{Y4def} and this special
choice of $B_3$ is a K3-fibration over $B_2$ and admits a heterotic
dual through fiberwise application of the standard equivalence of
F-theory on K3 and $E_8\times E_8$ heterotic strings on $T^2$.  We
recover the heterotic data by taking the stable degeneration limit,
wherein $Y_4$ splits into a pair of $dP_9$-fibrations glued together
along the elliptically fibered Calabi-Yau 3-fold $Z_{Het}$.  Each
$dP_9$ fibration, along with the corresponding $G$-flux, determines
one of the two heterotic bundles.  As we focus attention on
just one $E_8$ factor in this paper, namely one that is broken to
$E_6$ by the $SU(3)$ bundle $V$, we restrict our attention to the
$dP_9$ fibration that contains the surface of $E_6$ singularities.
This 4-fold, which we denote by $Y_4'$, can be described as a
hypersurface in a $\mathbb{P}^2_{1,2,3}$-bundle, $W_5'$, defined by
the equation
\begin{equation}y^2 = x^3 + fx Z^4v^4 + gZ^6v^6 + Z^2W v^3\left[a_0 Z^3v^3 + a_2 Zvx + a_3y\right]\label{dp9fibration}\end{equation}
where the objects that appear are sections of the indicated bundles
\begin{equation}\begin{array}{c|c}
\text{Section} & \text{Bundle} \\ \hline
v & \CO(\sigma) \\
x & \CO(2[\sigma + s + \cs]) \\
y & \CO(3[\sigma + s + \cs]) \\
f & \CO(4\cs) \\
g & \CO(6\cs) \\
a_0 & \CO(6\cs-t)=\CO(\eta) \\
a_2 & \CO(4\cs-t) = \CO(\eta-2\cs) \\
a_3 & \CO(3\cs-t) = \CO(\eta-3\cs) \\
Z & \CO(s) \\
W & \CO(s+t). \\
\end{array}\label{dp9sections}\end{equation}
The sections $a_m$ are related to the $b_m$ of \eqref{Y4def} by restriction to $B_2$
\begin{equation}a_m = b_m|_{B_2}\end{equation}
and hence reflect the `leading' behavior of the $b_m$'s near the surface of $E_6$ singularities at $Z=0$.  

The divisor $W=0$ is nothing other than the heterotic Calabi-Yau $Z_{Het}$ \eqref{zhet}.  To describe the heterotic bundle, let us follow the spectral divisor $\CC_F$ \eqref{specdivdef} through the stable degeneration limit to obtain the 3-fold given by\footnote{Recall that here, in the case of heterotic F-theory duality, we used $Z=0$ rather than $z=0$ for the GUT stack in $B_3$.} 
\begin{equation}a_0 Z^3v^3 + a_2 Zvx + a_3y = 0 \label{cylinderdef}\end{equation}
This 3-fold is often referred to as the `cylinder', which we denote by $\CC_{\text{cyl}}$, and it is a 3-sheeted cover of $B_3$ inside $Y_4'$ that is singular where the sheets come together at $Z=0$.  The 3-fold $\CC_{\text{cyl}}$ is a natural object for capturing the behavior of our $dP_9$ fibration that arises as follows\footnote{
If one prefers, the equation
$y^2 = x^3 + fxZ^4v^4 + gZ^6v^6$
less the components that coincide with the section can be used to followed $\CC_F$ through the stable degeneration limit.  Though the spectral divisor was irreducible, this 3-fold is not; it has acquired an extra component at $W=0$ corresponding to $Z_{Het}$ itself.  If we discard this component, what remains is a 3-fold that can be defined in $W_5'$ as
\begin{equation}\CC_{cyl}:\qquad \begin{array}{rcl}y^2 &=& x^3 + fxZ^4v^4 + gZ^6v^6 \\
0 &=& a_0 Z^3v^3 + a_2 Zvx + a_3y.
\end{array}\end{equation}}.

Our presentation of $Y_4'$ distinguishes two curve classes of each
$dP_9$ fiber: the elliptic fiber $x_9$ of $dP_9$ and the base, $e_9$.
The remaining curves in $dP_9$ or, more specifically, the classes in
$H_2(dP_9,\mathbb{Z})$ that are orthogonal to $x_9$ and $e_9$, will be
permuted as we move along various paths in $B_2$.  The lattice of
curve classes $C$ in $dP_9$ that satisfy $C\cdot x_9=C\cdot e_9=0$ is
isomorphic to $H_2(dP_8,\mathbb{Z})$ which, itself, is a copy of the
$E_8$ root lattice.  As a result, a generic $dP_9$ fibration will
interchange all 240 $E_8$ roots (that is all curve class $C\in
H_2(dP_9,\mathbb{Z})$ with $C\cdot x_9=C\cdot e_9=0$ and $C^2=-2$)
with one another according to actions of the Weyl group of $E_8$.  Of
the 240 roots of $E_8$, though, 72 are distinguished in $Y_4'$ as the
roots of $E_6$.  This means that the Weyl group of the $SU(3)$
commutant of $E_6$ inside $E_8$ controls the fibration in the case 
we consider, there the $dP_9$ fibration is non-generic so as to give
a surface of $E_6$ singularities at $Z=0$.  The action of the Weyl group is
governed by the decomposition of $E_8$ roots under $E_8\rightarrow
E_6\times SU(3)$
\begin{equation}240\rightarrow (72,1) + (1,6) + \left[(27,3) + \text{cc}\right]\end{equation}
The complete Weyl group action can therefore be determined by following a trio of roots that share a common $\mathbf{27}$ weight and mix under the action of $SU(3)$.  Given any such root, $C$, we can construct an exceptional line $\ell = x_9 + e_9 - C$ that is orthogonal to $e_9$ but meets $x_9$ in exactly one point.  The cylinder $\CC_{\text{cyl}}$ is the sum of three exceptional lines in the highest weight state $(1,0,0,0,0,0)$ of the $\mathbf{27}$ that mix as a triplet under $SU(3)$.  Each exceptional line meets $x_9$ exactly once and a distinguished node of the $E_6$ singular fiber exactly once.

From this general description, we expect $\CC_{\text{cyl}}|_{W=0}$ to comprise a set of 3 points on the elliptic fiber above each generic point in the base $B_2$.  This is a 3-sheeted cover of $B_2$ inside $Z_{Het}$ that is nothing other than the heterotic spectral cover $\CC_{Het}$:
\begin{equation}\CC_{\text{cyl}}|_{Z_{Het}} = \CC_{Het}.\end{equation}
Restricting to $Z_{Het}=\{W=0\}$ sets $Z=1$ in \eqref{cylinderdef} and recovers the well-known equation for
an $SU(3)$ spectral cover in the heterotic string.
Likewise, we can check that $\CC_{\text{cyl}}$ is in fact a $\mathbb{P}^1$-fibration over $\CC_{Het}$ though this $\mathbb{P}^1$ fibration is singular in $Y_4'$ where the sheets come together.  Nevertheless, we can define a projection
\begin{equation}p_{cyl}:\CC_{\text{cyl}}\rightarrow\CC_{Het}\end{equation}
and use this to describe the `cylinder' map relating heterotic bundle data to F-theory $G$-flux.  The idea is as follows.  Given a bundle $\CN_{Het}$ on $\CC_{Het}$ that completes the heterotic bundle data $(\CC_{Het},\CN_{Het})$, we write \eqref{Nhetdef}
\begin{equation}c_1(\CN_{Het}) = c_1(\CO(\lambda\gamma + r/2))\end{equation}
for a half-integral divisor class $\gamma$ on $\CC_{Het}$.  From $\gamma$ we obtain a $G$-flux by pulling $\gamma$ back to $\CC_{\text{cyl}}$ and pushing it forward to $Y_4'$ via the embedding
\begin{equation}\iota_{cyl}:\CC_{cyl}\rightarrow Y_4'\end{equation}
as in
\begin{equation}G_4 = \iota_{cyl*}p_{cyl}^*\lambda\gamma 
\label{G4cylindermap}\end{equation}
where we take the Poincare dual of the resulting holomorphic surface class to obtain the $(2,2)$-form $G_4$.

We would like to study this map a little more carefully in a way that doesn't require splitting $c_1(\CN_{Het})$ into a pair of half-integral divisor classes.  This will allow us to work at the level of line bundles and make a more direct connection to $C_3$.
Let $\CN_{cyl}$ be the bundle on $\CC_{cyl}$ obtained as the pull-back of $\CN_{Het}$
\begin{equation}\CN_{cyl} = p_{cyl}^*\CN_{Het}.\end{equation}
Further, let $A_{cyl}$ be the connection of $\CN_{cyl}$ and $A_{r,cyl}$ the connection of the bundle $\CO(r_{cyl})$ with $r_{cyl}$ the ramification divisor of the covering $p_{cyl}:\CC_{cyl}\rightarrow B_3$.  
We take
\begin{equation}C_3 = \iota_{cyl*}\left(A_{cyl} - \frac{1}{2}A_{r,cyl}\right)+\ldots\label{HetFC3}\end{equation}
which leads to 
\begin{equation}G_4 = \iota_{cyl*}\left(c_1(\CN_{cyl}) - \frac{1}{2}c_1(\hat L_r)\right)+\ldots\label{HetFG4}\end{equation}
This is completely equivalent to \eqref{G4cylindermap} because $r_{cyl}|_{\CC_{Het}} = r$.

We now turn to the restriction of $\CC_{\text{cyl}}$ to the singular fibration over $B_2$, $Z=0$.  Because $\CC_{\text{cyl}}$ is singular here, we must pass to the resolved geometry $\tilde{Y}_4'$ to obtain a proper description.  There is a lot we can say on general grounds, though, from the discussion of $\CC_{\text{cyl}}$ above.  Firstly, the three exceptional lines that comprise $\CC_{\text{cyl}}$ above generic points in $B_2$ intersect the nodes of the $E_6$ singular fiber according to the highest weight of the $\mathbf{27}$, $(1,0,0,0,0,0)$.  This means that $\CC_{\text{cyl}}$ meets a distinguished node of the singular fiber in 3 points above any generic point in $B_2$ so that $\CC_{\text{cyl}}|_{Z=0}$ is a 3-sheeted cover of $B_2$.  Further, we know that $\CC_{\text{cyl}}$ is in fact a $\mathbb{P}^1$-fibration over $\CC_{Het}$ so $\CC_{\text{cyl}}|_{Z=0}$ should be isomorphic to $\CC_{Het}$.

We verify all of these things explicitly by studying the resolution $\tilde{Y}_4'$ of $Y_4'$ in Appendix \ref{app:resolutions}.  Indeed, $\CC_{\text{cyl}}|_{Z=0}$ is completely equivalent to the restriction of our original spectral divisor, $\CC_F$, to $Z=0$ that is described in section \ref{subsubsec:higgscomment}.  It is a surface $\CC_{loc}$ 
\begin{equation}\CC_{loc} = \CC_{\text{cyl}}|_{Z=0}\end{equation}
that can be realized as a complete intersection in the $\mathbb{F}_1$-fibration $E_4\rightarrow B_2$ specified by \eqref{F1fibration}
\begin{equation}\begin{array}{c|c}
\text{Section} & \text{Bundle on }E_4\\ \hline
W & \CO(f) \\
X & \CO(f+2\cs) \\
u & \CO(b) \\
q & \CO(b+f+3\cs)
\end{array}\end{equation}
The defining equations of $\CC_{loc}$ inside $E_4$ are
\begin{equation}\begin{split}Wq^2 &= u^2\left(X^3+W^2Xf + W^3g\right) \\
0 &= b_3q + b_2uX + b_0uW
\end{split}\end{equation}
The identification \eqref{3foldmap} establishes that the first equation defines a Calabi-Yau 3-fold isomorphic to $Z_{Het}$.  It is then trivial to see that the second defines a surface that is isomorphic to $\CC_{Het}$.  In fact, we can use the $\mathbb{P}^1$-fibration of $\CC_{\text{cyl}}$ to translate the heterotic bundle data from $\CC_{Het}$ to $\CC_{loc}$
\begin{equation}\begin{split}\gamma_{loc} &= \gamma_{Het} \\
\CN_{loc} &= \CN_{Het}.
\end{split}\end{equation}
If we like, we can now describe the $G$-flux without any direct reference to the heterotic side at all.  We simply define
$\CN_{cyl}$ by
\begin{equation}\CN_{cyl} = p_{cyl,loc}^*\CN_{loc}\end{equation}
where
\begin{equation}p_{cyl,loc}:\CC_{\text{cyl}}\rightarrow \CC_{\text{loc}}.\end{equation}
The 3-form $C_3$ and $G$-flux $G_4$ are then given as before by \eqref{HetFC3} \eqref{HetFG4}.  We can also go the other way.  Given a line bundle $\CN_{cyl}$ on $\CC_{cyl}$, we can restrict it to $\CC_{loc}$ to obtain a line bundle $\CN_{loc}$ that completes the spectral data of the $E_6$ gauge theory.  This reproduces the `local/global' map of section \ref{subsec:FM5prelim} in this case because the bundle $\CL_{\hat{r}}$ \eqref{Lrhatdef} is simply
\begin{equation}\CL_{\hat{r}} = p_{cyl}^*\CO_{\CC_{Het}}(r_{Het})\end{equation}
when a heterotic dual exists (see Appendix \ref{app:resolutions}).  This is the motivation for the general prescription of section \ref{subsec:FM5prelim}, which can be applied in the absence of a heterotic dual.

\subsection{Heterotic WS Instantons and their M5 Counterparts}
\label{subsec:hetFinsts}

We now turn to the relation between heterotic worldsheet instantons and M5 instantons in F-theory.  The heterotic instanton of interest wraps a curve $\Sigma$ in the base $B_2$ of $Z_{Het}$.  The M5-instanton to which this corresponds is just\footnote{Technically the M5 is the full $K3$-fibration
over $\Sigma$, $\pi^* \Sigma$. Focusing on one of the $dP_9$-fibrations is equivalent to the fact in the heterotic
string that we have focused on one of the bundles $V$ responsible for breaking one $E_8$ factor,
ignoring the bundle $\tilde V$ which breaks the other $E_8$ factor.} the $dP_9$ fibration over $\Sigma$
\begin{equation}M5 = \rho^*\Sigma\end{equation}
where $\rho$ is the $dP_9$ fibration
\begin{equation}\rho:Y_4'\rightarrow B_2.\end{equation}
The relation between these instantons is essentially a `miniature' version of the standard heterotic/F-theory duality.  On the F-theory side, we have a $dP_9$ fibration $M5$ which can be viewed as an elliptic fibration over a surface $D3$ that itself is $\mathbb{P}^1$-fibered
\begin{equation}\pi_{M5}:M5\rightarrow D3\qquad \qquad \nu:D3\rightarrow \Sigma.\end{equation}
We also have a `cylinder', $c_{cyl}$, obtained as the restriction of $\CC_{cyl}$ to $M5$
\begin{equation}c_{cyl} = \CC_{cyl}|_{M5}\end{equation}
together with a projection $p_{M5}$ and embedding $\iota_{M5}$
\begin{equation}p_{M5}:c_{cyl}\rightarrow D3\qquad \iota_{M5}:c_{cyl}\rightarrow M5.\end{equation}
Inside the $M5$ we have\footnote{In section \ref{sec:computing} we will define an elliptic surface $\cE \equiv \pi^{-1}_{3,F} \Sigma$ and give a curve $c_{loc}$ in it. The surface $\cE$ and curve $c_{Het}$ in $\cE$ we define here are isomorphic to the `other' $\cE$ and its curve $c_{loc}$ in the case of duality. We abuse notation due to isomorphism.} at $W=0$ the elliptic fibration over $\Sigma$, $\CE=\pi_{M5}^*\Sigma$
\begin{equation}\pi_{M5,het}:\CE\rightarrow\Sigma\end{equation}
and the restriction of $c_{cyl}$ to $\CE$ is nothing more than the curve $c_{Het}$
\begin{equation}c_{cyl}|_{\CE} = c_{Het}.\end{equation}
In fact, $c_{cyl}$ is a $\mathbb{P}^1$ fibration over $c_{Het}$ whose projection\footnote{Note that this is just the restriction of the projection map $p_{cyl}$ to the M5.} $p_{M5,cyl}$
\begin{equation}p_{M5,cyl}:c_{cyl}\rightarrow c_{Het}\end{equation}
allows us to define a `cylinder' map relating line bundles on $c_{Het}$ to configurations for $C_3$ on $M5$.  The story is exactly as before; we simply restrict everything to $M5$.  Given a line bundle $\CN_{c,Het}$ on $c_{Het}$, we obtain a line bundle on $c_{cyl}$
\begin{equation}\CN_{M5,cyl} = p_{M5,cyl}^*\CN_{c,Het}\end{equation}
and we can go the other way by restriction.  Letting $A_{M5,cyl}$ be a connection on $\CN_{M5,cyl}$ and $A_{\hat{R}}$ a connection on $p_{M5,cyl}^*(\CO_{\CC_{Het}}(r_{Het})|_{\CE})=\CL_{\hat{r}}|_{M5}$, we obtain a 3-form on $M5$

\begin{equation}C_3 = \iota_{M5,cyl*}\left(A_{M5,cyl} - \frac{1}{2}A_{\hat{R}}\right)\label{C3M5map}\end{equation}
to which Cartan correction terms \eqref{cartancorrectionterm} must also be added in the end.
Because $A_{M5,cyl}$ and $A_{\hat{R}}$ are pulled back from connections $A_{c,Het}$ and $A_{c,\hat{R}}$ on $c_{Het}$, we can rewrite $C_3$ as
\begin{equation}\begin{split}C_3|_{M5} &= \iota_{M5,cyl*}p_{M5,cyl}^*\left(A_{c,het} - \frac{1}{2}A_{c,\hat{R}}\right) \\
&= \iota_{M5,cyl*}p_{M5,cyl}^*\left(A_{\Gamma}\right)\end{split}\label{restrictedcylindermapC3}\end{equation}
where $A_{\Gamma}$ is a connection on the bundle $\CL_{\Gamma}$ defined in \eqref{LGammaDef}.  Correspondingly,
\begin{equation}\begin{split}G_4|_{M5} &= \iota_{M5,cyl*}\left([c_1(\CN_{M5,cyl})]-\frac{1}{2}R_{cyl}\right) \\
&= \iota_{M5,cyl*}p_{M5,cyl}^*\left(c_1(\CN_{c,Het})-\frac{1}{2}R\right) \\
&= \iota_{M5,cyl*}p_{M5,cyl}^*c_1(\CL_{\Gamma}).
\end{split}\label{restrictedcylindermapG4}\end{equation}
We saw before that $\CL_{\Gamma}$ had to be a degree zero bundle to have any hope of a nonzero superpotential coupling.  Here, we see that this condition is equivalent to the requirement that $G_4|_{M5}$ be trivial.  This is of course a very familiar condition for M5 instantons; a nontrivial flux $G_4|_{M5}$ sources $M2$ brane charge on the M5 worldvolume so a nontrivial superpotential coupling can only be generated if suitable Wilson surface operators, corresponding to insertions of wrapped $M2$ brane states, are introduced to cancel it \cite{Donagi:2010pd,Marsano:2011nn}.  As we do not introduce any such operators, $G_4|_{M5}\ne 0$ forces our instanton corrections to vanish.

We can summarize the basic elements of the mapping as follows
\begin{equation}\begin{array}{ccc}\text{Object on F-theory Side}& & \text{Object on Het Side} \\ \hline
M5 &\leftrightarrow & \CE \\
c_{cyl} &\leftrightarrow & c_{Het} \\
\CN_{cyl} & \leftrightarrow & \CN_{c,Het} \\
C_3 & \leftrightarrow & \CL_{\Gamma} \\
G_4 & \leftrightarrow & c_1(\CL_{\Gamma}).
\end{array}\end{equation}
In the first three lines, the objects are related through pullback or projection by the $\mathbb{P}^1$ fibration $\nu:D3\rightarrow \Sigma$ of $D3$ or the corresponding $\mathbb{P}^1$ fibration $p_{M5,cyl}:c_{cyl}\rightarrow c_{Het}$ of the restricted cylinder $c_{cyl}$.  The last two lines emphasize something nice that we find upon restriction to $M5$ that does not happen in the $dP_9$ fibration, $Y_4'$.  To establish the heterotic/F-theory dictionary between heterotic bundle data and F-theory flux data, we had to split the heterotic bundle according to
\begin{equation}c_1(\CN_{Het}) = c_1(\CO(\lambda\gamma_{Het} + r_{Het}/2))\end{equation}
where $\lambda\gamma_{Het}$ and $r_{Het}/2$ are both half-integral divisor classes on $\CC_{Het}$ in general.  The splitting was important not just for convenience but because it is $\lambda\gamma_{Het}$, not $c_1(\CN_{Het})$, that maps to $G$-flux{\footnote{The half-integrality of $\lambda\gamma_{Het}$ matches with the half-integrality of $G_4$ that follows from \eqref{quantrule} and the fact that $c_2(Y_4,\mathbb{Z})$ is typically odd.}}.  The half-integrality of $\lambda\gamma_{Het}$ can be cumbersome to deal with, though, as it tempts us to work at the level of Chern classes, where the split into $\lambda\gamma_{Het}$ and $r_{Het}/2$ makes sense, as opposed to the level of line bundles where it doesn't.  Working with Chern classes has the potential to lose important information.

When we restrict the heterotic/F-theory duality map to $M5$, we find a more pleasing situation.  The restriction of $\lambda\gamma_{Het}+r_{Het}/2$ is $\lambda\Gamma_{Het} + R_{Het}/2$ where $R_{Het}$ is the ramification divisor of the covering $p_{c,Het}:c_{Het}\rightarrow \Sigma$.  The curve $c_{Het}$ is a 3-sheeted covering of $\Sigma=\mathbb{P}^1$ so $R$ is guaranteed to be even and the objects $\CO(\lambda\Gamma_{Het})\equiv \CL_{\Gamma}$ and $\CO(R_{Het}/2)$ make sense as honest line bundles.  This allows us to write
\begin{equation}\CN_{c,Het} = \CL_{\Gamma}\otimes \CO(R_{Het}/2).\label{CNhetdecomp}\end{equation}
As we saw, the connection $A_{\Gamma}$ on $\CL_{\Gamma}$ maps directly to $C_3|_{M5}$ via the restricted cylinder map \eqref{restrictedcylindermapC3} \eqref{restrictedcylindermapG4}.  The extra information that we retain about $C_3$ that is not found in $G_4$ is crucial; we know that $G_4|_{M5}=0$ for any M5-instanton that generates a nonzero superpotential coupling without operator insertions.
All of the nontrivial bundle behavior depends on the specific configuration of $C_3$ on $M5$.  This will be important in section \ref{subsec:pfaff2form}.  As we emphasized at great length in section \ref{subsec:hetpfaff}, however, the decomposition \eqref{CNhetdecomp} depends on the choice of square root for $\CO(R_{Het})$.  There are $2^{2g_{c_{Het}}}$ such choices in general which are in 1-1 correspondence with the choices of spin structure on $c_{Het}$.  This potentially leads to an ambiguity when we try to define $C_3$ in the F-theory dual.  Ultimately, the physics should not depend on this ambiguity and indeed we will see that the partition function of the M5 instanton is insensitive to it.

\subsection{Heterotic Right-Movers and M5 Fermi Zero Modes}

Before getting to the $C_3$-dependence, let us briefly discuss the
matching of heterotic right-moving fermi zero modes with the fermi
zero modes of the corresponding M5 instanton.  Recall that the
`non-universal' right-moving fermi zero modes of the heterotic
worldsheet instanton were counted by
\begin{equation}2\times\left[h^{0}(\Sigma,N_{\Sigma/B_2}) + h^1(\Sigma,N_{\Sigma/B_2}) + h^0(\Sigma,\CO) + h^1(\Sigma,\CO)\right]\label{hetfzero}\end{equation}
while the corresponding zero modes of an M5 instanton are counted by
\begin{equation}2\times\left[h^0(D3,N_{D3/B_3})+ h^1(D3,N_{D3/B_3}) + h^2(D3,N_{D3/B_3}) + h^1(D3,\CO)+h^2(D3,\CO)\right]\label{M5fzero}
\end{equation}
To relate these formulae, we first recall that the base $B_3$ is a $\mathbb{P}^1$-fibration \eqref{B3type}
\begin{equation}B_3 = \mathbb{P}(\CO\oplus N)\label{B3P1def}\end{equation}
This means that the base $D3$ of our M5 instanton is a $\mathbb{P}^1$ fibration over $\Sigma$ given by
\begin{equation}D3 = \mathbb{P}(\CO\oplus N|_{\Sigma})\qquad \nu:D3\rightarrow \Sigma\end{equation}
and, furthermore, that the normal bundle $N_{D3/B_3}$ is the pullback of the normal bundle of $\Sigma$ in $B_2$
\begin{equation}N_{D3/B_3} = \nu^*N_{\Sigma/B_2}.\end{equation}
We can now use Leray to relate cohomologies on $D3$ to cohomologies on $\Sigma$
\begin{equation}\begin{split}
H^0(D3,N_{D3/B_3}) &= H^0(\Sigma,R^0\nu_*N_{D3/B_3}) \\
H^1(D3,N_{D3/B_3}) &= H^0(\Sigma,R^1\nu_*N_{D3/B_3}) + H^1(\Sigma,R^0\nu_*N_{D3/B_3}) \\
H^2(D3,N_{D3/B_3}) &= H^1(\Sigma,R^1\nu_*N_{D3/B_3}).
\end{split}\end{equation}
For $N_{D3/B_3}=\nu^*N_{\Sigma/B_2}$, the direct image sheaves are particularly simple
\begin{equation}R^0\nu_*N_{D3/B_3} = N_{\Sigma/B_2}\qquad R^1\nu_*N_{D3/B_3} = 0\end{equation}
where $N$ is the bundle from \eqref{B3P1def}.  This leads to
\begin{equation}\begin{split}H^0(D3,N_{D3/B_3}) &= H^0(\Sigma,N_{\Sigma/B_2}) \\
H^1(D3,N_{D3/B_3}) &= H^1(\Sigma,N_{\Sigma/B_2}) \\
H^2(D3,N_{D3/B_3}) &= 0.
\end{split}\end{equation}
Repeating this exercise with $N_{D3/B_3}$ replaced by $\CO$ we have
\begin{equation}\begin{split}H^1(D3,\CO) &= H^1(\Sigma,\CO) \\
H^2(D3,\CO) &= H^2(\Sigma,\CO)
\end{split}\end{equation}
so that the M5 zero modes \eqref{M5fzero} are counted by 
\begin{equation}2\times\left[h^{0}(\Sigma,N_{\Sigma/B_2}) + h^1(\Sigma,N_{\Sigma/B_2}) + h^0(\Sigma,\CO) + h^1(\Sigma,\CO)\right]\end{equation}
which is equivalent to \eqref{hetfzero}. Concluding, the M5-instanton Fermi zero modes map to right moving
fermi zero modes of heterotic worldsheet instantons in the case of duality.

\subsection{Heterotic Pfaffian and the Chiral 2-form Partition Function}
\label{subsec:pfaff2form}

We finally turn to the contributions from left-movers to our heterotic instanton and the chiral 2-form to the M5.  
Recall that left-moving fermions contribute $\pfaff$, the Pfaffian of the Dirac operator on $\Sigma$ coupled to the bundle $V_{\Sigma}=V|_{\Sigma}$.  Because $V_{\Sigma}$ is the pushforward of a line bundle on the covering curve $c_{Het}$
\begin{equation}V_{\Sigma} = p_{c,Het*}\CN_{c,Het}\end{equation}
the study of fermions on $\Sigma$ coupled to the non-Abelian bundle $V_{\Sigma}$ can be lifted to the study of fermions on $c_{Het}$ coupled to an Abelian one.  In particular, we saw that the zero modes of $D_{-,V_{\Sigma}}$ on $\Sigma$ lifted to elements of
\begin{equation}H^p(c_{Het},\CN_{c,Het}\otimes p_{c,Het}^*K_{\Sigma}^{1/2}) = H^p(c_{Het},K_{c,Het}^{1/2}\otimes \CL_{\Gamma})\end{equation}
on $c_{Het}$ where we introduce $\CL_{\Gamma}$ by taking advantage of the decomposition \eqref{CNhetdecomp}
\begin{equation}\CN_{c,Het} = \CL_{\Gamma}\otimes \CO(R_{Het}/2)\label{CNhetsplit}\end{equation}
along with the fact that
\begin{equation}K_{c,Het}^{1/2} = \CO(R/2)\otimes p_{c,Het}^*K_{\Sigma}^{1/2}\end{equation}
is a theta characteristic 
(i.e. square root of the canonical bundle of) of $c_{Het}$.  This allows us to compute the Pfaffian of interest as
\begin{equation}\pfaff\sim \pfaffHet.\end{equation}
We saw in section \ref{subsec:hetpfaff} that $\pfaffHet=0$ unless $\text{deg}(\CL_{\Gamma})=0$, in which case it varies with $\CL_{\Gamma}$ as a holomorphic section of a line bundle on $\CJ_0(c_{Het})$, the space of degree 0 bundles on $c_{Het}$.  On general grounds, the Pfaffian of free fermions coupled to a flat bundle $\CL_{\Gamma}$ on a curve $c_{Het}$ is proportional to one of the $2^{2g_{c_{Het}}}$ $\theta$ functions on $\CJ_0(c_{Het})$; the trick is determining the right one.  The ambiguity here is an artificial one that arose from our insistence on splitting $\CN_{c,Het}$ according to \eqref{CNhetsplit}.  Looking instead at $H^p(c_{Het},\CN_{c,Het}\otimes p_{c,Het}^*K_{\Sigma}^{1/2})$, we were able to conclude that
\begin{equation}\pfaff\sim \theta_R(\CN_{c,Het}\otimes p_{c,Het}^*K_{\Sigma}^{1/2})\end{equation}
where $\theta_R$ is the Riemann $\theta$ function, that is the distinguished holomorphic section on $\CJ_{g-1}(c_{Het})$ whose vanishing locus is precisely the set of degree $g_{c_{Het}}-1$  bundles that admit holomorphic sections.  The split \eqref{CNhetsplit} simply means that $\pfaff\sim \pfaffHet$ is a translate of the Riemann $\theta$ function by the spin structure $K_{c_{Het}}^{1/2}$
\begin{equation}\pfaff\sim \theta_R(K_{c,Het}^{1/2}\otimes \CL_{\Gamma}) = \theta_{-K_{c_{Het}}^{1/2}}(\CL_{\Gamma}).\end{equation}
The ambiguity in our definitions of $K_{c,Het}^{1/2}$ and $\CL_{\Gamma}$ is absent in the first term but present in the second because we use $K_{c_{Het}}^{1/2}$ to translate $\theta_R$.

Given this discussion, one might wonder why $\CL_{\Gamma}$ is ever introduced at all.  Indeed, it brings the added complication of keeping track of the spin structure and ensuring that any consistent choice we can make does not impact the physics.  Of course, we saw in section \ref{subsec:hetFinsts} that $\CL_{\Gamma}$ is crucially important in the context of heterotic/F-theory duality because it is this quantity that determines the 3-form $C_3$ on the dual M5 instanton.
When we construct the map
\begin{equation}C_3 = \iota_{M5,cyl*}p_{M5,cyl}^*A_{\Gamma}\end{equation}
we are really defining a map from the space of flat bundles on $c_{Het}$ to the space of `flat $C_3$'s' on $M5$.  That is, we are defining a map
\begin{equation}f_{cyl}:\CJ_0(c_{Het})\rightarrow
  \CJ(M5)\end{equation} where we recall that $\CJ(M5) =
H^3(M5,\mathbb{R})/H^3(M5,\mathbb{Z})$ is the intermediate Jacobian.
This map can be complicated, but
$f_{cyl}$ maps $\theta$ divisors to $\theta$ divisors so the
$2^{2g_{c_{Het}}}$ different $\theta$ functions,
$\theta_{-K_{c,Het}^{1/2}}(\CL_{\Gamma})$ on $\CJ_0(c_{Het})$ are
identified with the $2^{2g_{c_{Het}}}$ different $\theta$ functions
$\theta_a(C_3)$ on $\CJ(M5)$.  That the heterotic Pfaffian maps to a
$\theta$ function on $\CJ(M5)$ is exactly what we need.  The heterotic
left-movers map to the chiral 2-form under heterotic/F-theory duality
so we have
\begin{equation}\pfaff\leftrightarrow \CZ_{\phi}\end{equation}
The 2-form partition function $\CZ_{\phi}$ must be proportional to a
$\theta$ function on $\CJ(M5)$ in order to ensure that the full M5
partition function \eqref{M5partfcn} is gauge invariant. Though the
`right' $\theta$ function can be determined from the spin structures
of $M5$ and $Y_4'$ via Witten's procedure \cite{Witten:1996hc}, we see
that the task is much easier in models with a heterotic dual.

Heterotic duality actually teaches something that goes beyond a
practical tool for picking the `right' $\theta$ function.  We see
clearly in this setting that the choice of $\theta$ function is
intricately connected to the specification of $C_3$ itself.  This is a
level of detail about the 3-form that is easy to miss if we only
specify a $G$-flux by giving the homology class of a holomorphic
surface.  We get a handle on the heterotic Pfaffian, or equivalently
the chiral 2-form partition function, when we are able to talk about
the line bundle $\CN_{c,Het}$ or, equivalently, its pullback to
$c_{cyl}$, $\CN_{M5,cyl}$.  This line bundle is the object that
unequivocally controls the vanishing of $\pfaff\sim \CZ_{\phi}$ and
the spin structure ambiguity arises because a choice of
$K_{c,Het}^{1/2}$ is needed to construct $C_3$ from $\CN_{M5,cyl}$.
In a sense, then, $\CN_{M5,cyl}$ is the object on which the physics
most directly depends.  We see something similar in the $E_6$ gauge
theory where the spectrum is determined by cohomologies of $\CN_{loc}$
without any ambiguity from spin structures.  It is only when we try to
work in terms of quantities derived from $\gamma$ that depend on the
separation \eqref{Nhetdef} that we run into trouble.  Of course any
trouble can be fixed by taking care to keep track of the requisite
spin structures in the problem.  It seems
far simpler to work directly with the bundle $\CN_{cyl}$, though, as
it is the object that directly affects the physics.  Any spin
structure dependence can be absorbed into a twist of the map from the
connection $A_{cyl}$ of $\CN_{cyl}$ to $C_3$.

\section{Instanton Prefactors in F-theory}
\label{sec:instantons in F}

We have discussed at length how one might think of instanton
prefactors in F-theory from an M-theory and a heterotic point of view,
and also the relationship between the two. At a fundamental level, the
instanton prefactor in F-theory is determined by configurations of
$C_3$ in the defining $d=3$ M-theory compactification, which are
elements of the intermediate Jacobian $\CJ(M5)$. If the M-theory flux
$G_4$ is cohomologically trivial on $M5$, the prefactor is the
partition function $\CZ_\phi$ and is given by a theta divisor on the
intermediate Jacobian. In the case of an M5-instanton which is dual to
a heterotic worldsheet instanton, the prefactor is dependent via
duality on the fermionic left-movers which couple to the gauge bundle
$V$ that breaks $E_8$ to the GUT group. Recalling that $\CL_A\equiv
K_{c_{Het}}^{1/2} \otimes \cL_\Gamma$, these modes are counted by the
Hodge numbers $h^i(c_{Het},\cL_A)$
where $c_{Het}$ is a spectral cover of the instanton curve
$\Sigma$. 
From sections \ref{subsec:hetpfaff} and
\ref{subsubsec:chiral2form}, in fact, we know how to determine the partition function
of these modes and, correspondingly, the partition function of the M5 chiral 2-form, both
of which correspond to $\theta$ functions.
While these issues of $\theta$ functions are important,
they are typically not important for understanding how the prefactor 
varies as a function of the 
available moduli.  The reason for this is that the full domain of the
$\theta$ function, either $\CJ_{g-1}(c_{Het})$ for our heterotic
instantons or $\CJ(M5)$ for our $M5$ instantons, is typically not
accessible, since the connection $A_\Gamma$ and three-form $C_3$ are
restricted from global objects defined outside of the instanton
geometry.  Moreover, we would like to express the prefactor in terms
of extrinsic moduli which explicitly determine the seven-brane
structure of the compactification. This depends upon the embedding of
$c_{Het}$ (or $c_{loc}$ in what follows) into the geometry, as opposed
to the theta functions that are intrinsic to $c_{Het}$. The
relationship between the `extrinsic' and `intrinsic' viewpoints is
discussed in \cite{Curio:2009wn} for heterotic worldsheets.

Study of this line bundle cohomology allows one to determine
how the M5-instanton prefactor depends on the moduli of $V$ via the
algebro-geometric techniques of \cite{Buchbinder:2002ic}, and is
equivalent to the study of the vector bundle cohomology
$h^i(\Sigma,V|_\Sigma \otimes \cO_\Sigma(-1))$. In this picture, the
moduli in the prefactor appear in the defining equation of $c_{Het}$
and also explicitly in the defining equation of $Y_4$; thus, it is a
simple matter to see the seven-brane dependence of the prefactor once
it is computed. However, the appearance of seven-brane physics is
rather unintuitive when computed via the study of line bundle
cohomology on $c_{Het}$, since this curve exists away from the GUT
stack and only in the case of a heterotic dual.  In particular the
type IIb intuition of zero modes localized near branes is lacking, a
fact which we would like to rectify.

In this section we note that in the heterotic case, the dependence of the
prefactor on the bundle $V$ can be described via the study of line bundle cohomology
on another, isomorphic curve $c_{loc}$. This has an analogue in F-theory that
naturally appears in the GUT geometry. We propose this as the right way to
compute the moduli dependence of the F-theoretic prefactor.
The curve $c_{loc}$ is a
multi-sheeted cover of the curve $\Sigma$ in $B_3$ where the GUT stack
intersects the instanton, and thus the associated zero modes admit a
natural IIb interpretation in terms of 3-7 strings. Moreover, the
relevant line bundle cohomology on $c_{loc}$ is isomorphic to the
discussed line bundle cohomology on $c_{Het}$ when a dual exists, and
thus the prefactors are equivalent. However, the curve $c_{loc}$ has
the advantage that it exists even in the absence of a heterotic
dual. We will make a precise conjecture for how line bundle cohomology
on $c_{loc}$ is
related to the worldvolume theory of the M5 brane and theta divisors
on the intermediate Jacobian $\CJ(M5)$. We will also
discuss the concrete steps necessary to set up the calculation of the prefactor, which
will be utilized in section \ref{sec:examples}.

We will first discuss how to compute the prefactor and why it makes sense from type IIb intuition
and also heterotic duality. We will then make a mathematically precise statement regarding the
relationship to the M5 brane worldvolume theory. We will finish with a concrete discussion of 
how to set up the computation for a given F-theory base, $B_3$.

\subsection{Intuition and a Proposal: What to Compute}

\label{subsec:conjecture}
Recall that when a heterotic dual exists the dependence of the M5-instanton prefactor
on the moduli of $V$ can be determined by the study of $\pfaff$.
In principle, one could study $\pfaff\sim \theta_R(\CN_{c,Het}\otimes
p_{c,Het}^*K_{\Sigma})$ for \emph{any} choice of line bundle $\CN_{c,Het}$ of
the right degree, but the line bundle $\CN_{c,Het}$ we study must arise from the
restriction of a line bundle $\cN_{Het}$ on $\CC_{Het}$. That is,
\begin{equation}\CN_{c,Het} = \CN_{Het}|_{c_{Het}}.\end{equation}
Similarly, the 3-form $C_3$ on $M5$ must arise as the restriction of a
well-defined configuration for $C_3$ on $Y_4'$. 
We are interested in the vanishing
locus and the behavior of $\pfaff\sim \CZ_{\phi}$ as one varies the
complex structure moduli of $\CC_{Het}$ and $Y_4'$.  We reviewed the
procedure of \cite{Buchbinder:2002ji,Buchbinder:2002ic,Buchbinder:2002pr} for
computing this dependence, at least on the heterotic side, in section
\ref{subsec:pfaffbdo}.  This approach is based on transforming the
non-Abelian problem to an Abelian one via
\begin{equation}\pfaff\sim \pfaffHet\end{equation}
and computing $\pfaffHet$ by standard techniques.  The end result is
that the heterotic Pfaffian is determined by the partition function of
free fermions on $c_{Het}$ coupled to a degree zero line bundle
$\CL_{\Gamma}$.  The techniques of \cite{Buchbinder:2002ic} computes
this via study of the bundle $\cL_A \equiv \cL_\Gamma \otimes
K_{c_{Het}}^{1/2}$, and when this bundle has a section the prefactor
has a zero. By duality, this Pfaffian should capture the behavior
of the chiral 2-form partition function as well.

This leads to a natural question: why should the partition function of
an M5 instanton have anything to do with fermions on some auxiliary
curve $c_{Het}$ that emerges in the stable degeneration limit?  One
encounters a similar question when studying the spectrum of the $E_6$
gauge theory in the context of heterotic/F-theory duality.  There, the
heterotic computations are cohomologies on a curve in $Z_{Het}$, which
sits at $W=0$.  This occurs away from the $E_6$ singular locus at
$Z=0$, where the $E_6$ charged degrees of freedom are localized.  The
resolution is that the proper computation on the F-theory side
involves cohomologies on curves in $\CC_{loc}$, which sits inside the
$E_6$ singular fibration over $Z=0$.  When a heterotic dual exists, it
happens that the spectral divisor $\CC_F$ becomes a $\mathbb{P}^1$
fibration, $\CC_{cyl}$, and this provides a second copy of
$\CC_{loc}=\CC_{Het}$ at $W=0$ as the second section.  The F-theory
computations can be translated from $Z=0$ to $W=0$ using this
$\mathbb{P}^1$ fibration and, in fact, the copy of
$\CC_{loc}=\CC_{Het}$ at $W=0$ is easier to work with because we don't
have to pass to the resolution of $Y_4$ to see it. For emphasis, the
conclusion is that the charged degrees of freedom are localized on
the GUT stack, but when a heterotic dual exists one could instead
study an isomorphic curve and line bundle away from the GUT stack
inside the heterotic threefold.

In a similar way, a copy of the curve $c_{Het}$ also sits above the
$E_6$ singular locus at $Z=0$.  It is just the curve
$c_{loc}=p_{loc}^*\Sigma$, inside the Higgs bundle spectral cover that
emerges from the restriction $\CC_{loc} = \CC_{cyl}|_{Z=0}$. As
a reminder, $\CC_{loc}$ is a multi-sheeted cover of $B_2$ and 
$c_{loc}$ is the multi-sheeted cover of $\Sigma\subset B_2$
derived from $\CC_{loc}$ by restriction.  While
fermions on some curve $c_{Het}$ at $W=0$ do not have an obvious
physical interpretation, fermions on $c_{loc}$, which sits above the
curve $\Sigma$ where $M5$ meets the $E_6$ branes, do: they should
correspond to an analog of 3-7 strings!  This suggestion is not unique
to us; the curve $c_{Het}$ has even been referred to as $\Sigma_{37}$
in \cite{Donagi:2010pd} and related cohomologies were also discussed
in \cite{Blumenhagen:2010ja}.  Our precise description
of the compact relation of $\CC_{loc}=\CC_{cyl}|_{\pi^*B_2}$, however,
allows us to identify the compact curve $c_{loc}$ for generic $E_6$ (or $SO(10)$ or $SU(5)$)
models in a way that reproduces $c_{loc}=c_{Het}$ when a heterotic
dual exists.  Further, if we identify $p_{loc,*}\CN_{loc}$ as the
gauge bundle $V$ associated to the local model on $B_2$, the
cohomologies on $c_{loc}$ are simply counting fermions localized
on the intersection $B_2\cap D3 = \Sigma$ that couple to
$V_{\Sigma}=V|_{\Sigma}$
\begin{equation}H^p(c_{loc}, \cL_A) \equiv
  H^p(c_{loc},p_{c,loc}^*K_{\Sigma}^{1/2}\otimes \CN_{c,loc}) =
  H^p(\Sigma,K_{\Sigma}^{1/2}\otimes
  V_{\Sigma}).\label{M5cohomsgeneral}\end{equation}
This is exactly how we would expect 3-7 modes to appear from the perspective of the worldvolume theory on $B_2$.

With our proper, compact description of the curve $c_{loc}$ in hand,
we propose that the cohomologies \eqref{M5cohomsgeneral} 
control the
vanishing of the M5 partition function for any M5 instanton in
the $E_6$ models we consider.  More generally, if we have multiple non-Abelian
gauge groups that can be described in the local framework of $SU(n)$
Higgs bundles, we expect the vanishing to be controlled by
cohomologies on all of the relevant\footnote{Including those
corresponding to the associated bundles $\wedge^k V$.} $c_{loc}$'s  provided the M5 avoids
any loci where non-Abelian singularities intersect. Specifically, the
proposed relationship between modes on $c_{loc}$ and the prefactor $A$ is
\begin{equation}
  h^i(c_{loc}, \cL_A) \ne 0 \qquad \Rightarrow \qquad A=0
\end{equation}
which is similar to the statement regarding modes on $c_{Het}$ and the
Pfaffian prefactor.
 The motivation
for this proposal is clear.  If the vanishing of the M5 instanton
partition function is controlled by cohomologies on $c_{loc}$, it
suggests that the physics responsible for this vanishing is captured
by the local geometry near $B_2$ and, correspondingly, should not
depend on whether $B_2$ is sitting inside a $\mathbb{P}^1$ fibration
or not. Thus, we have a prescription for computing the dependence of the
instanton prefactor on seven-brane moduli, even in the absence of a heterotic
dual. This gives a very powerful tool, computationally equivalent 
to the computations in \cite{Buchbinder:2002ic}, that
could be applied in a number of phenomenological models.  After all,
everything we say in this paper about $E_6$ can be trivially extended
to the phenomenologically interesting cases of $SO(10)$ and $SU(5)$.

We emphasize that while this prescription is physically sensible from the
type IIb point of view and isomorphic to the prescription for heterotic worldsheet
instantons in the case of duality, the relationship to the M5 brane worldvolume
theory could be put on more solid footing, and hence the result is still conjectural.
We will now address some of this issue.

\subsubsection{Mathematical Relationship to the M5-brane Theory}

Let us state in a mathematically precise way what this conjecture
would mean from the point of view of the M5-brane worldvolume theory,
at least for the $E_6$ models discussed in this paper.  The M5
worldvolume is an elliptic fibration with section over a surface $D3$
in $B_3$ that exhibits an $E_6$ singularity along the curve $\Sigma$
where $D3$ meets $B_2$.  We can describe $M5$ by an equation of the
form \eqref{Y4def}
\begin{equation}y^2 = x^3 + fx z^4v^4 + gz^6v^6 +  z^2v^3\left[\tilde b_0z^3v^3 + \tilde b_2 zv x + \tilde b_3 y\right]\label{Y4defagain}\end{equation}
where we have restricted the sections $b_q$ on $B_3$ should be restricted to $D3\subset B_3$, giving the
sections $\tilde b_q$.  Inside $M5$, we then consider the restriction of the spectral divisor $\CC_F$ \eqref{specdivdef}
\begin{equation}c_F = \CC_F|_{M5}\end{equation}
Recall that $\CC_F$ is a distinguished 3-sheeted cover of $B_3$ inside $Y_4$ that is given in terms of the data in \eqref{Y4defagain} by
\begin{equation}c_{F}:\qquad \tilde b_0z^3v^3 + \tilde b_2 zv x +
  \tilde b_3 y = 0\label{specdivdefagain}\end{equation} less the
components 
that coincide with the section.  Correspondingly, the cover $c_F$ of
$D3$ inside $M5$ is also distinguished and its restriction to
$\pi^*\Sigma$ gives us the curve $c_{loc}$
\begin{equation}c_{loc} = c_F|_{\pi^*\Sigma}\end{equation}
which is a 3-sheeted cover of $\Sigma$.  This curve is isomorphic to $c_{Het}$ when a heterotic dual exists.  The role of $c_F$ is the usual one: line bundles on $c_F$ simultaneously determine the 3-form $C_3$ on $M5$ and bundle data $\CN_{c,loc}$ in the local model.

Our conjecture about vanishing loci is a statement about $\theta$ divisors in $\CJ_{g-1}(c_{loc})$ and $\CJ(M5)$.  More specifically, let $\CA$ denote the subset of $\text{Pic}(c_F)$ given by
\begin{equation}\CA = \{\CN_{c,F}\in\text{Pic}(c_F)\,\,|\,\, \CN_{c,F}|_{c_{loc}}\otimes p_{c,loc}^*{K_{\Sigma}^{1/2}}\in \CJ_{g-1}(c_{loc})\}\end{equation}
We have the obvious map
\begin{equation}h_{loc}:\CA\rightarrow \CJ_{g-1}(c_{loc})\qquad h_{loc}:\CN_{c,F}\mapsto \CN_{c,F}|_{c_{loc}}\otimes p_{c,loc}^*K_{\Sigma}^{1/2}\label{hlocmap}\end{equation}
which determines the local data corresponding to a bundle $\CN_{c,F}$.  We also have the analog of the map \eqref{C3M5map} for determining $C_3$
\begin{equation}h_{M5}:\CA\rightarrow \CJ(M5)\qquad h_{M5}:\CN_{c,F}\mapsto \iota_{M5,c*}\left(A_{M5,c}-\frac{1}{2}A_{\hat{R}}\right)\label{hM5map}\end{equation}
Here $\iota_{M5,c}:c_F\rightarrow M5$ is the embedding map, $A_{M5,c}$ is a holomorphic connection on $\CN_{c,F}$, and $A_{\hat{R}}$ is a holomorphic connection on the bundle $\CL_{\hat{R}}=\CL_{\hat{r}}|_{M5}${\footnote{We can define $\CL_{\hat{R}}$ intrinsically on $M5$ as follows.  Start with the bundle $\CO_{c_F}(R_F)$ where $R_F$ is the ramification divisor of the covering $c_F\rightarrow D3$.  Now, consider the normal bundle $N_{c_{loc}/c_F}$ and let $L'$ denote the bundle obtained by removing the part of $N_{c_{loc}/c_F}$ that is pulled back from $D3$.  We take $\CL_{\hat{R}} = \CO_{c_F}(R_F)\otimes L^{',-1}$ and it has the property that $\CL_{\hat{R}}|_{c_{loc}}=\CO_{c_{loc}}(r_{loc})$ with $r_{loc}$ the ramification divisor of the covering $c_{loc}\rightarrow\Sigma$.}}.
Our conjecture is that the M5 instanton partition function vanishes for any choice of $\CN_{c,F}$ that maps to $\Theta_R\subset \CJ_{g-1}(c_{loc})$ under $h_{loc}$.  If we let $\CA_{\text{vanish}}$ denote the conjectured vanishing locus
\begin{equation}\CA_{\text{vanish}}=\{\CN_{c,F}\in \CA\,\,|\,\, h_{loc}(\CN_{c,F})\in \Theta_R\subset \CJ_{g-1}(c_{loc})\}\end{equation}
then our claim is that there exists a theta divisor $\Theta_{M5,vanish}\subset \CJ(M5)$ such that
\begin{equation}h_{M5}(\CA_{\text{vanish}})\subset \Theta_{M5,vanish}\end{equation}
The $\theta$ function corresponding to $\Theta_{M5,vanish}$ should be the `right' one that controls the partition function of the chiral 2-form on $M5$.

What we describe here is a `strong form' of the conjecture.  It should be noted that physics only requires a weaker version in which we specify $\CC_F$ as well as its restriction $c_F$ and consider only the subsets of $\CA$ and $\CA_{\text{vanish}}$ that correspond to bundles on $c_F$ that descend from bundles on $\CC_F$.

\subsection{Setting Up the Computation}
\label{sec:computing}
In the end, to compute the instanton prefactor we compute
moduli-dependent line bundle cohomology on a spectral curve $c_{loc}$,
which is the restriction of the local
model spectral cover $\CC_{loc}$ to the instanton worldvolume. We
will utilize the fact that  $\CC_{loc}$ is a compact spectral cover 
inside an auxiliary elliptic Calabi-Yau threefold $\pi_{3,F}: Z_{3,F} \rightarrow B_2$, as
discussed in section \ref{subsubsec:higgscomment}, which in turn
gives $c_{loc}$ as a curve inside the elliptic surface $\cE\equiv \pi_{3,F}^*\Sigma$. 
The advantage of this approach is that it simplifies computations. We will see that
one ends up computing the cohomology a line bundle on $c_{loc}$ via a Koszul sequence
from a line bundle on $\cE$.
Though
this computation is in an F-theory compactification which
may or may not have a heterotic dual, the mathematical computation is \emph{identical} to
those performed in \cite{Buchbinder:2002pr}, to which
we refer the reader for detailed examples.  Instead, here we will describe
how to obtain the topological data necessary to set up a computation along the
lines of \cite{Buchbinder:2002pr}. In particular, one must know
that structure of the elliptic surface $\cE$, the class of $c_{loc}$ inside $\cE$,
and how the bundle $\cL_A$ is obtained via a restriction from a line
bundle on $\cE$. The computation of cohomology on $c_{loc}$ as
a divisor in $\cE$ is a \emph{choice}, albeit a convenient one, and we
emphasize that it is the cohomology on $c_{loc}$ which governs the physics,
regardless of ambient space.

Recall from \ref{subsubsec:higgscomment} that $\pi_{3,F}:Z_{3,F}\rightarrow B_2$ is the ambient elliptic Calabi-Yau threefold in
which $\cC_{loc}$ is a divisor.   The class of $\cC_{loc}$ is
\begin{equation}
  [\cC_{loc}] = n\sigma_{3,F} + \pi_{3,F}^{-1} \eta
\end{equation}
where $\eta$ is a curve in $B_2$ of class $6\cs + N_{B_2|B_3}$ and $\sigma_{3,F}$ is
the section of $Z_{3,F}$. 
$\Sigma$ is a $\bP^1$ in $B_2$, and is the base of an elliptic surface $\cE\equiv \pi_{3,F}^{-1} \Sigma$. For instanton physics, we are interested in the curve $c_{loc}$.
Recalling that $c_{loc} = \CC_{loc}\cdot \cE$ inside $Z_{3,F}$, the class of 
$c_{loc}$ inside $\cE$ is then
\begin{equation}
\label{eqn:cloc class}
  [c_{loc}] = [n\sigma_{3,F} + \pi_{3,F}^{-1}]\cdot \Sigma = ns + rF
\end{equation}
where $s\equiv \sigma_{3,F}|_{\pi_{3,F}^{-1}\Sigma}$ is the section of the elliptic surface $\cE$,
$F$ is the fiber class of $\cE$ and $r\equiv \eta \cdot_{B_2} \Sigma\in \bZ$.
From \eqref{eqn:clocgeneral} we write the defining equation of $c_{loc}$    
\begin{equation}
f_{c_{loc}} = \sum_{q=1}^n \,\,\tilde b_q \,\, v^{n-q} x^{n_x} y^{n_y}
\label{eqn:fclocgeneral}
\end{equation}
which just reduces to 
\begin{equation}
f_{c_{loc}} = \tilde b_0 \, v^3 + \tilde b_2 \, vx + \tilde b_3 \, y
\end{equation}
in the $E_6$ case ($n=3$) that we consider in examples. The sections $\tilde b_q$ are obtained
via restriction of $b_q$ to $\cE$ and are sections of $\cO_\Sigma(r-q\chi)$.

Let us now discuss the relevant line bundles on $\CC_{loc}$ and $c_{loc}$. In both cases, these
bundles will be obtained via restriction from bundles on an ambient space of the same name,
abusing notation.
The line bundle  $\cN_{loc}$ on $\CC_{loc}$ can be obtained via restriction
from a line bundle on $Z_{3,F}$ of first Chern class 
\begin{equation}
c_1(\cN_{loc}) = \frac{1}{2}(n\sigma_{3,F} + \pi_{3,F}^{-1}\eta + \pi_{3,F}^{-1}\cs) + \lambda (n\sigma_{3,F} - (\pi_{3,F}^{-1}\eta - n \pi_{3,F}^{-1}\cs)).
\end{equation}
Restricting to $c_{loc}$ gives a bundle $\cN_{c,loc}\equiv \cN_{loc}|_{c_{loc}}$
which can be obtained via restriction from a line
bundle on $\cE$ of first Chern class
\begin{equation}
c_1(\cN_{c,loc}) = \frac{1}{2}(ns + (r+\chi)F) + \lambda (ns - (r-n\chi)F).
\end{equation}
where $\chi\equiv \cs\cdot_{B_2}\Sigma$. 
In terms of
data on $\cE$, the bundle $\cL_A$ in (\ref{M5cohomsgeneral}) relevant for the study of
the prefactor can be written $\cL_A\equiv \cN_{c,loc} \otimes \cO_\cE(-F)|_{c_{loc}}$, so we have
\begin{equation}
c_1(\cL_A) = (\lambda + \frac{1}{2})n\, s + [r(\frac{1}{2} - \lambda) + \chi(\frac{1}{2} + n\lambda)-1)]F
\label{eqn:LAchern}
\end{equation}
and one must compute $h^i(c_{loc},\cL_A|_{c_{loc}})$ in order to determine the prefactor. This
is done efficiently via the long exact sequence in cohomology associated to the Koszul
sequence
\begin{equation}
0 \rightarrow \cL_A\otimes \cO_\cE(-c_{loc}) \xrightarrow{f_{c_{loc}}} \cL_A \rightarrow \cL_A|_{c_{loc}} \rightarrow 0
\label{eqn:prefactor koszul}
\end{equation}
as in \cite{Buchbinder:2002pr}.
Though the restriction is explicit, remember that we abuse notation and use the name
$\cL_A$ for both the bundle on $\cE$
and its restriction to $c_{loc}$. 

Armed with these definitions and an F-theory compactification on a
particular base $B_3$ and an $E_6$ GUT along $B_2$, one can scan for
divisors in $B_3$ or $Y_4$ which satisfy the constraints \eqref{eqn:D3 W cont 1} and \eqref{eqn:D3 W cont 2} 
sufficient for an instanton on that divisor to correct the
superpotential. Knowing the instanton divisor, the numbers $r$ and
$\chi$ can be computed directly and together with an appropriately
quantized $\lambda$ one has all of the information necessary to
compute the dependence of the prefactor on the moduli of $c_{loc}$.
We will make this explicit in the examples of section \ref{sec:examples}. One nice
feature of this computation is that all of the requisite information
can be determined directly from $B_3$, allowing for a simple scan over
F-theory bases.

As emphasized, the computations we study are in the
$E_6$ case where $n=3$, but this formalism can also be applied to compute the components of the
instanton prefactor in $SU(5)$ GUTs with $n=5$. For cases $n>3$, however, one should in principle also study
the zero modes of the ``associated bundles''. By this we mean that the $\cL_A$ cohomology on
$c_{loc}$ here is equivalent to the bundle cohomology $V$ on $\Sigma$, but for higher rank bundles
one should also study $\wedge^k V$ cohomology on $\Sigma$.

\section{A Prefactor Scan in the Kreuzer-Skarke Database}
\label{sec:scan}

In a series of seminal works Kreuzer and Skarke classified all
three-dimensional\cite{Kreuzer:1998vb} and
four-dimensional\cite{Kreuzer:2000xy} reflexive polytopes. Associated to every
$d$-dimensional reflexive polytope is a $d$-dimensional toric variety
which admits a $(d-1)$-dimensional Calabi-Yau hypersurface. To do so, the anticanonical bundle
of the toric variety must have a section. While we will not be using this method to construct
Calabi-Yau hypersurfaces, the fact that the anticanonical bundle admits a section will be useful
for constructing F-theory compactifications.

In this brief section we will perform a scan of a large class of
threefold bases studying the statistics of possible instanton
corrections to the superpotential. We will do so by systematically collecting the
topological data necessary to set up the computation of the instanton
prefactor, as outlined in section \ref{sec:computing}. Rather than taking the Calabi-Yau hypersurface in the
$d$-dimensional toric variety $B_d$ associated to a $d$-dimensional
polytope, we will use the toric varieties $B_3$ in the Kreuzer-Skarke
list \cite{Kreuzer:1998vb} to be three-fold bases of F-theory
compactifications.  That the anticanonical bundle $K_{B_3}^{-1}$
admits a section ensures that $f$ and $g$ in the Weierstrass equation
(or alternatively the sections in the Tate form) have sections and
therefore the F-theory compactification exists.

We perform a simple scan of the Kreuzer-Skarke database as follows:
\begin{itemize}
  \item Find all fine triangulations of the 4319 reflexive polytopes in \cite{Kreuzer:1998vb}.
  \item Study the toric variety associated to each fine triangulation and keep each smooth toric variety
        as a candidate $B_3$.
  \item For each smooth $B_3$, scan the toric divisors for divisors $D3$ with $\chi(D3,N_{D3|B_3})=0$. These
        are the candidate instanton divisors in $B_3$.
  \item For each candidate instanton divisor $D3$, scan through the toric divisors for candidate GUT divisors $B_2$ and keep
        those $B_2$ which intersect $D3$ at a $\bP^1$ in $B_3$. This is $\Sigma$.
      \item For each pair $(D3,B_2)$ compute $r$ and $\chi$ as defined
        in section \ref{sec:computing}. Keep only those with $r>0$
        since otherwise $c_{loc}$ is not defined. Use $r$ and $\chi$
        to determine $\cL_A$ for $n=3$ and $\lambda =
        \frac{3}{2}$. Together, this data gives the class $c_{loc}$
        and the line bundle $\cL_A$ and therefore one could compute
        the prefactor at this point. Write $\cL_A \equiv
        \cO_\cE(Ms+NF)|_{c_{loc}}$.
  \item Add $B_3$ to the list of toric varieties which realize an instanton with prefactor determined by these values of $r$ and $\chi$.
\end{itemize}
There are a number of ways that one might generalize or strengthen this scan. One is to consider instanton or GUT divisors
which are not toric divisors, defined by the vanishing of a single homogeneous coordinate. Another is to consider different values for
$n$ or $\lambda$. For $n=3$ and $\lambda = \frac{1}{2}$, it is easy to show that $\cL_A|_{c_{loc}}$ almost always has a section and therefore 
the prefactor vanishes identically. For $n=3$ and $\lambda > \frac{3}{2}$ it is much more likely that one has
to add $\ov{D3}$-branes for
the sake of tadpole cancellation.

Having discussed possible generalizations and caveats, let us turn to
a discussion of the results of this scan presented in Table
\ref{table:scan}. This table contains the results for Kreuzer-Skarke
database except for polytope $4309$, which we discuss on its own. Each
row has a unique set of integers $r$, $\chi$, $M$, $N$ which determines
the class of $c_{loc}$ and the line bundle $\cL_A$ on it. That is,
$r$, $\chi$, $M$, and $N$ are the topological quantities which are the
inputs necessary to the compute the prefactor. The first row
corresponds to a prefactor which has already been computed in the
heterotic literature and it is identically zero. The second is the
prefactor of the F-theory GUT in section \ref{sec:ex two} and was also
previously computed in the heterotic literature. This prefactor
vanishes if and only if there exists a point of $E_8$ enhancement in
the instanton worldvolume. The third example is a new prefactor,
computed in section \ref{sec:example three} of this paper. It is
identically zero. To the authors' knowledge, none of the other
examples have been computed in the literature. It is worth noting that
the computational complexity of each example goes up as one progresses
down the rows. This can roughly be seen from the fact that $r$ is
increasing, which means that there are more moduli appearing in
$f_{c_{loc}}$, and the fact that $\chi$ is increasing, which roughly
means that the size of a moduli-dependent matrix is increasing. The
determinant of this matrix is the prefactor.

\begin{table}
\centering
\begin{tabular}{cccc|c|c}
  $r$ & $\chi$ & $M$ & $N$ & Multiplicity & Comments \\ \hline
 $1$ & $0$ & $6$ & $-2$ & $5078$ & $\Pfaff=0$, Ex. 3 of \cite{Buchbinder:2002pr}\\
  $5$ & $1$ &  $6$ &  $-1$ & $28267$ & pts of $E_8$, Ex. 2 of \cite{Buchbinder:2002pr}\\
  $6$ & $1$ & $6$ & $-2$ & $32634$ & $\Pfaff=0$, Ex. 3 of section \ref{sec:examples}. \\
  $7$ & $1$ & $6$ & $-3$ & $3217$ & not computed\\
 $8$ & $1$ & $6$ & $-4$ & $55$ & not computed\\
$11$ & $2$ & $6$ & $-2$ & $30670$ & not computed \\
 $12$ & $2$ & $6$ & $-3$ & $5996$ & not computed\\
 $13$ & $2$ & $6$ & $-4$ & $168$ & not computed\\
 $16$ & $3$ & $6$ & $-2$ & $4284$ & not computed\\
 $17$ & $3$ & $6$ & $-3$ & $2827$ & not computed\\ 
 $18$ & $3$ & $6$ & $-4$ & $155$ & not computed\\
 $19$ & $3$ & $6$ & $-5$ & $7$ & not computed\\
 $23$ & $4$ & $6$ & $-4$ & $33$ & not computed\\ 
\end{tabular}
\caption{Results of a prefactor scan in the Kreuzer-Skarke list of toric threefolds, not including threefolds derived from polytope 4309. The prefactors associated to the
first two rows of data are computed elsewhere in the literature. This scan is for $E_6$ GUTs ($n=3$) and $\lambda=\frac{3}{2}$. }
\label{table:scan}
\end{table}

\begin{table}
\centering
\begin{tabular}{cccc|c|c}
  $r$ & $\chi$ & $M$ & $N$ & Multiplicity & Comments \\ \hline
 $1$ & $0$ & $6$ & $-2$ & $24576$ & $\Pfaff=0$, Ex. 3 of \cite{Buchbinder:2002pr}\\
  $5$ & $1$ &  $6$ &  $-1$ & $98304$ & pts of $E_8$, Ex. 2 of \cite{Buchbinder:2002pr}\\
  $6$ & $1$ & $6$ & $-2$ & $73728$ & $\Pfaff=0$, Ex. 3 of section \ref{sec:examples}. \\
$11$ & $2$ & $6$ & $-2$ & $73728$ & not computed \\
 $16$ & $3$ & $6$ & $-2$ & $24576$ & not computed\\
\end{tabular}
\caption{Results of a prefactor scan which utilizing the toric varieties of all fine triangulations of Kreuzer-Skarke
polytope $4309$ as F-theory base manifolds $B_3$. The prefactors associated to the
first two rows of data are computed elsewhere in the literature. This scan is for $E_6$ GUTs ($n=3$) and $\lambda=\frac{3}{2}$. }
\label{table:scan 4309}
\end{table}

The primary reason for presenting these results, though, is that there
are over $100,000$ prefactors, and only thirteen of them are
unique as polynomial functions. To be more specific, every row corresponds to the
computation of a matrix determinant and yields a moduli-dependent
polynomial function $f(\phi)$ which \emph{is} the instanton prefactor
by holomorphy. There are only
thirteen explicit computations to perform, and given those
polynomials one could substitute in the appropriate moduli in the over
$100,000$ examples. We find this degeneracy interesting.

This has clear implications for moduli stabilization. First, the
prefactors are highly structured and there are not very many unique
functions. Second, many of them are identically zero which would
greatly affect K\" ahler moduli stabilization. Such a prefactor can be
the determinant of a highly non-trivial matrix in moduli, as
exemplified in section \ref{sec:example three}, and determining that there is a zero
mode and therefore an identically zero prefactor requires an explicit
construction of the moduli map. Though this occurred while studying a
seven-brane dependent instanton prefactor in F-theory, this is a more
generic mathematical phenomenon and should occur in other
compactifications as well. Therefore, one should not simply assume an
$\cO(1)$ prefactor, particularly given the fact that identically zero
prefactors were very common in our scan. Third, there are $28,267$
examples where the prefactor vanishes if and only if there exists a
point of $E_8$ in the instanton worldvolume.  It would be interesting
to compute the remaining prefactors in the list and to study these
ideas further.

Finally, the analogous results for polytope $4309$ are given in table
\ref{table:scan 4309}. We see that compactifications utilizing the
toric variety associated to a fine triangulation of this polytope as
F-theory bases $B_3$ give a very high multiplicity of instanton
prefactors. These results have been excluded from table
\ref{table:scan} because they would skew the results. Similar
conclusions regarding the structure of prefactors follow, however, and
it is interesting to note the obvious symmetry of multiplicities.

One may wonder why polytope $4309$ gives rise to such high multiplicities. 
Examining the polytope vertices and GLSM charges in table \ref{table:4309 data},
we see that there is a high degree of symmetry in the vertices which give rise to
many $\bP^1$-like GLSM relations. This symmetry likely increases the number
of triangulations, and also increases the number of intersection curves
$\Sigma\subset B_3$ which are $\bP^1$. Geometrically, this polytope and its
associated toric varieties are rather unique in the $d=3$ Kreuzer-Skarke list,
and it would be interesting to study them further.

\begin{table}[htb]
\centering
\begin{tabular}{cccccccccccccc}
$x_{1}$&$x_{2}$&$x_{3}$&$x_{4}$&$x_{5}$&$x_{6}$&$x_{7}$&$x_{8}$&$x_{9}$&$x_{10}$&$x_{11}$&$x_{12}$&$x_{13}$&$x_{14}$ \\ \hline
$1$&$0$&$0$&$0$&$0$&$0$&$0$&$0$&$0$&$0$&$0$&$0$&$0$&$1$ \\
$0$&$1$&$0$&$0$&$0$&$0$&$0$&$0$&$0$&$0$&$0$&$0$&$1$&$0$ \\
$0$&$0$&$1$&$0$&$0$&$0$&$0$&$0$&$0$&$0$&$0$&$1$&$0$&$0$ \\
$0$&$0$&$0$&$1$&$0$&$0$&$0$&$0$&$0$&$0$&$1$&$0$&$0$&$0$ \\
$0$&$0$&$0$&$0$&$1$&$0$&$0$&$0$&$0$&$1$&$0$&$0$&$0$&$0$ \\
$0$&$0$&$0$&$0$&$0$&$1$&$0$&$0$&$1$&$0$&$0$&$0$&$0$&$0$ \\
$0$&$0$&$0$&$0$&$0$&$0$&$1$&$1$&$0$&$0$&$0$&$0$&$0$&$0$ \\
$0$&$0$&$0$&$0$&$0$&$0$&$0$&$1$&$0$&$0$&$0$&$0$&$1$&$-1$ \\
$0$&$0$&$0$&$0$&$0$&$0$&$0$&$0$&$1$&$0$&$0$&$1$&$0$&$-1$ \\
$0$&$0$&$0$&$0$&$0$&$0$&$0$&$0$&$0$&$1$&$0$&$1$&$1$&$-2$ \\
$0$&$0$&$0$&$0$&$0$&$0$&$0$&$0$&$0$&$0$&$1$&$-1$&$-1$&$1$
\end{tabular}

\vspace{1cm}

\begin{tabular}{cccccccccccccccc}
$v_{1}$&$v_{2}$&$v_{3}$&$v_{4}$&$v_{5}$&$v_{6}$&$v_{7}$&$v_{8}$&$v_{9}$&$v_{10}$&$v_{11}$&$v_{12}$&$\
v_{13}$&$v_{14}$\\ \hline
$1$&$0$&$0$&$-1$&$2$&$1$&$1$&$-1$&$-1$&$-2$&$1$&$0$&$0$&$-1$ \\
$0$&$1$&$0$&$1$&$-1$&$0$&$-1$&$1$&$0$&$1$&$-1$&$0$&$-1$&$0$ \\
$0$&$0$&$1$&$1$&$-1$&$-1$&$0$&$0$&$1$&$1$&$-1$&$-1$&$0$&$0$\\
\end{tabular}
\caption{The first table gives the homogeneous coordinates and associated GLSM charges for all toric varieties corresponding to polytope 4309. We do not give the Stanley-Reisner ideal, since this data is determined by the triangulation. The second table gives the point matrix of the polytope. The point
matrix exhibits the symmetry $v_i + v_{15-i}=0$, trivially giving the first $7$ rows of the table
of GLSM charges.}
\label{table:4309 data}
\end{table}

\section{Points of $E_8$ and Superpotential Zeroes}
\label{sec:E8}
In the example of section
\ref{sec:ex two}, we will see that the prefactor vanishes if and only
if there exists a point of $E_8$ enhancement in the worldvolume of the
M5-instanton. Before getting to the example, in this section we will
explore the connection between points of $E_8$ enhancement and
prefactor zeroes on more general grounds. Points of $E_8$ enhancement
are known to give rise to interesting phenomenological features
\cite{Heckman:2009mn}, but it is a priori unclear how a
compactification would arrive at such a high codimension locus in
complex structure moduli space; to date, there is no reason to expect
stabilization at this point. We will
demonstrate via simple geometric arguments that there is often a
correspondence between points of $E_8$ enhancement and zeroes of the
M5-instanton corrections to the superpotential, which often play
an important role in moduli stabilization.

Before noting special features which occur in the geometry in the
presence of a point of $E_8$ enhancement, let us recall the basic
geometric setup. The prefactor is computed by computing the line bundle
cohomology describing instanton zero modes on a distinguished spectral
curve $c_{loc}$. In the discussion here, we will consider this curve as
as sitting in an elliptic surface $\cE$, and thus the discussion here
applies to both the F-theory and heterotic cases. The curve $c_{loc}$
has a defining equation given by \eqref{eqn:fclocgeneral} 
\begin{equation}
\label{eqn:fc e8 section}
f_{c_{loc}} = \sum_{q=1}^n \tilde b_q v^{n-q} x^{n_x} y^{n_y}
\end{equation}
where $q=2n_x+3n_y$. The sections $\tilde b_q$ in the defining
equations for the spectral curve $c_{loc}$ are obtained via
restricting the sections $b_q$ in the defining equation for the
spectral cover $\cC_{loc}$ to $\cE$. Since $b_q$ is a section of
$K_{B_2}^{\otimes q} \otimes \cO_{B_2}(\eta)$ for $\eta$ a curve in
$B_2$, $\tilde b_q$ is a section of $\cO_\Sigma(r-q\chi)$ where
$\Sigma$ is a $\bP^1$ and $\chi\equiv c_1\cdot_{B_2} \Sigma$, $r
\equiv \eta \cdot_{B_2} \Sigma$.  We are interested in line bundle
cohomology on $c_{loc}$, which we choose to compute via the Koszul
sequence \eqref{eqn:prefactor koszul} from the restriction of a line
bundle on $\cE$. We write this bundle as
$\cO_\cE(a\,s+b\,F)$, where $s$ is the section of $\cE$ and $F$ is the
class of the elliptic fiber.

The special features present in the case of a point of $E_8$ enhancement can be seen by considering
the intersections of $s$ and $F$ with $c_{loc}$. Inside $\cE$, $s\cdot c_{loc}$ is defined by the locus
\begin{equation}
  s\cdot c_{loc} = \{v=0\} \cdot \{f_{c_{loc}}=0\} = \{\tilde b_n=0\}.
\end{equation}
Since $\tilde b_n$ is a section of degree $r-n\chi$, it has $r-n\chi$ zeroes $u_k$ on $\Sigma=\bP^1$
and therefore $s\cdot c_{loc}$ is the collection of points
\begin{equation}
  s \cdot c_{loc} = \sum_{k=1}^{r-n\chi} \{e,u_k\}
\end{equation}
where $e$ is the identity of the elliptic fiber. Similarly, in $\cE$ above any point $u\in\Sigma$
there is an elliptic fiber $F_u$. The fiber $F_u$ above $u$ intersects $c_{loc}$ at $n$ points $q_i$ in
the elliptic fiber, so that
\begin{equation}
  F_u \cdot c_{loc} = \sum_{i=1}^n \{q_i,u\}.
\end{equation}
Intersecting $n$ copies of $s$ with $c_{loc}$, we obtain
\begin{equation}
\label{eqn:ns int c}
  ns \cdot c_{loc} = \sum_{k=1}^{r-n\chi} \{n\,e,u_k\} = \{n\,e, u_j\} + \sum_{k=1, k\ne j}^{r-n\chi}\{n\, e,u_k\},
\end{equation}
where we have separated off a distinguished root $u_j$ of $\tilde b_n$
for convenience. In the case that $u_j$ is a point of $E_8$
enhancement, $u_j$ is a root not only of $\tilde b_n$ but also of $\tilde b_q$ from
$2\dots n-1$. Evaluating the $\tilde b_q$'s at $u_j$ gives $f_{c_{loc}} = \tilde b_0\,\,  v^n$ and
therefore $F_{u_j} \cdot c_{loc} = \sum_{i=1}^n \{e,u_j\} =
\{n\,e,u_j\}$. Then
\begin{equation}
(ns - F_{u{j}}) \cdot c_{loc} = \sum_{k=1, k\ne j}^{r-n\chi}\{n\, e,u_k\},
\end{equation}
which means that $\cO_\cE(ns-F)|_{c_{loc}}$ has a section. 
Summarizing these arguments, in the presence
of a point of $E_8$ enhancement $\cO_\cE(ns-F)|_{c_{loc}}$ has a section. Of course, in \ref{eqn:ns int c}
we could split off up to $r-n\chi$ points. Similar arguments therefore lead to the conclusion that
in the presence of $l$ points of $E_8$ enhancement, $\cO_\cE(ns-aF)|_{c_{loc}}$ has a section for $a$ in
$1\dots l$. For convenience, let us give these bundles a name, 
$\cL_{E8,a} \equiv \cO_\cE(ns-aF)|_{c_{loc}}$. It is important to note that these bundles typically 
get a section at high codimension in the bundle moduli space. It is only in the case of $n=3$
bundles corresponding to $G_{4d}=E_6$ where an $\cL_{E8,a}$ bundle gets a section in codimension 
one.

Having discussed the line bundles $\cL_{E8,a}$ and under what circumstances
they obtain a section, let us discuss the relationship to the vanishing
of the prefactor in light of the example of section
\ref{sec:ex two}. There, by realizing that $\Pfaff_\Sigma\sim
f_{E8}^4$ where $f_{E8}=Res(\tilde b_2,\tilde b_3)$, it will be shown that the prefactor
vanishes if and only if there exists a point of $E_8$ enhancement in
the worldvolume of the instanton. How does one understand this at the
level of line bundles?  In this example the prefactor vanishes if
and only if the bundle $\cL_A=\cO_\cE(6s-F)|_{c_{loc}}$ has a section,
which occurs only on the $f_{E8}=0$ locus in moduli space.
On this locus, $\tilde b_2$ and $\tilde b_3$ have a common
zero which plays the role of $u_j$ in the above discussion, and
therefore $\cL_{E8,1} = \cO_\cE(3s-F)|_{c_{loc}}$ has a section,
where $n=3$ in this example. An important observation is that
\begin{equation}
\cL_A = \cO_\cE(6s-F)|_{c_{loc}} = \cO_\cE(3s)|_{c_{loc}} \otimes \cL_{E8,1},
\end{equation}
where it is convenient that $\cO_\cE(3s)|_{c_{loc}}$ always have a section.
Since this is true, $\cL_A$ has a section whenever $\cL_{E8,1}$
has a section, simply by multiplication of the sections of $\cL_{E8,1}$ and 
$\cO_\cE(3s)|_{c_{loc}}$. In this special $n=3$ case, where $\cL_{E8,1}$ has a section
in codimension one on $f_{E8}=0$, this can and does (in this example)
govern the entire prefactor. For higher $n$ cases, such as the $n=5$ case
with $G_{4d} = SU(5)$, vanishing loci triggered by points of $E_8$ enhancement
typically sit as a proper sublocus in the codimension one vanishing locus of
the prefactor.

Using these ideas, let us discuss a generalization of the circumstances
under which points of $E_8$ in the M5-instanton worldvolume will give rise to
a prefactor zero. Recall from \eqref{eqn:LAchern} that 
\begin{align}
   \cL_A = \cO_\cE(n(\frac{1}{2} + \lambda)s + (r(\frac{1}{2}-\lambda) + \chi(\frac{1}{2} 
 + n\lambda) -1)F)|_{c_{loc}} \equiv \cO_\cE(n(\frac{1}{2}+\lambda)s + BF)|_{c_{loc}}
\end{align}
where we have defined $B$ for convenience and $\lambda$ is half-integral (integral) for $n$
odd (even). Let us examine the conditions under which
$l$ points of $E_8$ in the instanton worldvolume will cause the prefactor to vanish.
Splitting the bundle as above, but generalizing the result, we have
\begin{equation}
   \cL_A = \cO_\cE(n(\lambda -\frac{1}{2}) s + (B+l)F)|_{c_{loc}} \otimes \cL_{E8,l}.
\end{equation}
In the presence of $l$ points of $E_8$ in the instanton worldvolume, $\cL_{E8,l}$ has a section,
and therefore the prefactor vanishes if the first bundle in the product has a section.
For simplicity
we restrict attention to the case of $n$ odd, since the $n=3$ case is the case of the examples
we study and $n=5$ is the case relevant for four-dimensional $SU(5)$ GUTs. If 
$\lambda = \frac{1}{2}$ the first factor is simply $\cO_\cE((B+l) F)|_{c_{loc}}$ which has a section
if $B+l \ge 0$. In the case of $\lambda > \frac{1}{2}$, the first
bundle in the product takes the form $\cO_\cE(Ps)|_{c_{loc}} \otimes \cO_\cE( ns + (B+l)F)|_{c_{loc}}$
where $P\equiv n(\lambda-\frac{1}{2})-n\ge 0$, and we can restrict our attention to 
$\cO_\cE(ns + (B+l) F)|_{c_{loc}}$ since  $\cO_\cE(Ps)|_{c_{loc}}$ has a section. For $(B+l)\ge 0$
this bundle has a section and for $(B+l)<0$ it is simply $\cL_{E8,B+l}$, which has a section
for $B+l \ge -l$, since we have $l$ points of $E_8$ present in the instanton worldvolume.

Summarizing the results for $n$ odd, covering $E_6$ and $SU(5)$ GUTs, we have that $l$ points
of $E_8$ will cause a prefactor zero if 
\begin{align}
  \label{eqn:E8 condition}
  \lambda = \frac{1}{2}:& \qquad \qquad B+l \ge 0 \nonumber \\
  \lambda > \frac{1}{2}:& \qquad \qquad B+l \ge -l \Rightarrow B+2l \ge 0.
\end{align}
In every example $B$ can be computed explicitly and these conditions can be checked. This
is useful because it is computationally intensive to explicitly compute a generic prefactor,
but these conditions allow one to determine cases where an $E_8$ point causes a prefactor
zero without performing the computation. We
see that, though the locus of point(s) of $E_8$ is typically of high codimension in
moduli space, it often sits inside the vanishing locus of the prefactor. Also, while we have
only addressed zeroes of the prefactor, we remind the reader that the prefactor often vanishes
to high degree and thus it is likely that superpotential zeroes associated with points of 
$E_8$ are also critical points. 

As a nontrivial check, let us apply our conditions to the example of section \ref{sec:ex one},
which is not the example obviously related to points of $E_8$.
There the relevant data is 
\begin{equation}
  \cL_A = \cO_\cE(6s-3F)|_{c_{loc}} \qquad \qquad \lambda = \frac{3}{2}
\end{equation}
so that $B=-3$. It is easy to see that $l=1$ does not satisfy the condition \ref{eqn:E8 condition}
and therefore the prefactor does not vanish in the presence of a single point of $E_8$ enhancement.
This is reflected by the explicit computation: a point of $E_8$ enhancement occurs in codimension
one in moduli space at $Res(\tilde b_2,\tilde b_3)=0$ since $n=3$, and this locus is not the prefactor locus
in this example. On the other hand, the conditions \eqref{eqn:E8 condition} are satisfied for
$l=2$, so that 2 points of $E_8$ in the instanton worldvolume cause the prefactor to vanish.
Since $\tilde b_2,\tilde b_3\in \Gamma(\cO_\Sigma(2))$ in the example, having two points of $E_8$ requires
that $\tilde b_2$ and $\tilde b_3$ are proportional to one another, which requires $\beta_i \sim \gamma_i$,
a condition which clearly causes the prefactor \eqref{eqn:Lambda Pfaff} to vanish. In this
example, $r-n\chi=2$, so two points of $E_8$ is the maximal number.

Let us make one final comment in this section about points of
$E_8$. It was shown from the study of heterotic worldsheet
instantons \cite{Curio:2009wn} that for $\chi=1$ there is a 
high codimension locus in
moduli space, called $\cR$, which always sits inside the 
vanishing locus of the prefactor. In the F-theory dual, this is caused
by the presence of $r-n$ points of $E_8$. Though this is for the
$\chi=1$ case, we expect that a simple analysis
along the lines of \cite{Curio:2009wn} would extend it to the $r-n\chi$ case,
which is the maximal number of points of $E_8$ which could appear in the instanton
worldvolume for geometries of this type. As mentioned, the zero caused by two points
of $E_8$ enhancement in the example of section \ref{sec:ex one} is the maximal number,
and so this point in moduli space is in the $\cR$ locus.

\section{Illustrative Examples}
\label{sec:examples}
In this section we will supplement the discussions of previous sections with explicit
examples which demonstrate the various ideas we have discussed. In studying these examples, it will
become clear that the instanton prefactor $A$ often factorizes nicely into powers of simpler
polynomials. Zeroes of these simpler polynomials often correspond to interesting seven-brane
physics.

We will present three examples. The first example is the F-theory dual of one of the heterotic
compactifications in \cite{Buchbinder:2002pr}. It is an $E_6$ GUT realized in a Calabi-Yau
fourfold with a base $B_3$ which is a generalized Hirzebruch variety. We consider an instanton
correction due to a euclidean D3-instanton, and in this example the instanton prefactor is the
fourth power of a polynomial $f_\Lambda$ which is cubic in moduli. The line bundle $\cL_\Lambda$ of
section \ref{sec:het worldsheets} obtains a section when $f_\Lambda=0$. Thus, in this example we have that $\cL_A$
obtains a section if and only if $\cL_\Lambda$ obtains a section. 

In the second example, we consider an F-theory compactification with $G_{4d}=E_6$ which does
not have a heterotic dual. The base threefold $B_3$ can be obtained by blowing up $\bP^3$
twice along $\bP^1$'s, and this procedure is discussed in detail in appendix \ref{app:B3}. For convenience,
we will use the description of $B_3$ as a toric variety. We will show that a superpotential 
correction due to a euclidean D3-instanton along a divisor in $B_3$ has a prefactor which
vanishes if and only if there exists a point of $E_8$ enhancement in the worldvolume of the
instanton. This exemplifies the discussion of section \ref{sec:E8}.

In the third example, we consider another F-theory compactification on the same base $B_3$
as the second example, but with the $E_6$ GUT stack localized on another divisor. In this
compactification we compute a new instanton prefactor, again without a heterotic dual, and
find that the prefactor is identically zero.

\subsection{Example One}

\label{sec:ex one}

In this example we study the prefactor of an M5-instanton correction which was originally
computed in the dual heterotic picture in \cite{Buchbinder:2002pr}, where it was called example $4$. It was also discussed from the heterotic point
of view in \cite{Curio:2008cm} and from the F-theory point of view in \cite{Cvetic:2011gp}. 

The global F-theory compactification we consider is an elliptically fibered Calabi-Yau fourfold
over a based $B_3$ specified by the toric data
\begin{align}
  \label{eqn:ex 1 toric}
  \begin{tabular}{cccccc|c}
    $x_0$ & $x_1$ & $x_2$ & $x_3$ & $z$ & $x_5$ & class \\ \hline
    $1$ & $1$ & $2$ & $0$ & $0$ & $2$ & $A$ \\
    $0$ & $0$ & $1$ & $1$ & $0$ & $2$ & $S$\\
    $0$ & $0$ & $0$ & $0$ & $1$ & $1$ & $B_2$
  \end{tabular}   \nonumber \\
   SRI = \langle x_0x_1,x_2x_3,zx_5\rangle,
\end{align}
where we have specified the GLSM charges of the homogeneous coordinates
and also the Stanley-Reisner ideal. We have defined the divisor
classes $A\equiv [x_1=0]$, $S\equiv [x_3=0]$, and $B_2 \equiv
[z=0]$. In this basis the non-zero intersection numbers are
\begin{equation}
\label{eqn:ex one int numbers}
B_2AS = 1 \qquad B_2^2A = -2 \qquad B_2^2S = 2 \qquad B_2S^2 = -2
\end{equation}
We see that
\begin{equation}
  c_1(B_3) = 6A + 4S + 2B_2
\end{equation}
and the defining equation for the Calabi-Yau with an $E_6$ GUT stack on $B_2\equiv\{z=0\}$ in 
$B_3$ is given by
\begin{equation}
y^2 = x^3 + f\,x\,z^4v^4 + g\,z^6v^6 + z^2v^4(b_0z^3v^3+b_2zvx+b_3y)
\end{equation}
where the objects appearing in this equations are sections of the bundles
\begin{equation}\begin{array}{c|c}
\text{Section} & \text{Bundle} \\ \hline
x_4 & \CO(B_2) \\
b_m & \CO([6-m]c_1(B_3)-[5-m]B_2) \\
f & \CO(c_1(B_3) - B_2)^{\otimes 4} \\
g & \CO(c_1(B_3)-B_2)^{\otimes 6}.
\end{array}\end{equation}
The GUT stack $B_2$ is the second Hirzebruch surface $\bF_2$ and the heterotic dual to
this F-theory compactification is an $E_8\times E_8$ compactification on a Calabi-Yau threefold
which is elliptically fibered over $B_2$ and with bundle data appropriately mapped. It is worth
noting that $B_2 \cdot S$ is the curve $\Sigma$ in $\bF_2$ with $\Sigma^2=-2$. This curve
is the curve $\Sigma$ relevant for the instanton physics we will discuss.
For convenience throughout we will define line bundles on $B_3$ as 
\begin{equation}
\cO(a,b,c) \equiv \cO(aA + bS + cB_2).
\end{equation}
Since they play an important role in determining the instanton physics, let us discuss
the sections $b_0$, $b_2$ and $b_3$ more explicitly. In this example,
$b_m \in H^0(B_3,\cO((6-m)6,(6-m)4,7-m))$, so that
\begin{align}\label{eqn:ex 1 bm sections}
  b_0 &\in H^0(B_3,\cO(36,24,7)) \nonumber \\ b_2 &\in H^0(B_3,\cO(24,16,5)) \nonumber \\ b_3 &\in H^0(B_3,\cO(18,12,4)).
\end{align}
It is clear that such non-trivial sections exist, and therefore an $E_6$ GUT on $B_2$ can
in fact be realized in this geometry. These sections determine the moduli dependence of
the $E_6$ GUT. They also also determine the compact Higgs bundle spectral cover $\cC_{loc}$
as in \eqref{F1fibration}. For convenience, let us note that
\begin{equation}
\eta = (36A + 24S+7B_2)|_{B_2} \qquad \text{and} \qquad \cs = (6A + 4S + B_2)|_{B_2}
\end{equation}
using adjunction and the definitions in \ref{sec:computing}.

Let us turn to a discussion of the instanton physics. Consider an M5-instanton wrapped
on $\pi^{-1} D3$ where $D3\equiv D_3 \equiv \{x_3=0\}$ inside $B_3$. 
It turns out that this instanton does not actually contribute to the superpotential,
since $h^i(D3,N_{D3|B_3})=(0,1,0)$, violating the necessary condition (\ref{eqn:D3 W cont 2}). In the dual heterotic picture where this example
was first studied, the non-contribution to the superpotential is related to the fact that
\cite{Buchbinder:2002pr} focused on left-moving zero modes which couple to the gauge bundle and
not the right-moving zero modes which do not. The study of left-movers leads to knowledge of the
prefactor, whereas the spectrum of right-movers determines whether the instanton
contributes to the superpotential or some other holomorphic coupling. 
In the language of type IIb, this instanton does not have the right spectrum of 
uncharged $3-3$ zero modes for a superpotential correction.

To determine the component of the prefactor $A$ which depends
on seven-brane moduli, we need to study instanton zero modes localized
on the curve $\Sigma = B_2 \cap D3$ in $B_3$ where the instanton
intersects the GUT stack. Topologically, $\Sigma$ is a $\bP^1$. As
discussed in section \ref{sec:instantons in F}, rather than computing vector bundle
cohomology on $\Sigma$ we compute an isomorphic line bundle cohomology
on a spectral curve $c_{loc}$ which is a triple-sheeted cover of
$\Sigma$.  Following section \ref{sec:computing},
we compute directly that $r=B_2\cdot S\cdot\eta = 2$ and $\chi=B_2\cdot S\cdot\cs=0$ and choose $n=3$ (for an
$E_6$ GUT) and $\lambda = \frac{3}{2}$. Using \eqref{eqn:cloc class} and \eqref{eqn:LAchern}, this gives the class of $c_{loc}$
as a divisor inside an elliptic surface $\cE$ and the relevant line bundle
$\cL_A$ for computing the prefactor to be 
\begin{equation}
[c_{loc}] = 3s + 2F \qquad \cL_A = \cO_{\cE}(6s-3F).
\end{equation}
In terms of the spectral cover, $c_{loc} =
\cC_{loc}|_{\pi^{-1}D3}$. At the level of defining equations, this
means that $f_{c_{loc}} = f_{\cC_{loc}}|_{x_3=0}$, and therefore
\begin{equation}
  f_{c_{loc}} = \tilde b_0 v^3 + \tilde b_2 vx + \tilde b_3 y
\end{equation}
where $\tilde b_m \equiv b_m|_{GUT\cap D3} = b_m|_{z=x_3=0}$. Remembering that $\cC_{loc}$ is 
naturally a divisor in an ambient elliptic threefold $Z_{3,F}$,
 $c_{loc}$ is therefore a divisor in
an ambient elliptic surface $\cE = \pi_{3,F}^{-1} \Sigma$  given by the restriction of the elliptic threefolds to $x_3=0$.
While it initially seems strange to be considering these ambient spaces which do not sit inside
the Calabi-Yau fourfold, the physics we study is determined entirely by line bundle cohomology
on $c_{loc}$, which does sit inside the Calabi-Yau. One is free to compute line bundle cohomology
on $c_{loc}$ via any allowed means, including via a Koszul sequence from the ambient space
$\cE$ which is an elliptic fibration over $\Sigma$.

Let us say more about the structure of $f_{c_{loc}}$ by studying the sections $\tilde b_m$.
Given $b_m$ as in \eqref{eqn:ex 1 bm sections}, the monomials in $b_m$ with neither a $z$ nor
an $x_3$ are the ones appearing in $\tilde b_m$ and take the form
\begin{align}
  \tilde b_0 \,\,\, \text{monomials} \,\, &\sim \,\, x_5^{7} x_2^{10}\,\, x_0^{i} x_1^{2-i} &i=0\dots 2 \nonumber \\
  \tilde b_2 \,\,\, \text{monomials} \,\, &\sim \,\, x_5^{5} x_2^{6}\,\, x_0^{i} x_1^{2-i} &i=0\dots 2 \nonumber \\
\tilde b_3 \,\,\, \text{monomials} \,\, &\sim \,\,  x_5^{4} x_2^{4}\,\, x_0^{i} x_1^{2-i} &i=0\dots 2
\end{align}
where the range of $i$ ensures that the monomials are global sections. From the
Stanley-Reisner ideal, it can be seen that $x_5$ and $x_3$ must be non-zero since $z=x_2=0$. Using
two of the scaling relations of the toric variety to set $x_5=x_3=1$, we can write down
unambiguously
\begin{align}
\tilde b_0  &= \alpha_1 \, x_0^2 + \alpha_2 \, x_0x_1 + \alpha_3 \,x_1^2\nonumber \\
\tilde b_2  &= \beta_1 \, x_0^2 + \beta_2 \, x_0x_1 + \beta_3 \,x_1^2\nonumber \\
\tilde b_3  &= \gamma_1 \, x_0^2 + \gamma_2 \, x_0x_1 + \gamma_3 \,x_1^2
\end{align}
in terms of moduli $\alpha_j,\beta_j,$ and $\gamma_k$ and the homogeneous
coordinates $(x_0,x_1)$ on $\Sigma$. While direct computation of $\chi(\Sigma)=2$ shows that
it is a $\bP^1$, it is also fairly easy to see this from looking at the GLSM charges and
Stanley-Reisner ideal in \eqref{eqn:ex 1 toric}.

We have calculated the quantities $\{n,\lambda,r,\chi\}$ as outlined in section \ref{sec:computing},
and written the defining equation $f_{c_{loc}}$. Given this data, it is possible to calculate the prefactor of an M5-instanton on $\pi^{-1} D3$
via a long exact sequence in cohomology involving $h^i(c_{loc},\cL_A|_{c_{loc}})$. For this example,
the computation was performed in the dual heterotic picture \cite{Buchbinder:2002pr} and was discussed in 
F-theory \cite{Cvetic:2011gp}. To avoid cluttering an already detailed story, we simply quote
the result and refer the reader to the literature for an explicit demonstration of the computation.
The prefactor is given by
\begin{equation}
  \label{eqn:Lambda Pfaff}
  \Pfaff \sim (\epsilon_{ijk}\,\, \alpha_i \beta_j \gamma_k)^4 \equiv f_\Lambda^4
\end{equation}
where $f_\Lambda$ is defined for convenience. A number
of questions are valid at this point: why does the instanton
prefactor, a degree twelve polynomial, factorize so nicely? What is
the significance of the polynomial $f_\Lambda$ and of the power to
which it is raised? 

   \subsubsection{Flat bundles, Intermediate Jacobians, and the Prefactor}
   In this subsection we will review the significance of the
   polynomial $f_\Lambda$, which determines the vanishing locus of the
   instanton prefactor in this example, and its relationship with the
   theta divisor on the intermediate Jacobian of the M5 instanton.

   Let us begin by reviewing the significance of the polynomial $f_\Lambda$, as discussed in
the heterotic case in \cite{Curio:2008cm}. A brief look at $f_\Lambda$ reveals that is it
\begin{equation}
  f_\Lambda = det\left( {\begin{array}{ccc}
  \alpha_1 & \alpha_2 & \alpha_3 \\
  \beta_1 & \beta_2 & \beta_3 \\
  \gamma_1 & \gamma_2 & \gamma_3 
\end{array}} \right) \equiv det M, 
\end{equation}
it is natural to ask the significance of this matrix $M$. If one were to compute the line bundle
cohomology of $\cO_\cE(3s-2F)|_{c_{loc}}$ via the Koszul sequence
\begin{equation}
  0 \rightarrow \cO_\cE(-4F) \xrightarrow{f_{c_{loc}}} \cO_\cE(3s-2F) \rightarrow \cO_\cE(3s-2F)|_{c_{loc}} \rightarrow 0
\end{equation}
and the corresponding long exact sequence in cohomology, one would find that $M$ appears as the
matrix map between $H^1(\cE,\cO_\cE(-4F))$ and $H^1(\cE,\cO_\cE(3s-2F))$. Defining $\Lambda \equiv
\cO_\cE(3s-2F)|_{c_{loc}}$, the statement then is that $f_\Lambda=0$ is the codimension $1$ locus 
in moduli space on which $\Lambda$ obtains a section. Therefore, by comparing to equation (\ref{eqn:Lambda Pfaff}) we see that the prefactor vanishes if and only if $\Lambda$ obtains a section. What
governs the relationship between the bundle $\Lambda$ and the bundle $\cL_A$ which actually
determines the prefactor? Decomposing as
\begin{equation}
  \cL_A \equiv \cO_\cE(6s-3F)|_{c_{loc}} = \cO(F)|_{c_{loc}} \otimes \Lambda^{\otimes 2} \equiv \cL_1 \otimes \Lambda^{\otimes 2 },
\end{equation}
will be useful for demonstrating the relationship. On the locus $f_\Lambda=0$, $\Lambda$ has a section and therefore
$\Lambda^{\otimes 2}$ also has a section. But $\cL_1$ itself has a section for all moduli, and
therefore when $f_\Lambda = 0$, $\cL_A$ also has a section and the prefactor vanishes. Thus,
we see that the polynomial substructure of the prefactor in this example actually corresponds
to the locus in moduli space where a related bundle $\Lambda$ obtains a section.

Let us state how this is related to the discussion in previous sections.
The bundle $\Lambda$ is a degree $0$ bundle and therefore specifies a point in $\CJ_0(c_{loc})$.
$\cL_A$, the bundle directly related to the prefactor, is a degree $g_{c_{loc}}-1$ bundle
but can be decomposed as
\begin{equation}
  \cL_A = K_{c_{loc}}^\frac{1}{2} \otimes \cL_\Gamma =  K_{c_{loc}}^\frac{1}{2} \otimes \Lambda^{\otimes \lambda}.
\end{equation}
The spin
structure defines a map from $\CJ_0(c_{loc})\leftrightarrow \CJ_{g-1}(c_{loc})$ via
tensor product. In general, when $\cL_A$ has a section, it corresponds to a point in the Riemann
theta divisor $\Theta_R$ in $\CJ_{g-1}(c_{loc})$. A choice of spin structure maps $\Theta_R$ to a 
particular theta divisor $\Theta_m$ on $\CJ_0(c_{loc})$, which in general does not have anything
to do with degree $0$ bundles which have a section. However, in this case it does and that bundle
is $\Lambda$.

Finally, let us make a speculative comment about the multiplicity of 
$f_\Lambda$, namely that $\Pfaff\sim f_\Lambda^4$.
Physically, one should expect that the order of vanishing is related to the number of zero modes.
A single zero mode would be enough to cause a prefactor zero, but in the presence of many modes
any will cause a zero, and thus the prefactor zero should be of high order. This makes sense from
a type IIb point of view, where it is known that increasing the flux quanta analogous to $\lambda$
increases the number of $D3-D7$ zero modes and also the order of vanishing of the instanton
prefactor. In this example, the Hodge numbers are $h^i(c_{loc},\cL_A)=(2,2)$, and thus there
are four zero modes. It is natural and tempting to think that this is related to the fourth power of
 $f_\Lambda$ in the prefactor, but one would need to study these ideas further to put them on 
firmer footing. 

\subsection{Example Two: Prefactors without a heterotic dual}
\label{sec:ex two}

In this section we study instantons in a global F-theory compactification which does not
admit a heterotic dual. Interestingly, the prefactor computed here in an intrinsically F-theoretic
geometry is mathematically identical to a prefactor computation in the heterotic string in \cite{Buchbinder:2002ic},
despite being in a different compactification. This exemplifies the multiplicities seen in section \ref{sec:scan}. 
Some of the prefactor structure was addressed in \cite{Curio:2008cm}, though we will be focused
on the F-theory physics here. We will also discuss other aspects of the geometry and example.

The global F-theory compactification we consider is an elliptically fibered Calabi-Yau fourfold
over a based $B_3$ specified by the toric data
\begin{align}
  \begin{tabular}{cccccc|c}
    $x_0$ & $x_1$ & $x_2$ & $x_3$ & $z$ & $x_5$ & class \\ \hline
    $1$ & $0$ & $0$ & $1$ & $1$ & $1$ & $H$ \\
    $0$ & $1$ & $0$ & $-1$ & $0$ & $-1$ & $E_1$\\
    $0$ & $0$ & $1$ & $0$ & $-1$ & $-1$ & $E_2$
  \end{tabular}   \nonumber \\
   SRI = \langle x_2x_3, x_3x_5, zx_5, x_0x_1x_2, x_0x_1z\rangle,
\end{align}
where we have specified the GLSM charges of the homogeneous coordinates and also the Stanley-Reisner
ideal. While this parameterization is most useful for comparing to the non-toric analysis
of appendix \ref{app:B3}, we now choose a different parameterization of the same toric variety. 
Adding the first row to the second and third rows, we have
\begin{align}
  \label{eqn:ex 2 toric}
  \begin{tabular}{cccccc|c}
    $x_0$ & $x_1$ & $x_2$ & $x_3$ & $z$ & $x_5$ & class \\ \hline
    $1$ & $0$ & $0$ & $1$ & $1$ & $1$ & $X$ \\
    $1$ & $1$ & $0$ & $0$ & $1$ & $0$ & $E_1$\\
    $1$ & $0$ & $1$ & $1$ & $0$ & $0$ & $E_2$
  \end{tabular}   \nonumber \\
   SRI = \langle x_2x_3, x_3x_5, zx_5, x_0x_1x_2, x_0x_1z\rangle,
\end{align}
where $X\equiv H-E_1-E_2$. In the basis $\{X,E_1,E_2\}$, the intersection
numbers are
\begin{align}
X X X = 1 \qquad
X X E_1 = -1 \qquad
X X E_2 = -1 \qquad
X E_1 E_1 = 1 \nonumber \\
X E_1 E_2 = 1\qquad
X E_2 E_2 = 1\qquad
E_1 E_1 E_1 = -2 \qquad
E_1 E_1 E_2 = 0 \nonumber \\
E_1 E_2 E_2 = -1 \qquad
E_2 E_2 E_2 = -1.
\end{align}
We see that
\begin{equation}
  c_1(B_3) = 4X + 3E_1 + 3E_2
\end{equation}
and the defining equation for the Calabi-Yau with an $E_6$ GUT stack on $B_2\equiv\{z=0\}$ in 
$B_3$ is given by
\begin{equation}
y^2 = x^3 + f\,x\,z^4v^4 + g\,z^6v^6 + z^2v^4(b_0z^3v^3+b_2zvx+b_3y)
\end{equation}
where the objects appearing in this equations are sections of the bundles
\begin{equation}\begin{array}{c|c}
\text{Section} & \text{Bundle} \\ \hline
x_4 & \CO(B_2) \\
b_m & \CO([6-m]c_1(B_3)-[5-m]B_2) \\
f & \CO(c_1(B_3) - B_2)^{\otimes 4} \\
g & \CO(c_1(B_3)-B_2)^{\otimes 6}.
\end{array}\end{equation}
The GUT stack satisfies $\int_{B_2} c_1^2=8$ and $\int_{B_2} c_2=4$ which is consistent with
$B_2$ being the first Hirzebruch surface $\bF_1$. 
For convenience throughout we will define line bundles on $B_3$ as 
\begin{equation}
\cO(a,b,c) \equiv \cO(aX + bE_1 + cE_2).
\end{equation}
Since they play an important role in determining the instanton physics, let us discuss
the sections $b_0$, $b_2$ and $b_3$ more explicitly. In this example,
$b_m \in H^0(B_3,\cO(19-3m,13-2m,18-3m))$, so that
\begin{align}\label{eqn:ex 2 bm sections}
  b_0 &\in H^0(B_3,\cO(19,13,18)) \nonumber \\ b_2 &\in H^0(B_3,\cO(13,9,12)) \nonumber \\ b_3 &\in H^0(B_3,\cO(10,7,9)).
\end{align}
It is clear that such non-trivial sections exist, and therefore an $E_6$ GUT on $B_2$ can
in fact be realized in this geometry. These sections determine the entire moduli dependence of
the $E_6$ GUT. They also also determine the compact Higgs bundle spectral cover $\cC_{loc}$
as in \eqref{F1fibration}. For convenience, note that 
\begin{equation}
\eta = (19X + 13E_1 + 18 E_2)|_{B_2}\qquad \text{and} \qquad \cs = (3X + 2E_1 + 3E_2)|_{B_2}
\end{equation}
in this compactification.

Let us turn to a discussion of the instanton physics. Consider an M5-instanton wrapped
on $\pi^{-1} D3$ where $D3\equiv D_1 \equiv \{x_1=0\}$ inside $B_3$. As discussed in section \ref{sec:FM5}, 
obtaining a superpotential correction requires $\chi(D3,N_{D3|B_3})=0$, and for this
instanton it is satisfied, since $h^i(D,N_{D3|B_3})=(0,0,0)$. In the language of M5-instantons,
these constraints are related to the arithmetic genus constraint.

To determine the component of the Pfaffian prefactor $A$ which depends
on seven-brane moduli, we need to study instanton zero modes localized
on the curve $\Sigma = B_2 \cap D3$ in $B_3$ where the instanton
intersects the GUT stack. Topologically, $\Sigma$ is a $\bP^1$. As
discussed in section \ref{subsec:conjecture}, rather than computing vector bundle
cohomology on $\Sigma$ we compute an isomorphic line bundle cohomology
on a spectral curve $c_{loc}$ which is a triple-sheeted cover of
$\Sigma$.  Following section \ref{sec:computing},
we compute directly that $r=\eta \cdot_{B_2} \Sigma=5$ and $\chi=\cs\cdot_{B_2}\Sigma=1$ and we have $n=3$ (for an
$E_6$ GUT) and choose $\lambda = \frac{3}{2}$. From equations \eqref{eqn:cloc class}
and \eqref{eqn:LAchern}, this gives the class of $c_{loc}$
as a divisor inside an elliptic surface $\cE$ and the relevant line bundle
$\cL_A$ for computing the prefactor to be 
\begin{equation}
[c_{loc}] = 3s + 5F \qquad \cL_A = \cO_{\cE}(6s-F)|_{c_{loc}}.
\end{equation}
In terms of the spectral cover, $c_{loc} =
\cC_{loc}|_{\pi^{-1}D3}$. At the level of defining equations, this
means that $f_{c_{loc}} = f_{\cC_{loc}}|_{x_1=0}$, and therefore
\begin{equation}
  f_{c_{loc}} = \tilde b_0 v^3 + \tilde b_2 vx + \tilde b_3 y
\end{equation}
where $\tilde b_m \equiv b_m|_{GUT\cap D3} = b_m|_{z=x_1=0}$. Remembering that $\cC_{loc}$ is 
naturally a divisor in an ambient elliptic threefold $Z_{3,F}$,
 $c_{loc}$ is therefore a divisor in
an ambient elliptic surface $\cE$  given by the restriction of the elliptic threefolds to $x_1=0$.
While it initially seems strange to be considering these ambient spaces which do not sit inside
the Calabi-Yau fourfold, the physics we study is determined entirely by line bundle cohomology
on $c_{loc}$, which does sit inside the Calabi-Yau. One is free to compute line bundle cohomology
on $c_{loc}$ via any allowed means, including via a Koszul sequence from the ambient space
$\cE$ which is an elliptic fibration over $\Sigma$.

Let us say more about the structure of $f_{c_{loc}}$ by studying the sections $\tilde b_m$.
Given $b_m$ as in \eqref{eqn:ex 2 bm sections}, the monomials in $b_m$ with neither a $z$ nor
an $x_1$ are the ones appearing in $\tilde b_m$ and take the form
\begin{align}
  \tilde b_0 \,\,\, \text{monomials} \,\, &\sim \,\, x_0^{13} x_3^j x_5^{6-j} x_2^{5-j} &j=0\dots 5\nonumber \\
  \tilde b_2 \,\,\, \text{monomials} \,\, &\sim \,\, x_0^{9} x_3^j x_5^{4-j} x_2^{3-j}  &j=0\dots 3\nonumber \\
\tilde b_3 \,\,\, \text{monomials} \,\, &\sim \,\, x_0^{7} x_3^j x_5^{3-j} x_2^{2-j}  &j=0\dots 2
\end{align}
where the range of $j$ has been chosen to ensure that the monomials are global sections. From the
Stanley-Reisner ideal, it can be seen that $x_0$ and $x_5$ must be non-zero since $z=x_1=0$. Using
two of the scaling relations of the toric variety to set $x_0=x_5=1$, we can write down
unambiguously
\begin{align}
\tilde b_0  &= \psi_1\, x_3^5 +\psi_2\, x_3^4x_2^1 +\psi_3\, x_3^3x_2^2 +\psi_4\, x_3^2x_2^3 +\psi_5\, x_3^1x_2^4 +\psi
_6\,x_2^5 \nonumber \\                                                                                 
\tilde b_2 &= \phi_1\, x_3^3 +\phi_2\, x_3^2x_2^1 +\phi_3\, x_3^1x_2^2 +\phi_4\, x_2^3  \nonumber \\              
\tilde b_3 &= \chi_1\, x_3^2 +\chi_2\, x_3^1x_2^1 +\chi_3\, x_2^2      
\end{align}
in terms of moduli $\psi_j,\phi_j,$ and $\chi_k$ and the homogeneous
coordinates $(x_2,x_3)$ on $\Sigma$. While direct computation of $\chi(\Sigma)=2$ shows that
it is a $\bP^1$, it is also fairly easy to see this from looking at the GLSM charges and
Stanley-Reisner ideal in \eqref{eqn:ex 2 toric}.

We have discussed the moduli dependence of $c_{loc}$ and also the bundle cohomology which one
must compute on it. From this data, it is possible to directly compute the instanton prefactor.  Interestingly,
though this is an F-theory compactification without a heterotic dual the same topological data
$\{n,\lambda,r,\chi\}=\{3,\frac{3}{2},5,1\}$ related to the prefactor appears in a heterotic
prefactor computation. 
The prefactor is given by \cite{Buchbinder:2002ic}
\begin{eqnarray}
&&\Pfaff \sim {f_{E8}}^{4} \equiv
(\chi_1^2 \chi_3 \phi_3^2 -
\chi_1^2 \chi_2 \phi_3 \phi_4 -
2\chi_1 \chi_3^2  \phi_3 \phi_1 - \nonumber \\
&&\chi_1 \chi_2 \chi_3  \phi_3 \phi_2 +
\chi_2^2 \chi_3  \phi_1 \phi_3 +
\phi_4^2 \chi_1^3 -              \nonumber \\
&&2 \phi_2 \phi_4 \chi_3 \chi_1^2  +
\chi_1 \chi_3^2 \phi_2^2 +
3 \phi_1 \phi_4 \chi_1 \chi_2 \chi_3 + \nonumber \\
&&\phi_2 \chi_1 \phi_4 \chi_2^2 +
\phi_1^2 \chi_3^3 -
\phi_2 \chi_2 \phi_1 \chi_3^2-
\phi_4 \phi_1 \chi_2^3)^{4} \; ,
\end{eqnarray}
and we again see that this very complicated expression factors into powers of a slightly less
complicated polynomial $f_{E8}$. We will demonstrate that the vanishing associated with
$f_{E8}$ admits a simple physical interpretation. Note also that the moduli $\psi_i$ are 
conspicuously absent the prefactor.

   \subsubsection{Points of $E_8$ and the Vanishing of the Prefactor}
   
   As in the first example, we see that the prefactor, which is a
   generically a complicated polynomial in algebraic moduli,
   factorizes into powers of significantly less complicated
   polynomials. In this example, we call this polynomial $f_{E8}$ and
   would like to discuss its physical significance. This polynomial is
   independent of the moduli which control the structure of $\tilde b_0$, but
   does depend on those moduli appearing in $\tilde b_2$ and $\tilde b_3$. Turning
   off $\tilde b_3$ and $\tilde b_2$ in succession would enhance the generic curve
   of $E_6$ singularities where the GUT stack intersects the instanton
   to $E_7$ and then $E_8$. The moduli appearing in $\tilde b_0$ are those
   which ensure that the singularity type of the curve does not
   enhance ``past $E_8$'' to a non-Kodaira singularity. Thus,
   only the moduli $\chi_i$ and $\phi_i$ which control the Higgsing of
   the $E_8$ curve appear in the Pfaffian.

Though we have a qualitative understanding of the moduli which appear in the Pfaffian, there is actually a simple geometric structure
which determines the entirety of the polynomial $f_{E8}$. It is none other than the determinant of \cite{Curio:2008cm}
\begin{equation}
M \equiv
\left( {\begin{array}{ccccc}
\phi_4 & \phi_3  & \phi_2 & \phi_1 & 0  \\
0   & \phi_4  & \phi_3 & \phi_2 & \phi_1\\
\chi_3 & \chi_2  & \chi_1 & 0   & 0  \\
0   & \chi_3  & \chi_2 & \chi_1 & 0  \\
0   & 0    & \chi_3 & \chi_2 & \chi_1
\end{array}} \right), 
\end{equation}
so that $f_{E8}= det(M)$ and therefore the prefactor $\Pfaff \sim det(M)^4$. Matrices of this form are well-known in elimination theory as Sylvester
matrices, and $M$ itself is the Sylvester matrix of the two polynomials $\tilde b_2$ and $\tilde b_3$. The determinant of the Sylvester matrix
of two polynomials is the resultant of those two polynomials, which has a zero if and only if the two polynomials have a common zero. 
We therefore have a precise algebraic understanding of $f_{E8}$. It is the resultant of $\tilde b_2$ and $\tilde b_3$, and thus the prefactor
vanishes if and only if $\tilde b_2$ and $\tilde b_3$ have a common zero.

The algebraic statement $f_{E8} = \text{Res}(\tilde b_2,\tilde b_3)$
has a concrete geometric realization in F-theory.  Suppose that we are
at a point in the moduli space which gives a zero of $f_{E8}$, so that
there is a point $(x_2^*,x_3^*)$ in $\Sigma$ along which $\tilde b_2$ and $\tilde b_3$
have a common zero. A simultaneous zero of $\tilde b_2$ and $\tilde b_3$ enhance
the singularity to $E_8$, which in this case occurs at a point. Thus, in this
example an F-theoretic understanding of the prefactor is
\begin{equation}
  \Pfaff \sim f_{E8}^4 = 0 \qquad \Leftrightarrow \qquad M5 \supset \text{pt of $E_8$ enhancement}.
\end{equation}
In addition to being beautiful, connections between superpotential zeroes and points of $E_8$
could also have phenomenological significance, as points of $E_8$ are known to give rise to 
interesting physics in $SU(5)$ GUT models \cite{Heckman:2009mn}.
While we see a direct correspondence in this example, the phenomenon is more generic,
and we refer the reader to section \ref{sec:E8} for a discussion. There we identify one
structure governing the relationship between superpotential zeroes and points of $E_8$. We will
show that the connection can also exist for other gauge groups, including $G_{4d}=SU(5)$.

\subsection{Example Three: A New Prefactor}
\label{sec:example three}

Let us briefly consider one final example of a prefactor computation in F-theory. Consider an F-theory
compactification on the same base $B_3$ as in section \ref{sec:ex two}. Again let the instanton
wrap $D3\equiv\{x_1=0\}$, but this time consider an $E_6$ GUT along $B_2\equiv\{x_0=0\}$\footnote{Throughout we have used $z=0$ as the GUT locus in $B_3$, but in this example we use $x_1=0$. The corresponding equations
for the geometries used throughout must replace $z$ by $x_1$ for this example only.}. Having changed the location
of the GUT stack, in this example we have
\begin{equation}
\eta = 19X+13E_1+13E_2 \qquad \cb = (3x+2E_1+2E_2)|_{B_2},
\end{equation}
and for $\lambda=\frac{3}{2}$ a simple computation as in the other two examples gives that 
$\{n,\lambda,r,\chi\} = \{3,\frac{3}{2},6,1\}.$ We then have that
\begin{equation}
[c_{loc}] = 3s + 6F \qquad \cL_A=\cO_\cE(6s-2F)|_{c_{loc}},
\end{equation}
which is enough data to set up a mathematical computation of line bundle
cohomology using the techniques of \cite{Buchbinder:2002pr}. Unlike the previous two examples, however,
this example has not been computed in the literature, and we would like to give the result. Note that
this geometry exemplifies the Pfaffian whose data is given in row 3 of \ref{table:scan}.

Per usual in $E_6$ models, the defining equation of $c_{loc}$ is
\begin{equation}f_{c_{loc}} = \tilde b_0 \,\, v^3 + \tilde b_2 \,\, vx + \tilde b_3 \,\, y \end{equation}
where $b_m\in H^0(B_3,\cO_{B_3}((19-3m)X + (13-2m) E_1 + (13-2m) E_2))$ in this example. 
Restricting to the intersection of the GUT stack and the instanton at $\{x_0=x_1=0\}$ we 
can write down the form of the monomials in $\tilde b_q$
\begin{align}
  \tilde b_0 \,\,\, \text{monomials} \,\, &\sim \,\, x_2^{13-i} x_3^i z^{13} x_5^{6-i}  &i=0\dots 6\nonumber \\
  \tilde b_2 \,\,\, \text{monomials} \,\, &\sim \,\,  x_2^{9-i} x_3^i z^{9} x_5^{4-i}  &i=0\dots 4\nonumber \\
\tilde b_3 \,\,\, \text{monomials} \,\, &\sim \,\,   x_2^{7-i} x_3^i z^{7} x_5^{3-i} &i=0\dots 3
\end{align}
where the range of $i$ has been chosen to ensure that the monomials are global sections.
From the Stanley-Reisner ideal it can be seen that when $x_0=x_1=0$ one can use the GLSM relations
to set $x_2=z=1$ leaving two free coordinates $x_3$ and $x_5$ which are the coordinates on $\Sigma=\bP^1$.
We can then write down
\begin{align}
\tilde b_0  &= \alpha_1 x_3^6 + \alpha_2 x_3^5 x_5+ \alpha_3 x_3^4 x_5^2 + \alpha_4 x_3^3 x_5^3 + \alpha_5 x_3^2 x_5^4 + \alpha_6 x_3^1 x_5^5 + \alpha_7 x_5^6  \nonumber \\                               \tilde b_2 &=   \beta_1 x_3^4 + \beta_2 x_3^3 x_5 + \beta_3 x_3^2 x_5^2 + \beta_4 x_3 x_5^3 + \beta_5 x_5^4\nonumber \\             
\tilde b_3 &=   \gamma_1 x_3^3 + \gamma_2 x_3^2 x_5 + \gamma_3 x_3 x_5^2 + \gamma_4 x_5^3.
\end{align}
and we thus know $f_{c_{loc}}$ explicitly.

Let us now perform the computation of the moduli-dependent prefactor. We are tasked with computing the cohomology
$H^i(c_{loc},\cL_A)$, and we will utilize the Koszul sequence
\begin{equation}
\label{eqn:ex 3 koszul}
0 \rightarrow \cO_\cE(3s-8F) \xrightarrow{f_{c_{loc}}}\cO_\cE(6s-2F)\rightarrow \cL_A \rightarrow 0.
\end{equation} 
Via a Leray spectral sequence\footnote{For a collection of useful results regarding line bundle cohomology
on $\cE$ and Leray spectral sequences see appendix C of \cite{Curio:2008cm}.}
it can be shown that 
\begin{equation}
H^0(\cE,\cO_\cE(3s-8F))=H^2(\cE,\cO_\cE(3s-8F))=H^0(\cE,\cO_\cE(6s-2F))=H^2(\cE,\cO_\cE(6s-2F))=0 
\end{equation}
and therefore the long exact sequence in cohomology corresponding to \ref{eqn:ex 3 koszul} simplifies to
\begin{equation}
0 \rightarrow H^0(c_{loc},\cL_A) \rightarrow H^1(\cE,\cO_\cE(3s-8F)) \xrightarrow{f} H^1(\cE,\cO_\cE(6s-2F)) \rightarrow H^1(c_{loc},\cL_A) \rightarrow 0.
\end{equation}
where $f$ is the moduli-dependent matrix map induced by $f_{c_{loc}}$ whose determinant is the prefactor. Via a Leray
spectral sequence, it can be shown that 
 $H^1(\cE,\cO_\cE(3s-8F)) = H^1(\Sigma,\pi_{3,F*}\cO_\cE(3s-8F))$ and
$H^1(\cE,\cO_\cE(6s-2F)) = H^1(\Sigma,\pi_{3,F*}\cO_\cE(6s-2F))$
where 
\begin{align}
\label{eqn:ex 3 direct image}
\pi_{3,F*}\cO_\cE(3s-8F) &= \cO_\Sigma(-8)\oplus\cO_\Sigma(-10)\oplus\cO_\Sigma(-11) \nonumber \\
\pi_{3,F*}\cO_\cE(6s-2F) &= \cO_\Sigma(-2)\oplus\cO_\Sigma(-4)\oplus\cO_\Sigma(-5)\cO_\Sigma(-6)\oplus\cO_\Sigma(-7)\oplus\cO_\Sigma(-8)
\end{align}
and we therefore see $h^1(\cE,\cO_\cE(3s-8F))=h^1(\cE,\cO_\cE(6s-2F))=26$ using \eqref{eqn:ex 3 direct image} and the fact that
$h^1(\bP^1,\cO(-i)) = i-1$. This means the $f$ is a $26\times 26$ matrix.

We have worked out a few more details of the computation than in previous examples, since this is a new example. We refer the reader
to \cite{Buchbinder:2002ic} for more details. Using the same techniques as in that work,
the moduli-dependent matrix $f$ can be constructed explicitly, and (in a particular basis) it is given by
\[ \scriptsize \hspace{-2.5cm}f=\left( \begin{array}{cccccccccccccccccccccccccccccccccccccccccccc}
\alpha_1& \alpha_2& \alpha_3& \alpha_4& \alpha_5& \alpha_6& \alpha_7& 0& 0& 0& 0& 0& 0& 0& 0& 0& 0& 0& 0& 0& 0& 0& 0& 0& 0& 0\\ \beta_1& \beta_2& \beta_3& \beta_4& \beta_5& 0& 0& \alpha_1& \alpha_2& \alpha_3& \alpha_4& \alpha_5& \alpha_6& \alpha_7& 0& 0& 0& 0& 0& 0& 0& 0& 0& 0& 0& 0\\ 0& \beta_1& \beta_2& \beta_3& \beta_4& \beta_5& 0& 0& \alpha_1& \alpha_2& \alpha_3& \alpha_4& \alpha_5& \alpha_6& \alpha_7& 0& 0& 0& 0& 0& 0& 0& 0& 0& 0& 0\\ 0& 0& \beta_1& \beta_2& \beta_3& \beta_4& \beta_5& 0& 0& \alpha_1& \alpha_2& \alpha_3& \alpha_4& \alpha_5& \alpha_6& \alpha_7& 0& 0& 0& 0& 0& 0& 0& 0& 0& 0\\ \gamma_1& \gamma_2& \gamma_3& \gamma_4& 0& 0& 0& 0& 0& 0& 0& 0& 0& 0& 0& 0& \alpha_1& \alpha_2& \alpha_3& \alpha_4& \alpha_5& \alpha_6& \alpha_7& 0& 0& 0\\ 0& \gamma_1& \gamma_2& \gamma_3& \gamma_4& 0& 0& 0& 0& 0& 0& 0& 0& 0& 0& 0& 0& \alpha_1& \alpha_2& \alpha_3& \alpha_4& \alpha_5& \alpha_6& \alpha_7& 0& 0\\ 0& 0& \gamma_1& \gamma_2& \gamma_3& \gamma_4& 0& 0& 0& 0& 0& 0& 0& 0& 0& 0& 0& 0& \alpha_1& \alpha_2& \alpha_3& \alpha_4& \alpha_5& \alpha_6& \alpha_7& 0\\ 0& 0& 0& \gamma_1& \gamma_2& \gamma_3& \gamma_4& 0& 0& 0& 0& 0& 0& 0& 0& 0& 0& 0& 0& \alpha_1& \alpha_2& \alpha_3& \alpha_4& \alpha_5& \alpha_6& \alpha_7\\ 0& 0& 0& 0& 0& 0& 0& \beta_1& \beta_2& \beta_3& \beta_4& \beta_5& 0& 0& 0& 0& 0& 0& 0& 0& 0& 0& 0& 0& 0& 0\\ 0& 0& 0& 0& 0& 0& 0& 0& \beta_1& \beta_2& \beta_3& \beta_4& \beta_5& 0& 0& 0& 0& 0& 0& 0& 0& 0& 0& 0& 0& 0\\ 0& 0& 0& 0& 0& 0& 0& 0& 0& \beta_1& \beta_2& \beta_3& \beta_4& \beta_5& 0& 0& 0& 0& 0& 0& 0& 0& 0& 0& 0& 0\\ 0& 0& 0& 0& 0& 0& 0& 0& 0& 0& \beta_1& \beta_2& \beta_3& \beta_4& \beta_5& 0& 0& 0& 0& 0& 0& 0& 0& 0& 0& 0\\ 0& 0& 0& 0& 0& 0& 0& 0& 0& 0& 0& \beta_1& \beta_2& \beta_3& \beta_4& \beta_5& 0& 0& 0& 0& 0& 0& 0& 0& 0& 0\\ 0& 0& 0& 0& 0& 0& 0& \gamma_1& \gamma_2& \gamma_3& \gamma_4& 0& 0& 0& 0& 0& \beta_1& \beta_2& \beta_3& \beta_4& \beta_5& 0& 0& 0& 0& 0\\ 0& 0& 0& 0& 0& 0& 0& 0& \gamma_1& \gamma_2& \gamma_3& \gamma_4& 0& 0& 0& 0& 0& \beta_1& \beta_2& \beta_3& \beta_4& \beta_5& 0& 0& 0& 0\\ 0& 0& 0& 0& 0& 0& 0& 0& 0& \gamma_1& \gamma_2& \gamma_3& \gamma_4& 0& 0& 0& 0& 0& \beta_1& \beta_2& \beta_3& \beta_4& \beta_5& 0& 0& 0\\ 0& 0& 0& 0& 0& 0& 0& 0& 0& 0& \gamma_1& \gamma_2& \gamma_3& \gamma_4& 0& 0& 0& 0& 0& \beta_1& \beta_2& \beta_3& \beta_4& \beta_5& 0& 0\\ 0& 0& 0& 0& 0& 0& 0& 0& 0& 0& 0& \gamma_1& \gamma_2& \gamma_3& \gamma_4& 0& 0& 0& 0& 0& \beta_1& \beta_2& \beta_3& \beta_4& \beta_5& 0\\ 0& 0& 0& 0& 0& 0& 0& 0& 0& 0& 0& 0& \gamma_1& \gamma_2& \gamma_3& \gamma_4& 0& 0& 0& 0& 0& \beta_1& \beta_2& \beta_3& \beta_4& \beta_5\\ 0& 0& 0& 0& 0& 0& 0& 0& 0& 0& 0& 0& 0& 0& 0& 0& \gamma_1& \gamma_2& \gamma_3& \gamma_4& 0& 0& 0& 0& 0& 0\\ 0& 0& 0& 0& 0& 0& 0& 0& 0& 0& 0& 0& 0& 0& 0& 0& 0& \gamma_1& \gamma_2& \gamma_3& \gamma_4& 0& 0& 0& 0& 0\\ 0& 0& 0& 0& 0& 0& 0& 0& 0& 0& 0& 0& 0& 0& 0& 0& 0& 0& \gamma_1& \gamma_2& \gamma_3& \gamma_4& 0& 0& 0& 0\\ 0& 0& 0& 0& 0& 0& 0& 0& 0& 0& 0& 0& 0& 0& 0& 0& 0& 0& 0& \gamma_1& \gamma_2& \gamma_3& \gamma_4& 0& 0& 0\\ 0& 0& 0& 0& 0& 0& 0& 0& 0& 0& 0& 0& 0& 0& 0& 0& 0& 0& 0& 0& \gamma_1& \gamma_2& \gamma_3& \gamma_4& 0& 0\\ 0& 0& 0& 0& 0& 0& 0& 0& 0& 0& 0& 0& 0& 0& 0& 0& 0& 0& 0& 0& 0& \gamma_1& \gamma_2& \gamma_3& \gamma_4& 0\\ 0& 0& 0& 0& 0& 0& 0& 0& 0& 0& 0& 0& 0& 0& 0& 0& 0& 0& 0& 0& 0& 0& \gamma_1& \gamma_2& \gamma_3& \gamma_4
\end{array} \right). \]
Given this expression, the matrix can be plugged into one's favorite computer algebra system and a few hours
later the prefactor is found to be
\begin{equation}
  A = det(f) = 0.
\end{equation}
Thus, there exists a zero mode for all $\alpha_i$, $\beta_i$ and
$\gamma_i$ moduli which one would not have known about without
constructing the maps explicitly. This zero mode causes the prefactor
to vanish for all of these moduli. It is worth noting that this
prefactor occurred very frequently in the scan we performed in section
\ref{sec:scan} and clearly has difficult implications for moduli
stabilization with the associated instanton. Though this occurred
while studying a seven-brane dependent instanton prefactor in
F-theory, this mathematical phenomenon should occur in other
compactifications as well and one should not simply assume an $\cO(1)$
prefactor.

\vspace{.5cm}
\noindent \textbf{Acknowledgments} \\
We thank Lara Anderson, Andres Collinucci, I\~ naki Garc\'
ia-Etxebarria, James Gray, Thomas Grimm, Jonathan Heckman, Max
Kerstan, Liam McAllister, Paul McGuirk, Natalia Saulina, Sakura Sch\"
afer-Nameki, Sav Sethi, Gary Shiu, Angel Uranga, Timo Weigand, and
Cumrun Vafa for conversations regarding this work. J.H. is
particularly grateful to Denis Klevers for many long and useful
conversations and Dave Morrison for a discussion of the draft.
R.D. thanks his long time collaborator Martijn Wijnholt.  J.H. and
M.C. are supported by DOE grant DOE-EY-76-02-3071.  J.H. also
acknowledges support from NSF grant PHY11-25915 and a DOE graduate
fellowship. J.M. is supported by DOE grant DE-FG02-90ER-40560 and NSF
grant PHY-0855039.  R.D. acknowledges partial support from NSF grants
DMS-0908487 and RTG-0636606.  M.C. is also supported by the Fay R. and
Eugene L. Langberg Endowed Chair and the Slovenian Research Agency
(ARRS).

\appendix

\section{Calabi-Yau Resolutions and the Spectral Divisor}
\label{app:resolutions}

In this Appendix, we describe a crepant resolution of the 4-fold $Y_4$ \eqref{Y4def} and the $dP_9$ fibration $Y_4'$ \eqref{dp9fibration}.  Further, we provide a detailed description of the spectral divisor $\CC_F$ \eqref{specdivdef}, its relation to the cylinder \eqref{cylinderdef}, as well as the emergence of the \emph{compact} Higgs bundle spectral cover \eqref{clocF1} and its relation to the heterotic spectral cover \eqref{chet}.

\subsection{Resolving 4-folds with a surface of $E_6$ singularities}

We now proceed to consider a 4-fold $\CY_4$ that is elliptically fibered with section over a base $B_3$ and exhibits a surface of $E_6$ singularities.  We realize $\CY_4$ as a hypersurface in a $\mathbb{P}^2_{1,2,3}$ bundle $\CW_5$ over $B_3$ with weighted homogeneous coordinates $[v,x,y]$ transforming as sections of the indicated bundles
\begin{equation}\begin{array}{c|c}
\text{Section} & \text{Bundle} \\ \hline
v & \CO(\sigma) \\
x & \CO(2\sigma + 2\CD) \\
y & \CO(3\sigma + 3\CD)
\end{array}\label{firstCD}\end{equation}
where $\CD$ is some divisor class in $B_3$.  Keeping $\CD$ general will allow us to describe $\CY_4$"s that are Calabi-Yau as well as $\CY_4$'s that are not Calabi-Yau, such as the $dP_9$ fibration $Y_4'$ \eqref{dp9fibration} that emerges in the stable degeneration limit of $Y_4$ \eqref{Y4def}.  We take our 4-fold $\CY_4$ to be the hypersurface defined by
\begin{equation}y^2 = x^3 + fxz^4v^4 + gz^6v^6 + z^2v^3\left[b_0 z^3 v^3 + b_2 zvx + b_3 y\right]\label{CY4def}\end{equation}
where the new objects are sections of the indicated bundles
\begin{equation}\begin{array}{c|c}
\text{Section} & \text{Bundle} \\ \hline
z & \CO(B_2) \\
b_m & \CO([6-m]\CD - [5-m]B_2) \\
f & \CO(4(\CD-B_2)) \\
g & \CO(6(\CD-B_2))
\end{array}\end{equation}
for $B_2$ a surface in $B_3$ along which $\CY_4$ exhibits an $E_6$ singularity.  The hypersurface $\CY_4$ is in the class $6(\sigma+\CD)$ in $\CW_5$ and has first Chern class
\begin{equation}c_1(\CY_4) = c_1(B_3) - \CD\label{c1CY4}\end{equation}
As usual, a crepant resolution of the singularities along $B_2$ can be obtained by performing a series of blow-ups in $\CW_5$ and passing from $\CY_4$ to its proper transform $\tilde{\CY}_4$.  We now describe this in detail.  While this work was in progress, the resolution of elliptically fibered Calabi-Yau's with a surface of $E_6$ singularities was also studied in \cite{Kuntzler:2012bu}.

\unnumsubsubsection{Blow-up 1}
\label{app:blowup1}

The first step is to blow up $\CW_5$ along the surface $x=y=z=0$ to get the once blown-up space $\CW_5^{(1)}$.  This gives an exceptional divisor $E_1$.  The holomorphic sections $x$, $y$, and $z$ factor in $\CW_5^{(1)}$ as
\begin{equation}\begin{split}
x &= x_1\delta_1 \\
y &= y_1\delta_1 \\
z &= z_1\delta_1
\end{split}\label{blowup1}\end{equation}
for $\delta_1$ the unique holomorphic section of $\CO(E_1)$ whose vanishing defines $E_1$.  The new objects in \eqref{blowup1} are holomorphic sections of the indicated bundles
\begin{equation}\begin{array}{c|c}
\text{Section} & \text{Bundle} \\ \hline
x_1 & \CO(2[\sigma+\CD] - E_1) \\
y_1 & \CO(3[\sigma+\CD] - E_1) \\
z_1 & \CO(B_2-E_1) \\
\delta_1 & \CO(E_1)
\end{array}\end{equation}
It will be helpful in the following to say something about the geometry of $E_1$.  Before the blow-up, $x=y=z=0$ is a copy of $B_2$ is $E_1$ is the projectivization of the normal bundle of this copy of $B_2$ inside $\CW_5$.  In particular, we have
\begin{equation}\begin{split}E_1 &= \mathbb{P}\left(\CO(2[\sigma+\CD])|_{B_2}\oplus \CO(3[\sigma+\CD])|_{B_2}\oplus\CO(B_2)|_{B_2}\right) \\
&= \mathbb{P}\left(\CO(2\CD)|_{B_2}\oplus \CO(3\CD)|_{B_2}\oplus \CO(B_2)|_{B_2}\right)
\end{split}\label{E1P2fibration}\end{equation}
where we have used the fact that $\sigma|_{x=y=0}=0$.  The restrictions of $x_1$, $y_1$, and $z_1$ give homogeneous coordinates on the $\mathbb{P}^2$ fiber of $E_1$.  From the description of $E_1$ in \eqref{E1P2fibration}, we can specify the bundles on $E_1$ of which $x_1$, $y_1$, and $z_1$ are sections by
\begin{equation}\begin{array}{c|c}
\text{Section} & \text{Bundle on }E_1 \\ \hline
x_1 & \CO_{E_1}(\sigma_1+2\CD) \\ 
y_1 & \CO_{E_1}(\sigma_1+3\CD) \\ 
z_1 & \CO_{E_1}(\sigma_1+B_2) \\ 
\end{array}\label{E1sections}\end{equation}
where $\sigma_1$ is the 'new' divisor class on $E_1$ that is not pulled back from $B_2$.  Using the implied restrictions
\begin{equation}\begin{split}\left[2(\sigma+\CD)-E_1\right]|_{E_1} &= \sigma_1 + 2\CD \\
\left[3(\sigma+\CD)-E_1\right]|_{E_1} &= \sigma_1+3\CD\\
\left[B_2-E_1\right]|_{E_1} &= \sigma_1 + B_2
\end{split}\end{equation}
and the fact that $\sigma|_{E_1}=0$, we see that $\sigma_1$ is nothing more than
\begin{equation}\sigma_1 = -E_1|_{E_1}\end{equation}
As such, the normal bundle of $E_1$ in $\CW_5^{(1)}$ is given by
\begin{equation}N_{E_1/\CW_5^{(1)}} = \CO_{E_1}(-\sigma_1) \end{equation}

\unnumsubsubsection{Blow-up 2}
\label{app:blowup2}

The second step in our procedure is to blow-up along the 3-fold $y_1=\delta_1=0$ in $\CW_5^{(1)}$ to get the twice blown-up space $\CW_5^{(2)}$.  This yields an exceptional divisor $E_2$.  The holomorphic sections $y_1$ and $\delta_1$ factor in $\CW_5^{(2)}$ as
\begin{equation}\begin{split}y_1 &= y_{12}\delta_2 \\
\delta_1 &= \delta_{12}\delta_2
\end{split}\end{equation}
where the new objects are sections of the indicated bundles
\begin{equation}\begin{array}{c|c}
\text{Section} & \text{Bundle} \\ \hline
y_{12} & \CO(3[\sigma+\CD] - E_1 - E_2) \\
\delta_{12} & \CO(E_1-E_2) \\
\delta_2 & \CO(E_2)
\end{array}\end{equation}
The exceptional divisor $E_2$ is the projectivization of the normal bundle of the 3-fold $y_1=\delta_1=0$ inside $\CW_5^{(1)}$.  This 3-fold, which we call $X_3^{(2)}$, is the divisor defined by $y_1=0$ inside $E_1$ and is in the class $\sigma_1 + 3\CD$.  From \eqref{E1sections}, we see that it is the following $\mathbb{P}^1$-bundle over $B_2$
\begin{equation}X_3^{(2)} = \mathbb{P}\left(\CO_{B_2}(2\CD)\oplus \CO_{B_2}(B_2)\right)\end{equation}
with $x_1$ and $z_1$ providing homogeneous coordinates on the $\mathbb{P}^1$ fiber.  The normal bundle of $X_3^{(2)}$ inside $\CW_5^{(1)}$ is simply
\begin{equation}\begin{split}N_{X_3^{(2)}/\CW_5^{(1)}} &= N_{X_3^{(2)}/E_1}\oplus N_{E_1/\CW_5^{(1)}}|_{X_3^{(2)}} \\
&= \CO_{E_1}(\sigma_1+3\CD)|_{X_3^{(2)}}\oplus \CO_{E_1}(-\sigma_1)|_{X_3^{(2)}}
\end{split}\end{equation}
and $E_2$ is the projectivization of this
\begin{equation}E_2 = \mathbb{P}\left(\CO_{E_1}(\sigma_1+3\CD)|_{X_3^{(2)}}\oplus \CO_{E_1}(-\sigma_1)|_{X_3^{(2)}}\right)\end{equation}
The sections $y_{12},\delta_{12}$ restrict to homogeneous coordinates on the $\mathbb{P}^1$ fiber.  In total $x_1,z_1,y_{12},\delta_{12}$ restrict to sections of the following bundles on $E_2$
\begin{equation}\begin{array}{c|c}
\text{Section} & \text{Bundle on }E_2 \\ \hline
x_1 & \CO_{E_2}(\sigma_1+2\CD) \\ 
z_1 & \CO_{E_2}(\sigma_1+B_2) \\ 
y_{12} & \CO_{E_2}(\sigma_2+\sigma_1+3\CD) \\
\delta_{12} & \sigma_2-\sigma_1
\end{array}\label{E2sections}\end{equation}
where $\sigma_2$ is the 'new' divisor class associated with the blown-up $\mathbb{P}^1$ fiber.  In particular,
\begin{equation}\sigma_1 = -E_1|_{E_2}\qquad \sigma_2 = -E_2|_{E_2}\end{equation}
Since $X_3^{(2)}$ is itself a $\mathbb{P}^1$-fibration over $B_2$, $E_2$ is a $\mathbb{P}^1$-fibration over a $\mathbb{P}^1$-fibration over $B_2$ or, equivalently, an $\mathbb{F}_n$-fibration over $B_2$.  Inside the $\mathbb{F}_n$, the divisor $\sigma_1$ gives the $\mathbb{P}^1$ fiber and the divisor $\sigma_2$ gives the $\mathbb{P}^1$ base.  in fact, we can verify that $E_2$ is an $\mathbb{F}_2$ fibration since $\sigma_2^2=-2$ in the fiber.
The normal bundle of $E_2$ inside $\CW_5^{(2)}$ is simply
\begin{equation}N_{E_2/\CW_5^{(2)}}=\CO(E_2)|_{E_2} = \CO_{E_2}(-\sigma_2)\end{equation}

\unnumsubsubsection{Blow-up 3}
\label{app:blowup3}

The third step is to blow-up $\CW_5^{(2)}$ along the 3-fold $x_1=\delta_2=0$ to get the 3-times blown-up space $\CW_5^{(3)}$.  This gives an exceptional divisor $E_3$.  The sections $x_1$ and $\delta_2$ factor in $\CW_5^{(3)}$ as
\begin{equation}\begin{split}x_1 &= x_{13}\delta_3 \\
\delta_2 &= \delta_{23}\delta_3
\end{split}\end{equation}
where the new objects are sections of the indicated bundles
\begin{equation}\begin{array}{c|c}
\text{Section} & \text{Bundle} \\ \hline
x_{13} & \CO(2[\sigma+\CD]-E_1-E_3) \\
\delta_{23} & \CO(E_2-E_3) \\
\delta_3 & \CO(E_3)
\end{array}\end{equation}
The exceptional divisor $E_3$ is the projectivization of the normal bundle of the 3-fold $x_1=\delta_2=0$ inside $\CW_5^{(2)}$.  This 3-fold, which we call $X_3^{(3)}$, is the divisor defined by $x_1$ inside $E_2$ and is in the class $\sigma_1+2\CD$.  We recall that the holomorphic section $z_1$ in \eqref{E2sections} cannot vanish along $x_1=0$ in $E_2$ so $(\sigma_1+2\CD)\cdot_{E_2}(\sigma_1+B_2)=0$.  This means that $\sigma_1|_{X_3^{(3)}}=-B_2|_{X_3^{(3)}}$ and, consequently, that $X_3^{(3)}$ is the following $\mathbb{P}^1$-bundle over $B_2$
\begin{equation}X_3^{(3)} = \mathbb{P}\left(\CO_{B_2}(3\CD-B_2)\oplus \CO_{B_2}(B_2)\right)\end{equation}
with $y_{12}$ and $\delta_{12}$ providing homogeneous coordinates on the $\mathbb{P}^1$ fiber. 
The normal bundle of $X_3^{(3)}$ in $\CW_5^{(2)}$ is simply
\begin{equation}\begin{split}N_{X_3^{(3)}/\CW_5^{(2)}} &= N_{X_3^{(3)}/E_2}\oplus N_{E_2/\CW_5^{(2)}}|_{X_3^{(3)}} \\
&= \CO_{E_2}(\sigma_1+2\CD)|_{X_3^{(3)}}\oplus \CO_{E_2}(-\sigma_2)|_{X_3^{(3)}} \\
&= \CO_{E_2}(2\CD-B_2)|_{X_3^{(3)}}\oplus \CO_{E_2}(-\sigma_2)|_{X_3^{(3)}}
\end{split}\end{equation}
and $E_3$ is the projectivization of this
\begin{equation}E_3 = \mathbb{P}\left(\CO_{E_2}(2\CD-B_2)|_{X_3^{(3)}}\oplus \CO_{E_2}(-\sigma_2)|_{X_3^{(3)}}\right)
\end{equation}
The sections $x_{13}$ and $\delta_{23}$ restrict to homogeneous coordinates on the $\mathbb{P}^1$ fiber.  In total, $y_{12},\delta_{12},x_{13},\delta_{23}$ restrict to sections of the following bundles on $E_3$
\begin{equation}\begin{array}{c|c}
\text{Section} & \text{Bundle on }E_3 \\ \hline
y_{12} & \CO_{E_3}(\sigma_2 + 3\CD-B_2) \\
\delta_{12} & \CO_{E_3}(\sigma_2+B_2) \\
x_{13} & \CO_{E_3}(\sigma_3+2\CD-B_2) \\
\delta_{23} & \CO_{E_3}(\sigma_3-\sigma_2)
\end{array}\label{E3sections}\end{equation}
where $\sigma_3$ is the 'new' divisor class associated with the blown-up $\mathbb{P}^1$ fiber.  In particular,
\begin{equation}\sigma_2 = -E_2|_{E_3}\qquad \sigma_3 = -E_3|_{E_3}\end{equation}
Since $X_3^{(3)}$ is itself a $\mathbb{P}^1$-fibration over $B_2$, $E_3$ is an $\mathbb{F}_n$-fibration over $B_2$.  Inside the $\mathbb{F}_n$, the divisor $\sigma_2$ gives the $\mathbb{P}^1$ fiber class and $\sigma_3$ the $\mathbb{P}^1$ base.  The normal bundle of $E_3$ inside $\CW_5^{(3)}$ is simply
\begin{equation}N_{E_3/\CW_5^{(3)}} = \CO(E_3)|_{E_3} = \CO_{E_3}(-\sigma_3)\end{equation}

\unnumsubsubsection{Blow-up 4}
\label{app:blowup4}

The fourth step is to blow-up $\CW_5^{(3)}$ along the 3-fold $y_{12}=\delta_3=0$ to get the 4-times blown-up space $\CW_5^{(4)}$.  This gives an exceptional divisor $E_4$.  The sections $y_{12}$ and $\delta_3$ factor in $\CW_5^{(4)}$ as
\begin{equation}\begin{split}y_{12} &= y_{124}\delta_4 \\
\delta_3 &= \delta_{34}\delta_4
\end{split}\end{equation}
where the new objects are sections of the indicated bundles
\begin{equation}\begin{array}{c|c}
\text{Section} & \text{Bundle} \\ \hline
y_{124} & \CO(3[\sigma+\CD]-E_1-E_2-E_4) \\
\delta_{34} & \CO(E_3-E_4) \\
\delta_4 & \CO(E_4)
\end{array}\end{equation}
The exceptional divisor $E_4$ is the projectivization of the normal bundle of the 3-fold $y_{12}=\delta_4=0$ inside $\CW_5^{(3)}$.  This 3-fold, which we call $X_3^{(4)}$, is the divisor defined by $y_{12}=0$ inside $E_3$ and is in the class $\sigma_2+3\CD-B_2$.  We recall that the holomorphic section $\delta_{12}$ in \eqref{E3sections} cannot vanish along $y_{12}=0$ in $E_3$ so $(\sigma_2+3\CD-B_2)\cdot_{E_3}(\sigma_2+B_2)=0$.  This means that $\sigma_2|_{X_3^{(4)}}=-B_2|_{X_3^{(4)}}$ and, consequently, that $X_3^{(4)}$ is the following $\mathbb{P}^1$-bundle over $B_2$
\begin{equation}X_3^{(4)} = \mathbb{P}\left(\CO_{B_2}(2\CD-B_2)\oplus \CO_{B_2}(B_2)\right)\end{equation}
with $x_{13}$ and $\delta_{23}$ providing homogeneous coordinates on the $\mathbb{P}^1$ fiber.  The normal bundle of $X_3^{(4)}$ in $\CW_5^{(3)}$ is simply
\begin{equation}\begin{split}N_{X_3^{(4)}/\CW_5^{(3)}} &=
N_{X_3^{(4)}/E_3}\oplus N_{E_3/\CW_5^{(3)}}|_{X_3^{(4)}} \\
&= \CO(E_3)(\sigma_2+3\CD-B_2)|_{X_3^{(4)}}\oplus \CO_{E_3}(-\sigma_3)|_{X_3^{(4)}} \\
&= \CO(E_3)(3\CD-2B_2)|_{X_3^{(4)}}\oplus \CO_{E_3}(-\sigma_3)|_{X_3^{(4)}}
\end{split}\end{equation}
and $E_4$ is the projectivization of this
\begin{equation}E_4 = \mathbb{P}\left( \CO(E_3)(3\CD-2B_2)|_{X_3^{(4)}}\oplus \CO_{E_3}(-\sigma_3)|_{X_3^{(4)}}\right)\end{equation}
The sections $y_{124}$ and $\delta_{34}$ restrict to homogeneous coordinates on the $\mathbb{P}^1$ fiber.  In total, $x_{13},\delta_{23},y_{124},\delta_{34}$ restrict to sections of the following bundles on $E_4$
\begin{equation}\begin{array}{c|c}
\text{Section} & \text{Bundle on }E_4 \\ \hline
x_{13} & \CO_{E_4}(\sigma_3+2\CD-B_2) \\
\delta_{23} & \CO_{E_4}(\sigma_3+B_2) \\
y_{124} & \CO_{E_4}(\sigma_4 + 3\CD-2S_2) \\
\delta_{34} & \CO_{E_4}(\sigma_4-\sigma_3)
\end{array}\label{E4sections}\end{equation}
where $\sigma_4$ is the 'new' divisor class associated with the blown-up $\mathbb{P}^1$ fiber.  In particular,
\begin{equation}\sigma_3 = -E_3|_{E_4}\qquad \sigma_4=-E_4|_{E_4}\end{equation}
Since $X_3^{(4)}$ is itself a $\mathbb{P}^1$-fibration over $B_2$, $E_4$ is an $\mathbb{F}_n$-fibration over $B_2$.  Inside the $\mathbb{F}_n$, the divisor $\sigma_3$ gives the $\mathbb{P}^1$ fiber class and $\sigma_4$ the $\mathbb{P}^1$ base.  The normal bundle of $E_4$ inside $\CW_5^{(4)}$ is simply
\begin{equation}N_{E_4/\CW_5^{(4)}} = \CO(E_4)|_{E_4} = \CO_{E_4}(-\sigma_4)\end{equation}

\unnumsubsubsection{The last two blow-ups}

The four blow-ups that we have described so far are sufficient to resolve a 4-fold with a surface of $SU(5)$ singularities along $B_2$ \cite{Krause:2011xj}.  Further, our detailed description of the exceptional divisors $E_1$, $E_2$, $E_3$, and $E_4$ will be sufficient to give a careful description of the compact Higgs bundle spectral cover $\CC_{loc}$ in $SU(5)$, $SO(10)$, and $E_6$ models.  Actually resolving our 4-fold $\CY_4$ with $E_6$ singularities, however, requires two further blow-ups that we describe briefly.  

First, we blow-up along $\delta_{23}=\delta_{34}=0$ to get the 5-times blown-up space $\CW_5^{(5)}$.  This gives an exceptional divisor $E_5$.  The sections $\delta_{23}$ and $\delta_{34}$ factor in $\CW_5^{(5)}$ as
\begin{equation}\begin{split}\delta_{23} &= \delta_{235}\delta_5 \\
\delta_{34} &= \delta_{345}\delta_5
\end{split}\end{equation}
where the new objects are sections of the indicated bundles
\begin{equation}\begin{array}{c|c}
\text{Section} & \text{Bundle} \\ \hline
\delta_{235} & \CO(E_2-E_3-E_5) \\
\delta_{345} & \CO(E_3-E_4-E_5) \\
\delta_5 & \CO(E_5)
\end{array}\end{equation}

Finally, we blow-up along $\delta_5=\delta_{235}=0$ to get the 6-times blown-up space $\CW_5^{(6)}$.  This gives an exceptional divisor $E_6$.  The sections $\delta_5$ and $\delta_{235}$ factor in $\CW_5^{(6)}$ as
\begin{equation}\begin{split}
\delta_5 &= \delta_{56}\delta_6 \\
\delta_{235} &= \delta_{2356}\delta_6
\end{split}\end{equation}
where the new objects are sections of the indicated bundles
\begin{equation}\begin{array}{c|c}
\text{Section} & \text{Bundle} \\ \hline
\delta_{56} & \CO(E_5-E_6) \\
\delta_{2356} & \CO(E_2-E_3-E_5-E_6) \\
\delta_6 & \CO(E_6)
\end{array}\end{equation}

\unnumsubsubsection{Resolved Geometry}

At the end of our blow-ups we have the 5-fold $\CW_5^{(6)}$.  Our original sections $x$, $y$, and $z$ factor in $\CW_5^{(6)}$ according to
\begin{equation}\begin{split}
x &= x_{13}\delta_{12}\delta_{2356}\delta_{345}^2\delta_4^2\delta_{56}^3\delta_6^4 \\
y &= y_{124}\delta_{12}\delta_{2356}^2\delta_{345}^2\delta_4^3\delta_{56}^4\delta_6^6 \\
z &= z_1\delta_{12}\delta_{2356}\delta_{345}\delta_4\delta_{56}^2\delta_6^3
\end{split}\end{equation}
For convenience, we summarize all of the relevant sections and their corresponding bundles as
\begin{equation}\begin{array}{c|c}
\text{Section} & \text{Bundle} \\ \hline
x_{13} & \CO(2[\sigma+\CD]-E_1-E_2) \\
y_{124} & \CO(3[\sigma+\CD]-E_1-E_2-E_4) \\
z_1 & \CO(B_2-E_1) \\
\delta_{12} & \CO(E_1-E_2) \\
\delta_{2356} & \CO(E_2-E_3-E_5-E_6) \\
\delta_{345} & \CO(E_3-E_4-E_5) \\
\delta_4 & \CO(E_4) \\
\delta_{56} & \CO(E_5-E_6) \\
\delta_6 & \CO(E_6)
\end{array}\end{equation}
We also give the Stanley-Reisner ideal
\begin{equation}\left\{\begin{array}{c}v\delta_{12},v\delta_{2356},v\delta_{345},v\delta_4,v\delta_{56},v\delta_6,\\
\delta_{12}y_{124},\delta_{12}\delta_4,\delta_{2356}x_{13},\delta_{2356}\delta_{345},\delta_{2356}\delta_{56}, \\
\delta_{345}z_1,\delta_{345}y_{124},\delta_{345}\delta_6,\delta_4z_1,\delta_{56}x_{13},\delta_{56}y_{124},\delta_{56}z_1,\\
\delta_6x_{13},\delta_6y_{124},\delta_6z_1,z_1x_{13}y_{124}\end{array}\right\}\label{theSRideal}\end{equation}

We now turn to the proper transform of $\tilde{\CY}_4$ of $\CY_4$ \eqref{CY4def}:
\begin{equation}\begin{split}\delta_4\left(\delta_{2356}y_{124}^2-\delta_{12}\delta_{345}^2\delta_{56}x_{13}^3-\delta_{12}^3\delta_{2356}^2\delta_{345}^2\delta_{56}^3\delta_6^4 f x_{13}(z_1v)^4 - \delta_{12}^4\delta_{2356}^3\delta_{345}^2\delta_{56}^4\delta_6^6 g(z_1v)^6\right) \\
= \delta_{12}\delta_{2356}v^3z_1^2\left[b_3y_{124}+\delta_{12}\delta_{345}\delta_{56}\delta_6 b_2x_{13}(z_1v) + \delta_{12}^2\delta_{2356}\delta_{345}\delta_{56}^2\delta_6^3b_0(z_1v)^3\right]
\end{split}\label{CY4tildedef}\end{equation}
This is smooth for generic $b_m$ and is in the class
\begin{equation}\tilde{\CY}_4 = 6[\sigma+\CD] - 2E_1-E_2-E_3-E_4-E_5-E_6\end{equation}
Since $c_1(\CW_5^{(6)})=c_1(\CW_5)-2E_1-E_2-E_3-E_4-E_5-E_6$ we have that
\begin{equation}c_1(\tilde{\CY}_4) = c_1(B_3)-\CD\end{equation}
indicating that the resolution is indeed crepant.  It is a simple matter to describe the Cartan divisors of $\tilde{\CY}_4$, which are 3-folds in $\CW_5^{(6)}$
\begin{equation}\begin{array}{c|c|c}
\text{Cart Div} & \text{Class in }\tilde{\CY}_4 & \text{Equations} \\ \hline
\CD_1 & E_4 & \delta_4=0 \\
& -(E_2-E_3-E_5-E_6) & b_3y_{124}+\delta_{12}\delta_{345}\delta_{56}\delta_6(z_1v)(b_2x_{13}+b_0\delta_{12}\delta_{2356}\delta_{56}\delta_6^2(z_1v)^2) \\
& + (E_1-E_2) & \\ \hline
\CD_2 & (E_2-E_3-E_5-E_6)& \delta_{2356}=0 \\
&-(E_1-E_2) &  \delta_4=0 \\ \hline
\CD_3 & E_6 & \delta_6=0 \\
& & \delta_4(\delta_{12}\delta_{345}^2\delta_{56}x_{13}^3-\delta_{2356}y_{124}^2) + vb_3\delta_{12}\delta_{2356}y_{124}(z_1v)^2=0 \\ \hline
\CD_4 & E_5-E_6 & \delta_{56}=0 \\
& & \delta_4y_{124}-vb_3\delta_{12}(z_1v)^2=0 \\ \hline
\CD_5 & E_3-E_4-E_5 & \delta_{345}=0 \\
& & \delta_4y_{124} - vb_3\delta_{12}(z_1v)^2=0 \\ \hline
\CD_6 & E_1-E_2 & \delta_{12}=0 \\
& & \delta_{2356}=0
\end{array}\end{equation}
and of course we have the extended node
\begin{equation}\CD_0 = B_2-E_1\end{equation}

To each Cartan divisor $\CD_i$ we associate a curve $\Sigma_i$ which is the irreducible component of the singular fiber to which it corresponds.  More specifically, $\Sigma_i$ is the intersection of $\CD_i$ with $\pi^*p_0$ for a generic point $p_0$ in the section.

It is now a simple matter to compute the intersection matrix
\begin{equation}\begin{array}{c|ccccccc}
\cdot_{\tilde{\CY}_4} & \Sigma_0 & \Sigma_1 & \Sigma_2 & \Sigma_3 & \Sigma_4 & \Sigma_5 & \Sigma_6 \\
\CD_0 & -2 & 0 & 0 & 0 & 0 & 0 & 1 \\
\CD_1 & 0 & -2 & 1 & 0 & 0 & 0 & 0 \\
\CD_2 & 0 & 1 & -2 & 1 & 0 & 0 & 0 \\
\CD_3 & 0 & 0 & 1 & -2 & 1 & 0 & 1 \\
\CD_4 & 0 & 0 & 0 & 1 & -2 & 1 & 0 \\
\CD_5 & 0 & 0 & 0 & 0 & 1 & -2 & 0 \\
\CD_6 & 1 & 0 & 0 & 1 & 0 & 0 & -2
\end{array}\end{equation}
which is the intersection matrix of the extended $E_6$ Dynkin diagram as expected.

\subsection{Spectral Divisor, Heterotic Spectral Cover, and Higgs Bundle Spectral Cover}

We now describe the spectral divisor $\CC_F$ in detail and its connection to the heterotic and Higgs bundle spectral covers, $\CC_{Het}$ and $\CC_{loc}$.  Among other things, we give a clear prescription for the compact surface $\CC_{loc}$ that emerges from the intersection of $\CC_F$ with $\pi^*B_2$ and its equivalence to $\CC_{Het}$ when a heterotic dual exists.

\unnumsubsubsection{The Spectral Divisor}
\label{app:subsubsec:spectraldiv}

First, we look at the spectral divisor $\CC_F$ \eqref{specdivdef}.  In $\CY_4$, it is the irreducible component of the hypersurface
\begin{equation}y^2 = x^3 + fxz^4v^4 + g z^6v^6\label{app:specdivdef}\end{equation}
This surface is a 3-sheeted cover of $B_3$ inside $\CY_4$ that is singular at $z=0$ where the sheets come together.  The true object of interest is the proper transform $\tilde{\CC}_F$ of $\CC_F$ inside $\tilde{\CY}_4$ since it is here that we define the line bundle $\CN_F$ for constructing $C_3$ and $G_4$.  The 3-fold $\tilde{\CC}_F$ is smooth and can be most easily described as a complete intersection in $\CW_5^{(6)}$
\begin{equation}\begin{split}\delta_{2356}y_{124}^2 &= \delta_{12}\delta_{345}^2\delta_{56}\left(x_{13}^3+\delta_{12}^2\delta_{2356}^2\delta_{56}^2\delta_6^4 f x_{13}(z_1v)^4 + \delta_{12}^3\delta_{2356}^3\delta_{56}^3\delta_6^6 g(z_1v)^6\right) \\
0 &= b_3y_{124} + \delta_{12}\delta_{345}\delta_{56}\delta_6b_2x_{13}(z_1v) + \delta_{12}^2\delta_{2356}\delta_{345}\delta_{56}^2\delta_6^3 b_0 (z_1v)^3
\end{split}\label{resolvedspecdiv}\end{equation}
which is in the class
\begin{equation}\tilde{\CC}_F = (3\sigma + 6\CD - E_1-E_2-E_4-2B_2)\cdot (6\sigma + 6\CD - 2E_1-E_2-E_3-2E_4-E_5-E_6)\end{equation}
We can write the class of $\tilde{\CC}_F$ as a restriction of a divisor class in $\CW_5^{(6)}$ to $\tilde{\CY}_4$ using
\begin{equation}\begin{split}[b_3y_{124}+\ldots]\cdot [\tilde{\CY}_4] &= \tilde{\CC}_F + [b_3y_{124}+\ldots]\cdot \delta_4 \\
[\delta_4]\cdot [\tilde{\CY}_4] &= [\delta_4]\cdot [b_3y_{124}+\ldots] + [\delta_4]\cdot [\delta_{2356}] \\
[\delta_{2356}]\cdot [\tilde{\CY}_4] &= [\delta_{2356}]\cdot [\delta_4] + [\delta_{2356}]\cdot [\delta_{12}] \\
[\delta_{12}]\cdot [\tilde{\CY}_4] &= [\delta_{12}]\cdot [\delta_{2356}]
\end{split}\end{equation}
This means that
\begin{equation}\begin{split}\tilde{\CC}_F &= \left([b_3y_{124}+\ldots] - [\delta_4]+[\delta_{2356}] - [\delta_{12}]\right)\cdot [\tilde{\CY}_4] \\
&= (3\sigma + 6\CD - 2E_1+E_2-E_3-2E_4-E_5-E_6-2B_2)|_{\tilde{\CY}_4}
\end{split}\end{equation}
It is also easy to verify that $\tilde{\CC}_F$ has the expected intersections with the nodes $\Sigma_i$ of the $E_6$ singular fiber
\begin{equation}\begin{array}{c|ccccccc}
\cdot_{\tilde{\CY}_4} & \Sigma_0 & \Sigma_1 & \Sigma_2 & \Sigma_3 & \Sigma_4 & \Sigma_5 & \Sigma_6 \\ \hline
\tilde{\CC}_F & 0 & 3 & 0 & 0 & 0 & 0 & 0
\end{array}\label{CFfiber}\end{equation}
Indeed, $\tilde{\CC}_F$ is supposed to behave near $B_2$ as the cylinder of the heterotic/F-theory duality map in that it is a 3-sheeted cover of $B_2$ whose sheets each intersect the $E_6$ singular fiber according to the highest weight of the $\mathbf{27}$ representation.  The highest weight of the $\mathbf{27}$ is $(1,0,0,0,0,0)$ in our labeling conventions for the nodes $\Sigma_i$ ($i=1,\ldots,6)$ so this works out correctly.

\unnumsubsubsection{The Higgs Bundle Spectral Cover, $\CC_{loc}$}
\label{app:subsubsec:higgs}

We now turn to the emergence of the Higgs bundle spectral cover, $\CC_{loc}$, from $\CC_F$ with special attention to describing the compact surface $\CC_{loc}$ in its entirety.  By $\CC_{loc}$, we mean something very special: the restriction of $\CC_F$ to $\pi^*B_2$ or, more properly, the restriction of $\tilde{\CC}_F$ to $\pi^*B_2$.  We saw before \eqref{CFfiber} that $\tilde{\CC}_F$ intersects only one node of the $E_6$ singular fiber, namely the node $\Sigma_1$ corresponding to the Cartan divisor $\CD_1 = E_4$.  The surface $\CC_{loc}$, then, is obtained as the following intersection in $\CW_5^{(6)}$
\begin{equation}\begin{split}\CC_{loc} &= \tilde{\CC}_F\cdot E_4 \\
&= (3\sigma + 6\CD - E_1-E_2-E_4-2B_2)\cdot (6\sigma + 6\CD - 2E_1-E_2-E_3-2E_4-E_5-E_6)\cdot E_4
\end{split}\end{equation}
which we describe in equations as
\begin{equation}\CC_{loc}:\quad \left\{\begin{array}{rcl}\delta_{2356}y_{124}^2 &=& \delta_{12}\delta_{345}^2\delta_{56}\left(x_{13}^3+\delta_{12}^2\delta_{2356}^2\delta_{56}^2\delta_6^4 f x_{13}(z_1v)^4 + \delta_{12}^3\delta_{2356}^3\delta_{56}^3\delta_6^6 g(z_1v)^6\right) \\
0 &=& b_3y_{124} + \delta_{12}\delta_{345}\delta_{56}\delta_6b_2x_{13}(z_1v) + \delta_{12}^2\delta_{2356}\delta_{345}\delta_{56}^2\delta_6^3 b_0 (z_1v)^3 \\
0 &=& \delta_4
\end{array}\right.
\label{cclocdefined}\end{equation}
As written, this is a bit cumbersome to work with.  We can make our lives easier, however, by considering the proper transform $\tilde{\CC}_F^{(4)}$ of $\CC_F$ under only the first 4 blow-ups of sections \ref{app:blowup1}-\ref{app:blowup4}.  Consider then the surface
\begin{equation}\CC_{loc}^{(4)} = \tilde{\CC}_F^{(4)}\cdot_{\CW_5^{(4)}} E_4\end{equation}
given by the following complete intersection in $\CW_5^{(4)}$
\begin{equation}\CC_{loc}^{(4)}:\quad \left\{\begin{array}{rcl}
\delta_{23}y_{124}^2 &=& \delta_{12}\delta_{34}^2x_{13}^3 + \delta_{12}^3\delta_{23}^2\delta_{34}^2 f x_{13}(z_1v)^4 + \delta_{12}^4\delta_{23}^3 \delta_{34}^2 g (z_1 v)^6 \\
0 &=& b_3y_{124} + b_2\delta_{12}\delta_{34} x_{13} (z_1 v) + b_0 \delta_{12}^2\delta_{23}\delta_{34} (z_1 v)^3 \\
0 &=& \delta_4\end{array}\right.\label{ccloc4}\end{equation}
The remaining two blow-ups are along 3-folds that meet $\CC_{loc}^{(4)}$ along the curve $\delta_{23}=\delta_{34}=b_3=\delta_4=0$.  These are blow-ups in codimension 1; they have no effect on $\CC_{loc}^{(4)}$ so it is isomorphic to $\CC_{loc}$
\begin{equation}\CC_{loc}^{(4)} = \CC_{loc}\end{equation}
We prefer to work with the description \eqref{ccloc4} of $\CC_{loc}$ inside $\CW_5^{(4)}$ because it is a simple presentation of $\CC_{loc}$ as a divisor inside the exceptional divisor $E_4$ that we described in section \ref{app:blowup4}.  Recall from that discussion that $E_4$ is an $\mathbb{F}_1$ fibration with homogeneous coordinates on the base given by restrictions of $[x_{13},\delta_{23}]$ and homogeneous coordinates on the fiber given by restrictions of $[y_{124},\delta_{34}]$.  These were associated to bundles on $E_4$ as
\begin{equation}\begin{array}{c|c}
\text{Section} & \text{Bundle on }E_4 \\ \hline
x_{13} & \CO_{E_4}(\sigma_3 + 2\CD - B_2) \\
\delta_{23} & \CO_{E_4}(\sigma_3+B_2) \\
y_{124} & \CO_{E_4}(\sigma_4+3\CD-2B_2) \\
\delta_{34} & \CO_{E_4}(\sigma_4-\sigma_3)
\end{array}\end{equation}
Let us introduce new notation to make things look simpler and clarify the structure of the $\mathbb{F}_1$ fibration.  By direct computation, it is easy to verify that the curve $\delta_{34}=0$ inside the $\mathbb{F}_1$ has self-intersection -1.  Renaming the classes $\sigma_4-\sigma_3=b$ and $\sigma_3+B_2=f$ and using $[W,X]$ for homogeneous coordinates on the base and $[u,q]$ for homogeneous coordinates on the fiber, we have
\begin{equation}\begin{array}{c|c|c}\text{Section} & \text{Restriction to }E_4 & \text{Bundle on }E_4 \\ \hline
x_{13} & X & \CO(f + 2(\CD-B_2)) \\
\delta_{23} & W & \CO(f) \\
y_{124} & q & \CO(b+f+3(\CD-B_2)) \\
\delta_{34} & u & \CO(b)
\end{array}\end{equation}
Finally, since $\delta_{12}=z_1=v=1$ everywhere along $E_4$, we can rewrite the first two equations of \eqref{ccloc4}, which define $\CC_{loc}$ inside $E_4$, as
\begin{equation}\begin{split}Wq^2 &= u^2\left(X^3 + W^2Xf + W^3g\right) \\
0 &= a_3q + a_2uX + a_0uW
\end{split}\end{equation}
where we replaced the $b_m$ by $a_m$
\begin{equation}a_m = b_m|_{B_2}\end{equation}
which represent the restriction of $b_m$ to $\delta_4=0$.  Above generic points in $B_2$ the first equation defines a curve in $\mathbb{F}_1$ in the class $2b+3f$.  This is an anti-canonical curve and hence has genus 0.  We see, then, that the first equation defines an elliptic fibration $\CZ_3$ over $B_2$.  The second gives a 3-fold cover of $B_2$ inside $\CZ_3$.  We can say more about $\CZ_3$ and $\CC_{loc}$, in fact.  We have that
\begin{equation}c_1(E_4) = c_1(B_2) + 2b+3f+5(\CD-B_2)\end{equation}
while $\CZ_3$ is a hypersurface of $E_4$ in the class
\begin{equation}2b+3f+6(\CD-B_2)\end{equation}
This means that
\begin{equation}c_1(\CZ_3) = c_1(B_2) - (\CD-B_2)\end{equation}
In all of the examples that we study in the rest of this Appendix (i.e. $\CY_4$ a Calabi-Yau 4-fold or a $dP_9$-fibration), we will have that
\begin{equation}(\CD-B_2)|_{B_2}=c_1(B_2)\label{DDproperty}\end{equation}
so we make this assumption from now on.  This means that $\CZ_3$ is an elliptically fibered Calabi-Yau 3-fold
\begin{equation}c_1(\CZ_3) = 0\end{equation}
In fact, we can use the replacements
\begin{equation}W=v^2,\quad
X = x, \quad
q = y, \quad
u = v
\label{replacements}\end{equation}
to set up a map from $\CZ_3$ to a Calabi-Yau hypersurface of a $\mathbb{P}^2_{1,2,3}$-bundle $\CX_4$ over $B_2$ with weighted homogeneous coordinates $[v,x,y]$ on the fiber.  More specifically, $[v,x,y]$ are sections of the indicated bundles
\begin{equation}\begin{array}{c|c}
\text{Section} & \text{Bundle} \\ \hline
v & \sigma_{loc} \\
x & 2(\sigma_{loc}+\cs) \\
y & 3(\sigma_{loc}+\cs)
\end{array}\end{equation}
while the 3-fold $\CZ_3$ is the hypersurface
\begin{equation}\CZ_3:\qquad y^2 = x^3 + fxv^4 + gv^6\label{cz3}\end{equation}
and the final equation of \eqref{ccloc4} maps to
\begin{equation}0=a_3y + a_2vx + a_0v^3\label{clocinz3}\end{equation}
This is the conventional form of an $SU(3)$ spectral cover inside $\CZ_3$.  We recognize this as a specific compactification of the non-compact spectral cover of local models.  The cover $\CC_{loc}$ can be viewed as a divisor inside the Calabi-Yau 3-fold $\CZ_3$ in the class
\begin{equation}\CC_{loc} = 3\sigma_{loc} + \eta_{loc}\qquad \eta_{loc} = 6\cs +B_2\end{equation}
where as usual we do not distinguish between divisors in the base, $B_2$, and their pullbacks to $\CZ_3$.

The realization \eqref{clocinz3} of $\CC_{loc}$ as a hypersurface inside the auxiliary elliptically fibered Calabi-Yau 3-fold $\CZ_3$ \eqref{cz3} is very useful for calculations.  Let us describe how restrictions of divisors on $\tilde{\CY}_4$ appear in this setting.  Following through our studies of $E_4$ in section \ref{app:blowup4}, it is easy to see that
\begin{equation}E_1|_{E_4} = E_2|_{E_4} = B_2\quad\implies E_1|_{\CC_{loc}}=E_2|_{\CC_{loc}}=B_2\label{E1E2dictionary}\end{equation}
The map \eqref{replacements} now tells us that
\begin{equation}E_3|_{E_4}=-\left(2\sigma_{loc}-B_2\right)\qquad E_4|_{E_4}=-\left(3\sigma_{loc}-B_2\right)\label{E3E4dictionary}\end{equation}
As a quick check, we can use this dictionary to compute $c_1(\CC_{loc})$
\begin{equation}\begin{split}c_1(\CC_{loc}) &= c_1(\CW_5^{(6)})|_{\CC_{loc}} \\
&\quad - \left[\left(3\sigma + 6\CD-E_1-E_2-E_4-2B_2\right)+\left(6\sigma+6\CD-2E_1-E_2-E_3-2E_4-E_5-E_6\right)+\left(E_4\right)\right]|_{\CC_{loc}} \\
&= -\left[3\sigma + 7\CD-\cb-E_1-E_2-E_4-2B_2\right]|_{\CC_{loc}} \\
&= -\left(3\sigma_{loc}+\eta_{loc}\right)|_{\CC_{loc}}
\end{split}\label{c1cloc}\end{equation}
where we have expressed $c_1(\CC_{loc})$ as the restriction of divisor classes in $\CZ_3$ in the last line using \eqref{E1E2dictionary} \eqref{E3E4dictionary} along with the trivial statement $\sigma|_{E_4}=0$ and the property \eqref{DDproperty} of $\CD$.  This result is exactly what we expect because it is simply $c_1(\CC_{loc}) = -[\CC_{loc}]|_{\CC_{loc}}$ where $[\CC_{loc}]$ is the class of $\CC_{loc}$ as a divisor inside the Calabi-Yau 3-fold $\CZ_3$.

The relations \eqref{E3E4dictionary} are particularly useful for constructing divisors $\gamma_F$ on $\tilde{\CC}_F$ that `extend' divisors on $\CC_{loc}$ for building $G$-flux.  Consider, for instance, the standard inherited local model bundle
\begin{equation}\CN_{loc} = \CO_{\CC_{loc}}(\lambda\gamma_{loc}+r_{loc}/2)\label{Nlocinherited}\end{equation}
where $r_{loc}$ is the ramification divisor of the covering $p_{loc}:\CC_{loc}\rightarrow B_2$ and $\gamma_{loc}$ is the divisor class
\begin{equation}\gamma_{loc} = 3\sigma_{loc} - (\eta_{loc}-3\cs) = 3\sigma_{loc} - (3\cs-t)\label{gammaloc}\end{equation}
Here $t$ is the usual shorthand notation for divisor associated to the inverse normal bundle
\begin{equation}\CO_{B_2}(-t) = N_{B_2/B_3} = \CO_{B_3}(B_2)|_{B_2}\end{equation}
When we construct the $G$-flux corresponding to $\CN_{loc}$, part of our task is to find a divisor $\gamma_F$ that satisfies
\begin{equation}\gamma_F|_{\CC_{loc}}=\gamma_{loc}\end{equation}
We can do this by noting that
\begin{equation}\eta-3\cs = \left[3\cb - 4B_2\right]|_{B_2}\end{equation}
and using \eqref{E3E4dictionary} to make the replacement
\begin{equation}\sigma_{loc}\rightarrow E_3-E_4\end{equation}
In this way, we arrive at
\begin{equation}\gamma_F = \left[3(E_3-E_4)-(3\cb-2B_2)\right]|_{\tilde{\CC}_F}\label{gammaFdef}\end{equation}
which is an important building block for the $C_3/G_4$ configuration corresponding to the local model data \eqref{Nlocinherited}.

We close this section by looking at another useful quantity, namely the normal bundle of $\CC_{loc}$ inside $\tilde{\CC}_F$
\begin{equation}N_{\CC_{loc}/\tilde{\CC}_F} = \CO(E_4)|_{\CC_{loc}} = \CO_{\CZ_3}(B_2-3\sigma_{loc})|_{\CC_{loc}}\label{nClocCF}\end{equation}
This allows us to relate the ramification divisor $r_{loc}$ of the covering $p_{loc}:\CC_{loc}\rightarrow B_2$ to the restriction of the ramification divisor $r_F$ of $p_F:\tilde{\CC}_F\rightarrow B_3$
\begin{equation}\begin{split}r_F|_{\CC_{loc}} &= \left[p_F^*\cb - c_1(\tilde{\CC}_F)\right]_{\CC_{loc}} \\
&= p_{loc}^*\left(\cs + B_2\right) - c_1(\CC_{loc})-N_{\CC_{loc}/\tilde{\CC}_F} \\
&= r_{loc}+3\sigma_{loc}
\end{split}\end{equation}
In light of this, let us define a bundle $\CL_{\hat{r}}$ on $\tilde{\CC}_F$ by
\begin{equation}\CL_{\hat{r}} = \CO_{\tilde{\CC}_F}(r_F)\otimes \CO(-3[E_3-E_4])|_{\CC_{loc}}\label{Lrhatdef}\end{equation}
We have that
\begin{equation}\CL_{\hat{r}}|_{\CC_{loc}} = \CO(r_{loc})\end{equation}
which will be helpful in our later discussion of flux quantization.

\unnumsubsubsection{$\CY_4$ a Calabi-Yau and $G$-flux Quantization}
\label{app:case1CY}

We now look at a few special choices for the class $\CD$ in \eqref{firstCD}.  To apply the results of this Appendix to $\CY_4$'s that are Calabi-Yau, we simply set
\begin{equation}\CD = \cb\end{equation}
Our object $\tilde{\CC}_F$ \eqref{resolvedspecdiv} is a generic spectral divisor of the Calabi-Yau 4-fold $\tilde{Y}_4$ and its intersection with $E_4$ yields a compactification of the local model Higgs bundle spectral cover obtained by embedding the total space of $K_{S_2}$ into the auxiliary Calabi-Yau 3-fold $\CZ_3$.  Note that there is some ambiguity in the choice of $\tilde{\CC}_F$ because we could absorb part or all of the $f$ and $g$ terms in \eqref{CY4tildedef} into the $b_m$'s.  If we absorb them completely, effectively setting $f$ and $g$ to zero in \eqref{CY4tildedef}, then we obtain the analog of the 'Tate divisor' of \cite{Dolan:2011iu,Marsano:2011nn,Marsano:2012yc} which can extend a factorization structure of the Higgs bundle spectral cover.  We do not consider split spectral covers here so there is no difference between the analog of the 'Tate divisor' or any other spectral divisor.

We devote the rest of this subsection to the $G$-flux quantization rule \eqref{quantrule} \cite{Witten:1996md}
\begin{equation}G_4 + \frac{1}{2}c_2(\tilde{\CY}_4)\in H^{2,2}(\tilde{Y}_4,\mathbb{Z})\end{equation}
Because of the related quantization rule in the local model that follows from \eqref{Nlocgammaloc}
\begin{equation}\lambda\gamma_{loc}+\frac{r_{loc}}{2}\in H_2(\CC_{loc},\mathbb{Z})\label{localquantrule}\end{equation}
and the general expectation that $\lambda\gamma_{loc}$ maps to $G$-flux, we expect that the odd part of $c_2(\tilde{Y}_4)$ is captured by $\iota_{F*}c_1(\CL_{\hat{r}})$ for some line bundle $\CL_{\hat{r}}$ on $\tilde{\CC}_F$ such that $\CL_{\hat{r}}|_{\CC_{loc}}=\CO_{\CC_{loc}}(r_{loc})$.  We have already found such a bundle \eqref{Lrhatdef}
\begin{equation}\CL_{\hat{r}}=\CO_{\tilde{\CC}_F}(r_F)\otimes \CO(-3[E_3-E_4])|_{\tilde{\CC}_F}\qquad \CL_{\hat{r}}|_{\CC_{loc}} = \CO_{\CC_{loc}}(r_{loc})\end{equation}
One might conjecture, then, that
\begin{equation}c_2(\tilde{Y}_4) + \iota_{F*}c_1(\CL_{\hat{r}})\quad\text{ is an even class?}\label{isiteven}\end{equation}
It turns out that this is almost true but not quite.  To investigate further, we must compute $c_2(\tilde{Y}_4)$ explicitly.  The computation is straightforward \cite{Marsano:2011hv} but messy and can be complicated to interpret because the result is expressed as a linear combination of surfaces of the form
\begin{equation}E_i\cdot_{\tilde{Y}_4}E_j,\qquad E_k\cdot_{\tilde{Y}_4}\cb,\qquad E_k\cdot_{\tilde{Y}_4}B_2\label{surfacecollection}\end{equation}
which, as in \cite{Marsano:2011hv}, is very redundant.
We can use the intersection relations that follow from \eqref{theSRideal} to relate several elements of \eqref{surfacecollection} to one another.  Further, there are additional intersection relations in $\tilde{Y}_4$ that are not the restriction of relations in the ambient 5-fold.  We phrase these as another collection of monomials as in \eqref{theSRideal} corresponding to holomorphic sections that do not simultaneously vanish in $\tilde{Y}_4$
\begin{equation}\{\delta_{12}x_{13},\delta_{12}\delta_{345},\delta_{12}\delta_{56},z_1y_{124},z_1x_{13}\}\label{newrelations}\end{equation}
The fact that $z_1=y_{124}=0$ and $z_1=x_{13}=0$ have no solutions in $\tilde{\CY}_4$ means that
\begin{equation}(E_1-2E_2+3E_3-2E_4)\cdot_{\tilde{\CY}_4}(B_2-E_1)=0\end{equation}
Using the relations from \eqref{theSRideal} and \eqref{newrelations}, then, we get a set 19 relations that can be used to relate 19 of the 21 surfaces of the form
\begin{equation}E_i\cdot_{\tilde{\CY}_4}E_j\end{equation}
to surfaces of the form
\begin{equation}E_i\cdot_{\tilde{Y}_4}\cb,\quad E_i\cdot_{\tilde{Y}_4}B_2,\quad E_1\cdot_{\tilde{Y}_4}E_2,\quad E_3\cdot_{\tilde{Y}_4}E_4\end{equation}
The 19 equations that we choose to apply are as follows:
\begin{equation}\begin{split}
E_1^2 &= B_2(E_1-2E_2)+2E_1E_2 \\
E_2^2 &= B_2(E_1-2E_2)+3\cb(E_2-E_1)+2E_1E_2 \\
E_3^2 &= B_2(E_1-2E_2)+ \cb(-3E_1+E_2+2E_3) + 2E_1E_2 \\
E_4^2 &= B_2(E_1-2E_2+2E_3-2E_4) + 3\cb(-E_1+E_2-E_3+E_4) + E_1E_2 + E_3E_4 \\
E_5^2 &= B_2(2E_1-4E_2-E_3+E_4+3E_5) + \cb(-6E_1+4E_2+2E_3-3E_5) + 3E_1E_2-E_3E_4 \\
E_6^2 &= B_2(2E_1-4E_2-E_3+E_4+E_5+2E_6) + \cb(-6E_1+4E_2+2E_3-E_5-2E_6) + 3E_1E_2-E_3E_4 \\
E_1E_3 &= B_2E_3 \\
E_1E_4 &= B_2E_4 \\
E_1E_5 &= B_2E_5 \\
E_1E_6 &= B_2E_6 \\
E_2E_3 &= B_2(E_1-2E_2+E_3) + 3\cb(E_2-E_1) + E_1E_2 \\
E_2E_4 &= B_2E_4 \\
E_2E_5 &= B_2(E_1-2E_2+E_5)+3\cb(E_2-E_1)+E_1E_2 \\
E_2E_6 &= B_2(E_1-2E_2+E_6) + 3\cb(E_2-E_1) + E_1E_2 \\
E_3E_5 &= -B_2E_5 + 2\cb E_5 \\
E_3E_6 &= -B_2E_6 + 2\cb E_6 \\
E_4E_5 &= B_2(-E_1+2E_2-2E_5) + 3\cb(E_1-E_2+E_5) -E_1E_2 \\
E_4 E_6 &= B_2(-E_1+2E_2-2E_6) + 3\cb(E_1-E_2+E_6) - E_1E_2 \\
E_5 E_6 &= B_2(E_1-2E_2+E_6)+\cb(-3E_1+3E_2-E_6) + E_1E_2
\end{split}\end{equation}
where all of the intersections here are taken in $\tilde{Y}_4$.  With these results, we can write $c_2(\tilde{Y}_4)$ as
\begin{equation}\begin{split}c_2(\tilde{Y}_4) &= c_2(Y_4) - 3E_1E_2 + 3E_3E_4 - \cb(4E_1+7E_3+6E_4+2E_6) \\
& + B_2(2E_2+3E_3+E_4-E_5+E_6)\\
&= E_1E_2+E_3E_4+\cb E_3+B_2(E_3+E_4+E_5+E_6)+\text{even}\end{split}\end{equation}
This is to be compared with
\begin{equation}\begin{split}\iota_{F*}c_1(\CL_{\hat{r}}) &= \CC_F\cdot_{\tilde{Y}_4}\left(\cb+\tilde{\CC}_F-3(E_3-E_4)\right) \\
&= E_1E_2+E_3E_4+\cb(E_2+E_3)+B_2(E_3+E_4+E_5+E_6)+\text{even}
\end{split}\end{equation}
This means that \eqref{isiteven} must be modified to
\begin{equation}c_2(\tilde{Y}_4) + c_1(\hat{\CL}_{\hat{r}}) + G_0\text{ is an even class}\label{G0twist}\end{equation}
where $G_0$ is a 'Cartan correction term' of the form
\begin{equation}G_0 = c_1\cdot_{\tilde{Y}_4}E_2\label{cartcorrectionterm}\end{equation}

The presence of the Cartan correction term in \eqref{G0twist} is actually very important for proper functioning of the local-to-global dictionary.  Let us return to the local model bundle \eqref{Nlocinherited} with $\gamma_{loc}$ as in \eqref{gammaloc}.  We found a divisor $\gamma_F$ \eqref{gammaFdef} in $\tilde{CC}_{F}$
\begin{equation}\gamma_F = \left[3(E_3-E_4)-(3\cb-2B_2)\right]|_{\tilde{\CC}_F}\end{equation}
with $\gamma_F|_{\CC_{loc}}=\gamma_{loc}$.  This gives a holomorphic surface inside $\tilde{Y}_4$ whose Poincare dual specifies part of the $G$-flux.  We must supplement this by two types of corrections, though.  First, we must add horizontal and vertical surfaces (that is surfaces containing the elliptic fiber or sitting in the section of $\tilde{Y}_4$) to ensure that the resulting $G$-flux satisfies the 'one leg on the fiber' condition.  Second, we have to add a Cartan flux so that the $G$-flux does not break $E_6$.  We illustrate these corrections on the second and third lines of the following
\begin{equation}\begin{split}\gamma_{loc}\rightarrow \CG &= \left[3(E_3-E_4)-(3\cb-2B_2)\right]\cdot_{\tilde{Y}_4}\tilde{\CC}_F \\
&\quad + (3\cb - 2B_2)\cdot_{\tilde{Y}_4}(3\sigma + 6\cb - 2B_2) \\
&\quad + E_2\cdot_{\tilde{Y}_4}(6S_2-9\cb)
\end{split}\label{inheritedGflux}\end{equation}
In the end, it is not $\gamma_{loc}$ that appears in $\CN_{loc}$ but a multiple $\lambda\gamma_{loc}$ with the integrality or non-integrality of $\lambda$ fixed by the quantization condition \eqref{localquantrule}.  Similarly, the $G$-flux is $\lambda\CG$ where we expect
\begin{equation}\lambda\CG + \frac{1}{2}c_2(\tilde{Y}_4)\in H^{2,2}(\tilde{Y}_4,\mathbb{Z})\end{equation}
As is familiar for this type of flux, the local model result is that $\lambda$ must be an odd half-integer.  For $\CG$ to be properly quantized, then, it must be that the odd parts of $\CG$ and $c_2(\tilde{Y}_4)$ agree.  Because the first two lines of \eqref{inheritedGflux} were constructed directly from $\gamma_{loc}$ we expect that the odd part of those lines is the same as the odd part of $\iota_{F*}c_1(\CL_{\hat{r}})$.  Indeed, it is easy to check that
\begin{equation}\begin{split}\left[3(E_3-E_4)-(3\cb-2B_2)\right]\cdot_{\tilde{Y}_4}\tilde{\CC}_F &+ (3\cb - 2B_2)\cdot_{\tilde{Y}_4}(3\sigma + 6\cb - 2B_2)  \\
& + \tilde{\CC}_F\cdot_{\tilde{Y}_4} c_1(\hat{\CL}_{\hat{r}})\text{ is an even class}
\end{split}\end{equation}
If we did not have to add the Cartan flux on the third line of \eqref{inheritedGflux} then $\CG$ would be properly quantized under the assumption \eqref{isiteven} that $\iota_{F*}c_1(\CL_{\hat{r}})$ captures the full odd contribution to $c_2(\tilde{Y}_4)$.  There is a Cartan correction \eqref{cartcorrectionterm} in the odd part of $c_2(\tilde{Y}_4)$ \eqref{G0twist}, though.  This exactly matches the odd part of the third line in \eqref{inheritedGflux} that we had to add in order to make $\CG$ truly $E_6$-preserving.

We should now summarize the construction of $G$-fluxes using the spectral divisor.  We start with a line bundle $\CN_F$ on $\tilde{\CC}_F$.  Let $A_F$ be the connection on this bundle and let $A_{\hat{r}}$ be the connection on $\CL_{\hat{r}}$ \eqref{Lrhatdef}.  We then let
\begin{equation}C_3 = \iota_{F*}\left(A_F - \frac{1}{2}A_{\hat{r}}\right) + \frac{1}{2}C_{3,0}\quad\leftrightarrow\quad G_4\sim \iota_{F*}\left(c_1(\CN_F)-\frac{1}{2}\CL_{\hat{r}}\right) + \frac{1}{2}G_{4,0}\label{Gfluxmap}\end{equation}
where
\begin{equation}G_{4,0} = \cb\cdot_{\tilde{Y}_4}E_2\label{cartancorrectionterm}\end{equation}
and $C_{3,0}$ is a 3-form with $dC_{3,0}=G_{4,0}$.  We could let $A_{K^{-1}}$ be a connection on the anti-canonical bundle of $B_3$ and take $C_{3,0}=A_{K^{-1}}\wedge c_1(\CO(E_2))$ where we implicitly mean that $A_{K^{-1}}$ should be pulled back to $\tilde{Y}_4$.

There is no guarantee that this procedure will yield an $E_6$-preserving $G$-flux.  In these cases, one can try to restore $E_6$ by adding suitable integrally quantized Cartan fluxes.  The $G$-flux in \eqref{inheritedGflux} is an example of this.

\unnumsubsubsection{$\CY_4$ a $dP_9$ fibration and the Cylinder}

This is the case of interest for the stable degeneration limit of section \ref{subsec:hetf}.  Here, we take $B_3$ to be a $\mathbb{P}^1$-fibration over $B_2$ obtained by projectivizing a line bundle $N$ \eqref{B3type}
\begin{equation}B_3 = \mathbb{P}(\CO\oplus N)\end{equation}
Following the main text, we use the notation $s$ for the divisor class of a section of this $\mathbb{P}^1$ bundle and let $[Z,W]$ denote homogeneous coordinates on the fiber
\begin{equation}\begin{array}{c|c}
\text{Section} & \text{Bundle} \\ \hline
Z & \CO(s) \\
W & \CO(s+t)
\end{array}\end{equation}
with $t$ a divisor on $B_2$.  With this notation, the surface of $E_6$ singularities is denoted by $s$ rather than $B_2$ to distinguish it from the other section, $s+t$, of $B_3$ which is also a copy of $B_2$.  We recall that
\begin{equation}\cb|_{B_2} = \cs + 2s + t\end{equation}
We can obtain the $dP_9$-fibration $Y_4'$ \eqref{dp9fibration} from the general form \eqref{CY4def} by choosing
\begin{equation}\CD = s+\cs\end{equation}
so that our initial sections are associated to the bundles
\begin{equation}\begin{array}{c|c}
\text{Section} & \text{Bundle} \\ \hline
v & \CO(\sigma) \\
x & \CO(2[\sigma+s+\cs]) \\
y& \CO(3[\sigma+s+\cs]) \\
b_m & \CO(s+[6-m]\cs) \\
f & \CO(4\cs) \\
g & \CO(6\cs)
\end{array}\end{equation}
We further write
\begin{equation}b_m = a_m W\label{bmWfactor}\end{equation}
with
\begin{equation}\begin{array}{c|c}
\text{Section} & \text{Bundle} \\ \hline
a_0 & 6\cs-t= \eta \\
a_2 & 4\cs-t = \eta-2\cs \\
a_3 & 3\cs-t = \eta - 3\cs
\end{array}\end{equation}
where we have defined
\begin{equation}\eta\equiv 6\cs-t\end{equation}
With this notation, $\CY_4$ \eqref{CY4def} is the $dP_9$-fibration \eqref{dp9fibration} from the text
\begin{equation}y^2 = x^3 + fxZ^4 v^4 + gZ^6v^6 + Z^2Wv^3\left[a_0Z^3v^3+a_2Zvx+a_3y\right]\end{equation}
The restriction to $W=0$ yields the heterotic 3-fold
\begin{equation}\CZ_{Het}:\,\,\, y^2 = x^3 + fx + g\end{equation}

Our construction of the spectral divisor $\CC_F$ goes through almost as in section \ref{app:subsubsec:spectraldiv}.  We just have to make one subtle change.  Because we chose the $b_m$'s to have $W=0$ as a common factor \eqref{bmWfactor}, we must remove this common factor from the second equation in \eqref{resolvedspecdiv}.  The resulting 3-fold is the proper transform $\tilde{\CC}_{cyl}$ of the cylinder \eqref{cylinderdef}
\begin{equation}\tilde{\CC}_{cyl}: \qquad \left\{\begin{array}{rcl}0&=&\delta_{2356}y_{124}^2 - \delta_{12}\delta_{345}^2\delta_{56}x_{13}^3 - \delta_{12}^3\delta_{2356}^2\delta_{345}^2\delta_{56}^3\delta_6^4 f x_{13} (Z_1v)^4 - \delta_{12}^4\delta_{2356}^3\delta_{345}^2\delta_{56}^4\delta_6^6 g (Z_1v)^6 \\
0 &=& a_3y_{124} + a_2\delta_{12}\delta_{345}\delta_{56}\delta_6 x_{13}(Z_1 v) + a_0\delta_{12}^2\delta_{2356}\delta_{345}\delta_{56}^2\delta_6^3 (Z_1 v)^3 
\end{array}\right.\label{thespecdiv}\end{equation}
which is the complete intersection
\begin{equation}\begin{split}\tilde{\CC}_{cyl} &= \left(3\sigma + (6(\cs-t) - E_1-E_2-E_4+3s\right)\\
&\qquad\cdot \left(6[\sigma + \cs+s] - 2E_1-E_2-E_3-2E_4-E_5-E_6\right)
\end{split}\end{equation}
in $\CW_5^{(6)}$.  The restriction of $\tilde{\CC}_{cyl}$ to $\CZ_{Het}$ is the heterotic spectral cover
\begin{equation}\CC_{Het}:\,\,\,a_0v^3 + a_2vx + a_3y=0\label{app:chet}\end{equation}
In fact $\tilde{\CC}_{cyl}$ is a $\mathbb{P}^1$-fibration over $\CC_{Het}$ with $[W,\delta_4]$ providing homogeneous coordinates on the $\mathbb{P}^1$ fiber.  As usual for such a vibration, $\tilde{\CC}_{cyl}$ admits two sections $W=0$ and $\delta_4=0$.  The first is $\CC_{Het}$ at $W=0$ while we know from section \ref{app:subsubsec:higgs} that the second, $\delta_4=0$, is the Higgs bundle spectral cover $\CC_{loc}$ \eqref{cclocdefined}.  Consistency requires that $\CC_{loc}$ be isomorphic to $\CC_{Het}$ and indeed this is easy to see from the discussion of section \ref{app:subsubsec:higgs}.  There, we showed that $\CC_{loc}$ could be described as a surface inside an elliptically fibered Calabi-Yau 3-fold $\CZ_3$.  The identification \eqref{cz3} implies that
\begin{equation}\CZ_3=\CZ_{Het}\end{equation}
and the defining equation of $\CC_{loc}$ inside $\CZ_3$ \eqref{clocinz3} is identical to \eqref{app:chet}.  We will continue to use the notation $\CC_{Het}$ for the section of $\tilde{\CC}_{cyl}$ at $W=0$ and $\CC_{loc}$ for the section at $\delta_4=0$.

We can describe the geometry of $\tilde{\CC}_{cyl}$ in even more detail by providing the normal bundles of $\CC_{loc}$ and $\CC_{Het}$ in $\tilde{\CC}_{cyl}$.  The normal bundle of $\CC_{loc}$ is captured by our computation in section \ref{app:subsubsec:higgs}
\begin{equation}N_{\CC_{loc}/\tilde{\CC}_{cyl}} = \CO_{\CZ_3}(-t-3\sigma_{loc})|_{\CC_{loc}}\end{equation}
where we have replaced $B_2$ in \eqref{nClocCF} with $-t= s|_s$.  The normal bundle of $\CC_{Het}$ is just the restriction of the normal bundle of $\CZ_{Het}$ inside $\tilde{\CY}_4$.  Recalling that $\CZ_{Het}$ is just given by $W=0$, this is the bundle pulled back from $\CO_{B_3}(s+t)|_{s+t} = \CO_{B_3}(t)|_{s+t}$
\begin{equation}N_{\CC_{Het}/\tilde{\CC}_{cyl}} = \CO_{\CZ_{Het}}(t)|_{\CC_{Het}}\end{equation}
In the end, then, $\CC_{cyl}$ is the following $\mathbb{P}^1$-fibration with base $\CC_{Het}$
\begin{equation}\CC_{cyl} = \mathbb{P}_{\CC_{Het}}\left(\CO_{\CC_{Het}}(t)\oplus \CO_{\CC_{Het}}(-t-3\sigma_{Het})\right)\end{equation}

Let us turn now to ramification divisor and the bundle $\CL_{\hat{r}}$ of \eqref{Lrhatdef}.  Since the normal bundle of $\CC_{Het}$ in $\tilde{\CC}_{cyl}$ is pulled back from $B_3$, the ramification divisor $r_{Het}$ of the covering $p_{Het}:\CC_{Het}\rightarrow B_2$ is the restriction of the ramification divisor $r_{cyl}$ of $p_{cyl}:\tilde{\CC}_{cyl}\rightarrow B_3$ to $\CC_{Het}$
\begin{equation}\begin{split}r_F|_{\CC_{Het}} &= \left[p_{cyl}^*\cb - c_1(\tilde{\CC}_{cyl})\right]_{\CC_{Het}} \\
&= p_{Het}^*\left(\cs+[W=0]\right)-c_1(\CC_{Het})-N_{\CC_{Het}/\tilde{\CC}_{cyl}} \\
&= r_{Het}
\end{split}\end{equation}
Because $\CO(-3[E_3-E_4])|_{W=0}=0$ we also have that the bundle $\CL_{\hat{r}}$ from \eqref{Lrhatdef} restricts to $\CO_{\CC_{Het}}(r_{Het})$
\begin{equation}\CL_{\hat{r}}|_{\CC_{Het}} = \CO_{\CC_{Het}}(r_{Het})\end{equation}
while the correction term \eqref{cartancorrectionterm} restricts trivially{\footnote{The twist of $\CL_{\hat{r}}$ relative to the ramification divisor of $\CC_{cyl}$ \eqref{Lrhatdef} reflects the fact that the normal bundle of $\CC_{loc}$ inside $\CC_{cyl}$, or equivalently $\CC_F$, is not pulled back from $B_2$.  Had the normal bundle been a simple pullback from $B_2$, the ramification divisor of $\CC_{cyl}$ would have been a simple pullback of the ramification divisor of $\CC_{loc}$.  Instead $\CL_{\hat{r}}$, which is the pullback of $\CO_{\CC_{loc}}(r_{loc})$, or equivalently $\CO_{\CC_{Het}}(r_{Het})$, to $\CC_{cyl}$, is twisted relative to $\CO_{\CC_{cyl}}(r_{cyl})$ by the contribution to $N_{\CC_{loc}/\CC_{cyl}}$ that is not pulled back from $B_2$.}}.  This is important because we still have that $\CL_{\hat{r}}$ restricts to $\CO_{\CC_{loc}}(r_{loc})$
\begin{equation}\CL_{\hat{r}}|_{\CC_{loc}} = \CO_{\CC_{loc}}(r_{loc})\end{equation}
Because $\CL_{\hat{r}}$ restricts to the same bundle $\CO_{\CC_{Het}}(r_{Het})=\CO_{\CC_{loc}}(r_{loc})$ in both sections, $\CC_{Het}$ and $\CC_{loc}$, it must be the pullback
\begin{equation}\CL_{\hat{r}} = p_{cyl}^*\CO_{\CC_{Het}}(r_{Het})\end{equation}

We can now write the map \eqref{Gfluxmap} from spectral data to $C_3/G_4$ as follows
\begin{equation}C_3\sim \iota_{cyl*}p_{cyl}^*\left(A_{cyl}-\frac{1}{2}A_{\hat{r}}\right)+\frac{1}{2}C_{3,0}\quad\leftrightarrow\quad G_4\sim \iota_{cyl*}p_{cyl}^*\left(c_1(\CN_{Het}) - \frac{1}{2}c_1(\CL_{\hat{r}})\right) + \frac{1}{2}G_{4,0}\label{cylindermapdetail}\end{equation}
Apart from the Cartan correction terms \eqref{cartancorrectionterm}, this is the usual cylinder map in which $p_{cyl}$ is used to pull bundle data back from $\CC_{Het}$ to $\CC_{cyl}$ and the embedding map $\iota_{cyl}$ is used to push this forward into $\tilde{Y}_4'$.  It is easy to verify that the Poincare dual holomorphic surface to $G_4$ is nothing other than
\begin{equation}c_1(\CN_{Het})-\frac{1}{2}r_{Het} = \lambda\gamma_{Het}\end{equation}
with $\gamma_{Het}$ as in \eqref{Nhetdef}.  In general, we will need to add additional Cartan fluxes to \eqref{cylindermapdetail} in order to ensure that $E_6$ is truly unbroken on the F-theory side as it is on the heterotic side.

\section{Details of the Base in Example Two}
\label{app:B3}

We now construct an interesting 3-manifold $B_3$ to serve as the base of our F-theory compactifications by performing a sequence of two blow-ups on $\mathbb{P}^3$.  Let us denote the homogeneous coordinates of our $\mathbb{P}^3$ by $w_1,w_2,w_3,w_4$ and consider the two $\mathbb{P}^1$'s given by
\begin{equation}\mathbb{P}^1_1:\,\,w_1=w_2=0\qquad \mathbb{P}^1_2:\,\,w_2=w_3=0\end{equation}
We will blow-up $\mathbb{P}^3$ by successively blowing up along
$\mathbb{P}^1_1$ and the proper transform of $\mathbb{P}^1_2$.  In the
following, we describe the space $B_3^{(0)}$ that results from the
first blow-up and our final 3-manifold $B_3$ that results from the
second blow-up.  Both blow-ups are along toric subvarieties of the
toric space $\mathbb{P}^3$ so we can describe the full geometry in the
language of toric geometry.  Indeed we will do this but because this
3-fold is so simple we would like to take some care to describe both
its properties and the nature of its toric divisors.  All of the toric
divisors $D_i$ in $B_3$ are del Pezzo surfaces or Hirzebruch surfaces
and we provide a complete description including a complete enumeration
of the curve classes in each $D_i$ to which the divisors $D_j$
descend.  Throughout we will use standard notation for the curve
classes of $dP_n$ and $\mathbb{F}_k$ surfaces.  In the case of $dP_n$,
we will use $h$ for the hyperplane and $e_i$ with $i=1,\ldots,n$ for
the exceptional curves.  For $\mathbb{F}_k$ surfaces, we will use $b$
for the base with $b^2=-k$ and $f$ for the fiber with $f^2=0$ and
$f\cdot b=1$.

\unnumsubsection{The first blown-up space $B_3^{(0)}$}

The first step is to blow up along the $\mathbb{P}^1_1$ to obtain the first blown-up space $B_3^{(0)}$.
\begin{equation}\begin{split}w_1 &= \tilde{w}_1\delta_1 \\
w_2 &= \tilde{w}_2\delta_1\end{split}\end{equation}
where $\delta_1$ is the unique holomorphic section of $\CO(E_1)$ whose vanishing defines the exceptional divisor $E_1$.  The resulting 3-manifold is toric and the toric divisors are in the classes $H$, $H-E_1$, and $E_1$.  We investigate the properties of these toric divisors in turn.

\unnumsubsubsection{$E_1$}

As the intersection of two hyperplanes in $\mathbb{P}^3$, the normal bundle of $\mathbb{P}^1_1$ is $\CO(1,1)$.  This means that the exceptional divisor $E_1$ that we get by blowing up along $\mathbb{P}^1_1$ is $\mathbb{F}_0=\mathbb{P}^1\times\mathbb{P}^1$.  What can we say about the restrictions of divisors from $B_3^{(0)}$ to $E_1$?  Well we know that $\tilde{w}_1=0$ restricts to a copy of the 'base' $b$ of $\mathbb{F}_0$ while $w_3=0$ restricts to a copy of the fiber.  This means that
\begin{equation}\begin{split}\CO(H-E_1)|_{E_1} &= \CO_{\mathbb{F}_0}(b) \\
\CO(H)|_{E_1} &= \CO_{\mathbb{F}_0}(f)\end{split}\end{equation}
and hence that
\begin{equation}\CO(E_1)|_{E_1} = \CO_{\mathbb{F}_0}(f-b)\end{equation}

\unnumsubsubsection{$H-E_1$}

This divisor is the proper transform of a hyperplane that contains the curve $\mathbb{P}^1_1$ along which we perform the blow-up.  Because the curve is of codimension 1 in this hyperplane, this proper transform is again a $\mathbb{P}^2$.  The restriction of the divisor $H$ is just the line $h$ of this $\mathbb{P}^1$ while the exceptional divisor $E_1$ also restricts to a line in the class $h$.  As a result, we have
\begin{equation}\begin{split}\CO(H)|_{H-E_1} &= \CO_{\mathbb{P}^2}(h) \\
\CO(E_1)|_{H-E_1} &= \CO_{\mathbb{P}^2}(h)
\end{split}\end{equation}

\unnumsubsubsection{$H$}

This divisor is a hyperplane that does not contain the curve $\mathbb{P}^1_1$ along which we perform the blow-up.  From the perspective of a hyperplane like this, say $w_4=0$, our blow-up is just the blow-up of the single point at $w_1=w_2=0$.  As a result, $H$ is a $dP_1$ surface.  The restriction of $E_1$ is the exceptional curve $e_1$ of the $dP_1$.  The restriction of $H-E_1$ is the proper transform of a line inside $w_4=0$ that contains the point $w_1=w_2=0$.  This is the curve $h-e_1$ of the $dP_1$.  Finally, the restriction of $H$ is clearly the hyperplane $h$ so we have
\begin{equation}\begin{split}\CO(E_1)|_{H} &= \CO_{dP_1}(e_1) \\
\CO(H-E_1)|_{H} &= \CO_{dP_1}(h-e_1) \\
\CO(H)|_{H} &= \CO_{dP_1}(h)
\end{split}\end{equation}

\unnumsubsection{The second blown-up space $B_3$}

Now we proceed to perform the second blow-up.  This is along the line $\tilde{w}_2=w_3=0$ which we denote by $\mathbb{P}^1_2$
\begin{equation}\mathbb{P}^1_2:\,\,\,\tilde{w}_2=w_3=0\end{equation}
After this blow-up, the original homogeneous coordinates can be written as
\begin{equation}\begin{split}w_1 &= \tilde{w}_1\delta_1 \\
w_2 &= \hat{w}_2\delta_1\delta_2 \\
w_3 &= \hat{w}_3\delta_2 \\
w_4 &= w_4
\end{split}\end{equation}
where the objects on the right hand side are sections of the indicated bundles
\begin{equation}\begin{array}{c|c}
\text{Section} & \text{Bundle} \\ \hline
\tilde{w}_1 & \CO(H-E_1) \\
\hat{w}_2 & \CO(H-E_1-E_2) \\
\hat{w}_3 & \CO(H-E_2) \\
w_4 & \CO(H) \\
\delta_1 & \CO(E_1) \\
\delta_2 & \CO(E_2)
\end{array}\end{equation}
Because the blow-ups are toric we can specify $B_3$ by toric data.  The matrix of GLSM charges is
\begin{equation}\begin{pmatrix}\tilde{w}_1 \\ \hat{w}_2 \\ \hat{w}_3 \\ w_4 \\ \delta_1 \\ \delta_2\end{pmatrix}\leftrightarrow\left(\begin{array}{ccc}1 & -1 & 0 \\
1 & -1 & -1 \\
1 & 0 & -1 \\
1 & 0 & 0 \\
0 & 1 & 0 \\
0 & 0 & 1
\end{array}\right)\leftrightarrow\begin{pmatrix}H-E_1 \\ H-E_1-E_2 \\ H-E_2 \\ H \\ E_1 \\ E_2\end{pmatrix}
\end{equation}
where we also indicate our labels for the divisor classes and our notation for the toric coordinates.  The Stanley-Reisner Ideal follows from the blow-up data
\begin{equation}SR\sim \{\tilde{w}_1\hat{w}_2,\tilde{w}_1\delta_2,\hat{w}_2\hat{w}_3,w_4\delta_1\delta_2,w_4\hat{w}_3\delta_1\}\end{equation}
We now describe all of the toric divisors in turn.

\unnumsubsubsection{$E_2$}

The line $\tilde{w}_2=w_3=0$ is in the class $(H-E_1)\cdot H$ in $B_3^{(0)}$.  In $B_3^{(0)}$, it is in the class $h-e_1$ inside the $dP_1$ surface $H$ and in the class $h$ in the $\mathbb{P}^2$ given by $H-E_1$.  As a result, its normal bundle in $B_3^{(0)}$ is $\CO(0,1)$.  The exceptional divisor that we get by blowing up $B_3^{(0)}$ along this line is therefore an $\mathbb{F}_1$ surface.  Both $\delta_1=0$ and $w_4=0$ restrict to copies of the fiber so we have $E_1|_{E_2}=f$ and $H|_{E_2}=f$.  The two sections of the $\mathbb{F}_1$ are $\tilde{w}_2=0$ and $\tilde{w}_3=0$ which are the restrictions of $H-E_1-E_2$ and $H-E_2$.  It is easy to verify that $(H-E_1-E_2)^2E_2=-1$ and $(H-E_2)^2E_2=1$ so that $(H-E_1-E_2)|_{E_2}=b$ and $(H-E_2)|_{E_2} = f+b$.  We can summarize this by saying
\begin{equation}\begin{split}\CO(H)|_{E_2} &= \CO_{\mathbb{F}_1}(f) \\
\CO(E_1)|_{E_2} &= \CO_{\mathbb{F}_1}(f) \\
\CO(E_2)|_{E_2} &= \CO_{\mathbb{F}_1}(-b)
\end{split}\end{equation}

\unnumsubsubsection{$E_1$}

Before the second blow-up, the exceptional divisor $E_1$ of $B_3^{(0)}$ is a $\mathbb{F}_0=\mathbb{P}^1\times\mathbb{P}^1$.  The line $\mathbb{P}^1_2$ meets $E_1$ in exactly one point so after the blow-up $E_1$ is the blow-up of $\mathbb{F}_0$ at a point, which yields a $dP_2$ surface.  Let us recall how classes of $\mathbb{F}_0$ and the exceptional curve $e$ of such a blow-up are related to the standard parametrization of curves inside $dP_2$.  Curves in the class $f$ or $b$ that do not pass through the blown-up point map to the class $h-e_1$ and $h-e_2$ of the $dP_2$ surface while the exceptional curve $e$ maps to the line $h-e_1-e_2$ of the $dP_2$
\begin{equation}\begin{split}b\rightarrow h-e_1 \\
f\rightarrow h-e_2 \\
e\rightarrow h-e_1-e_2
\end{split}\end{equation}
Now, the restriction of $H$ and $H-E_1$ to $E_1$ yield curves in the classes $f$ and $b$ that miss the blown-up point while the restriction of $E_2$ yields the exceptional curve $e$.  This leads to
\begin{equation}\begin{split}\CO(H)|_{E_1} &= \CO_{dP_2}(h-e_2) \\
\CO(E_1)|_{E_1} &= \CO_{dP_2}(e_1-e_2) \\
\CO(E_2)|_{E_2} &= \CO_{dP_2}(h-e_1-e_2)
\end{split}\end{equation}

\unnumsubsection{$H$}

Let us now turn to various toric divisors that descend from hyperplanes of the original $\mathbb{P}^3$.  We start with $H$ which is the class of the divisor $w_4=0$.  From the perspective of the hyperplane $w_4=0$, our first step blows up one point and our second step blows up another one.  As a result, $H$ is a $dP_2$ surface and the restrictions of the classes $H$, $E_1$, and $E_2$ are obvious
\begin{equation}\begin{split}\CO(H)|_H &= \CO_{dP_2}(h) \\
\CO(E_1)|_H &= \CO_{dP_2}(e_1) \\
\CO(E_2)|_H &= \CO_{dP_2}(e_2)
\end{split}\end{equation}

\unnumsubsection{$H-E_1$}

Before the second blow-up, $H-E_1$ was just a $\mathbb{P}^2$.  The second blow-up effectively blows up a single point of this $\mathbb{P}^2$ resulting in a $dP_1$ surface.  The restrictions of $H$ and $E_1$, which were hyperplanes of the $\mathbb{P}^2$ before the second blow-up, are hyperplanes $h$ of the $dP_1$ while the restoration of $E_2$ is the exceptional line
\begin{equation}\begin{split}
\CO(H)|_{H-E_1} &= \CO_{dP_1}(h) \\
\CO(E_1)|_{H-E_1} &= \CO_{dP_1}(h) \\
\CO(E_2)|_{H-E_1}) &= \CO_{dP_1}(e)
\end{split}\end{equation}

\unnumsubsection{$H-E_2$}

This divisor class is represented by $\hat{w}_3=0$ and is the proper transform of the $dP_1$ surface $w_3=0$ of $B_3^{(0)}$ that contains the curve $\mathbb{P}^1_2$ of the second blow-up.  Because $\mathbb{P}^1_2$ is of codimension 1 inside $w_3=0$, this proper transform is again a $dP_1$ surface and the restriction of the exceptional line is just a copy of $\mathbb{P}^1_2$ inside this $dP_1$, which is a hyperplane of the class $h$.  This means that
\begin{equation}\begin{split}\CO(H)_{H-E_2} &= \CO_{dP_1}(h) \\
\CO(E_1)|_{H-E_2} &= \CO_{dP_1}(e) \\
\CO(E_2)|_{H-E_2} &= \CO_{dP_1}(h)
\end{split}\end{equation}

\unnumsubsection{$H-E_1-E_2$}

This divisor class is represented by $\hat{w}_2=0$ and is the proper transform of $\mathbb{P}^2$ given by $\tilde{w}_2=0$ inside $B_3^{(0)}$.  This $\mathbb{P}^2$ contains the curve $\mathbb{P}^1_2$ so the proper transform is again a $\mathbb{P}^2$ and all of $H$, $E_1$, and $E_2$ restrict to lines
\begin{equation}\begin{split}
\CO(H)|_{H-E_1-E_2}=\CO_{\mathbb{P}^2}(h) \\
\CO(H)|_{H-E_1-E_2}=\CO_{\mathbb{P}^2}(h) \\
\CO(H)|_{H-E_1-E_2}=\CO_{\mathbb{P}^2}(h)
\end{split}\end{equation}

\unnumsubsection{Summary}

We summarize the toric divisors of our final $B_3$ and their properties in the following table
\begin{equation}\begin{array}{cc|ccc|c}
\text{Divisor Class, } \CD& \text{Surface Type} & \CO(H)|_{\CD} & \CO(E_1)|_{\CD} & \CO(E_2)|_{\CD} & N_{\CD/B_3} \\ \hline
H & dP_2 & \CO_{dP_2}(h) & \CO_{dP_2}(e_1) & \CO_{dP_2}(e_2) & \CO_{dP_2}(h) \\ \hline
H-E_1 & dP_1 & \CO_{dP_1}(h) & \CO_{dP_1}(h) & \CO_{dP_1}(e_1) & \CO_{dP_1} \\ \hline
H-E_2 & dP_1 & \CO_{dP_1}(h) & \CO_{dP_1}(e_1) & \CO_{dP_1}(h) & \CO_{dP_1} \\ \hline
H-E_1-E_2 & \mathbb{P}^2 & \CO_{\mathbb{P}^2}(h) & \CO_{\mathbb{P}^2}(h) & \CO_{\mathbb{P}^2}(h) & \CO_{\mathbb{P}^2}(-h) \\ \hline
E_1 & dP_2 & \CO_{dP_2}(h-e_2) & \CO_{dP_2}(e_1-e_2) & \CO_{dP_2}(h-e_1-e_2) & \CO_{dP_2}(e_1-e_2) \\ \hline
E_2 & \mathbb{F}_1 & \CO_{\mathbb{F}_1}(f) & \CO_{\mathbb{F}_1}(f) & \CO_{\mathbb{F}_1}(-b) & \CO_{\mathbb{F}_1}(-b) \\ \hline
\end{array}\end{equation}

\bibliographystyle{utphys}
\bibliography{refs}

\end{document}